\documentclass{article}

\usepackage{microtype}
\usepackage{graphicx}
\usepackage{booktabs} 
\usepackage{thmtools}
\usepackage{thm-restate}
\usepackage{enumitem,kantlipsum}

\usepackage{hyperref}

\usepackage[accepted]{icml2024}

\usepackage{amsmath}
\usepackage{amssymb}
\usepackage{mathtools}
\usepackage{amsthm}

\usepackage{dsfont}
\usepackage{subcaption}
\usepackage{makecell}
\usepackage{multirow} 

\newcommand\norm[1]{\left\lVert#1\right\rVert}

\DeclareMathOperator\supp{supp}
\newcolumntype{P}[1]{>{\centering\arraybackslash}p{#1}}

\usepackage[capitalize,noabbrev]{cleveref}

\theoremstyle{plain}
\newtheorem{theorem}{Theorem}[section]

\newtheorem{lemma}[theorem]{Lemma}

\theoremstyle{definition}
\newtheorem{definition}[theorem]{Definition}

\theoremstyle{remark}

\usepackage[textsize=tiny]{todonotes}

\icmltitlerunning{The Perception-Robustness Tradeoff in Deterministic Image Restoration}

\begin{document}

\twocolumn[
\icmltitle{The Perception-Robustness Tradeoff in Deterministic Image Restoration}



\icmlsetsymbol{equal}{*}

\begin{icmlauthorlist}
\icmlauthor{Guy Ohayon}{techcs}
\icmlauthor{Tomer Michaeli}{techee}
\icmlauthor{Michael Elad}{techcs}
\end{icmlauthorlist}

\icmlaffiliation{techcs}{Faculty of Computer Science, Technion, Haifa, Israel}
\icmlaffiliation{techee}{Faculty of Electrical and Computer Engineering, Technion, Haifa, Israel}

\icmlcorrespondingauthor{Guy Ohayon}{ohayonguy@campus.technion.ac.il}

\icmlkeywords{Machine Learning, ICML, Computer vision, Image restoration, Image reconstruction, Image restoration theory, Perceptual quality, Tradeoff with perceptual quality, Consistency, Consistency with the measurements, Faithfulness to the measurements, Adversarial attacks, Robustness, Stability, Fundamental tradeoffs, Posterior sampling, Deterministic restoration algorithms, Theory,Real-world image super-resolution}

\vskip 0.3in
]



\printAffiliationsAndNotice{} 

\begin{abstract}
We study the behavior of deterministic methods for solving inverse problems in imaging. These methods are commonly designed to achieve two goals: (1) attaining high perceptual quality, and (2) generating reconstructions that are consistent with the measurements. We provide a rigorous proof that the better a predictor satisfies these two requirements, the larger its Lipschitz constant must be, regardless of the nature of the degradation involved. In particular, to approach perfect perceptual quality and perfect consistency, the Lipschitz constant of the model must grow to infinity. This implies that such methods are necessarily more susceptible to adversarial attacks.
We demonstrate our theory on single image super-resolution algorithms, addressing both noisy and noiseless settings.
We also show how this undesired behavior can be leveraged to explore the posterior distribution, thereby allowing the deterministic model to imitate stochastic methods.
\end{abstract}    
\section{Introduction}
Inverse problems arise in numerous imaging applications. Indeed, the images we use on our phones, in medical diagnosis devices, and in remote sensing systems, are all created by algorithms that reconstruct and enhance raw, low-quality observations. 
Examples include image denoising, super-resolution, magnetic resonance imaging (MRI) reconstruction, and any image-to-image translation task.

Denoting a natural image as a random vector $X$, the degraded measurement as $Y$, and their joint distribution as $p_{X,Y}$, a deterministic image restoration algorithm\footnote{An algorithm that always produces the same output per input.} $\hat{X}$ (an estimator) often strives to generate outputs of high perceptual quality that are also consistent with the input measurements.
Perceptual quality is commonly measured by the extent to which human observers cannot distinguish between the estimator's outputs and natural images~\citep{pix2pix2017,pixel-recursive,NIPS2015_aa169b49,46167,10.1145/2897824.2925974,NIPS2016_8a3363ab,zhang2016colorful,zhang2017real}.
This can be quantified by the statistical distance between the distributions $p_{\hat{X}}$~and~$p_{X}$~\citep{Blau2018}.
Of course, perceptual quality by itself  does not necessarily indicate the usefulness of an algorithm.
For instance, an estimator that completely ignores the input and outputs samples from $p_{X}$ attains perfect perceptual quality ($p_{\hat{X}}=p_{X}$), but is obviously useless. Indeed, a good estimator is one whose outputs are also consistent with the measurements.
\begin{figure}
    \centering
    \includegraphics{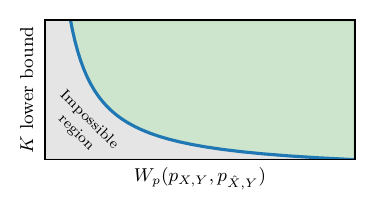}
    \caption{Qualitative illustration of~\cref{theorem:erratic_behavior}. 
    In any restoration task with a non-invertible degradation (\cref{definition:non-invertible-degradation}), the Lipschitz constant $K$ of a deterministic estimator $\hat{X}=f(Y)$ is lower bounded by a function that grows to infinity as the Wasserstein distance between $p_{\hat{X},Y}$ and $p_{X,Y}$ decreases to zero.
    }
    \label{fig:teaser}
\end{figure}

Consistency with the measurements is best understood in the case where the degradation is deterministic, such that $Y=D(X)$.
In this case, $\hat{X}$ is perfectly consistent if $D(\hat{X})=Y$, and as a measure, consistency is typically quantified by the discrepancy between \mbox{$D(\hat{X})$ and $Y$}~\cite{Lugmayr_2021_CVPR,Lugmayr_2022_CVPR}.
As with perceptual quality, consistency by itself is not a sufficient indication that an algorithm is useful. Indeed, the fact that the outputs are consistent with the measurements does not guarantee that they look natural. 
Moreover, consistency becomes a questionable concept in the setting of stochastic degradations: What can be regarded as a consistent reconstruction when many different measurements can emerge from the same source image?

To circumvent these complexities and address all types of degradations, we integrate perceptual quality and consistency into a single concept.
Specifically, we refer to the \emph{joint perceptual quality} of an estimator as the extent to which human observers cannot distinguish between pairs of restored outputs and degraded inputs, and pairs of natural images and degraded measurements.
This can be quantified by the statistical distance between $p_{\hat{X},Y}$ and $p_{X,Y}$.
In~\cref{section:consistency} we explain why attaining perfect perceptual quality and perfect consistency is equivalent to attaining perfect joint perceptual quality, making it a useful stand-alone measure.

In this paper we rigorously prove that, whenever the degradation is not invertible\footnote{A degradation is invertible when $X$ can be reconstructed from $Y$ with zero error. See~\cref{definition:non-invertible-degradation}.}, the Lipschitz constant of a deterministic estimator is bounded from below by a factor inversely proportional to the Wasserstein distance between $p_{\hat{X},Y}$ and $p_{X,Y}$ (see~\cref{fig:teaser} for a qualitative illustration).
This means that the Lipschitz constant must grow to infinity as the Wasserstein distance between $p_{\hat{X},Y}$ and $p_{X,Y}$ approaches zero.
An immediate practical implication of this result is that the higher the joint perceptual quality of an estimator (i.e., the better its perceptual quality and consistency), the more susceptible it is to input adversarial attacks, which is a clear disadvantage. 
On the positive side, we demonstrate that this behavior can be leveraged to explore the posterior distribution by slightly perturbing its input.

It is important to note that studying deterministic estimators of this type is not a pursuit of mere theoretical interest.
In fact, estimators that satisfy $p_{\hat{X},Y}\approx p_{X,Y}$\footnote{By $p_{\hat{X},Y}\approx p_{X,Y}$ we mean that the statistical distance (e.g., Wasserstein) between $p_{\hat{X},Y}$ and $p_{X,Y}$ is small.} appear frequently in the literature and are being used across many types of inverse problems, e.g., in tasks where the estimator is a Conditional Generative Adversarial Network (CGAN)~\cite{MirzaO14,pix2pix2017,park2019SPADE,Shaham_2021_CVPR,wang2018pix2pixHD,ma2018speckle,ohayon2021high,Man_2023_CVPR,yi2018sharpness,joung2020,sun2023deep,Agustsson_2023_CVPR,islam2019fast}.

Our work is closely related to a previous study by~\citet{Ohayon2023}. 
Here, we generalize and extend the results in~\citep{Ohayon2023} by considering any type of degradation, by replacing intuitive claims with a formal proof, and by demonstrating the predicted behaviors on a slew of estimators (see~\cref{subsection:det-deg} for a thorough discussion). To conclude, our main contributions are the following:
\begin{enumerate}[leftmargin=*,topsep=0pt]
\item We rigorously prove that any deterministic estimator $\hat{X}$ that satisfies $p_{\hat{X},Y}\approx p_{X,Y}$ must possess a high Lipschitz constant. Specifically, we show that the Lipschitz constant of $\hat{X}$ is bounded from below by a factor inversely proportional to the Wasserstein distance between \mbox{$p_{\hat{X},Y}$ and $p_{X,Y}$~(\cref{theorem:erratic_behavior})}.
\item Through experiments on popular deterministic image super-resolution methods~(\cref{section:real-world-experiments}), we demonstrate that a lower statistical distance between $p_{\hat{X},Y}$ and $p_{X,Y}$ indicates \emph{worse} average robustness to adversarial attacks.
Our measure of average robustness is effectively a \emph{lower bound} of the Lipschitz constant of an estimator. These experiments thereby validate~\cref{theorem:erratic_behavior}.
\item To demonstrate the practical relevance of~\cref{theorem:erratic_behavior}, we show in~\cref{section:adv_attacks} how an adversary can exploit the instability of a deterministic method (with high joint perceptual quality) to alter the results of an image decision-making pipeline. In~\cref{section:posterior-sampling-imitation} we also demonstrate how the erratic behavior of this estimator can practically be leveraged to explore the posterior distribution.
\end{enumerate}
The paper is organized as follows.
In~\cref{section:motivation} we provide background and motivation, explore the requirement $p_{\hat{X},Y}\approx p_{X,Y}$, and draw its connection to the literature.
In~\cref{section:problem_setting} we formulate the image restoration task and set mathematical notations.
In~\cref{section:erratic_behavior} we present our main result (\cref{theorem:erratic_behavior}) and illustrate its practical implications.
In~\cref{section:summary} we summarize our paper.
Finally, in~\cref{section:limitations} we discuss the limitations of this work and propose ideas for future research.
\section{Motivation and Related Works}\label{section:motivation}
\subsection{Deterministic Degradations}\label{subsection:det-deg}
In the setting of deterministic degradations, a fundamental property desired from an image restoration algorithm $\hat{X}$ is that its outputs be perfectly consistent with its inputs, i.e., $D(\hat{X})=Y$.
As~\citet{Ohayon2023} explain, this condition is equivalent to $\smash{p_{Y|\hat{X}}=p_{Y|X}}$.
Such a property stems from the definition of solving an inverse problem~\citep{kirsch1996introduction} and is quite intuitive, as it implies that the estimated solution $\hat{X}$ is a valid explanation for the measurements.
Another highly desired property of image restoration algorithms is to attain perfect perceptual quality, so that humans could not distinguish between the restored outputs and natural images.
This property is equivalent to attaining $p_{\hat{X}}=p_{X}$~\citep{Blau2018}.

Can a deterministic estimator satisfy $p_{Y|\hat{X}}=p_{Y|X}$ (perfect consistency) and $p_{\hat{X}}=p_{X}$ (perfect perceptual quality) simultaneously?
By making a simple use Bayes' rule,~\citet{Ohayon2023} showed that the answer is negative.
An estimator $\hat{X}$ satisfies these two conditions if and only if it is a posterior sampler, one that satisfies  $p_{\hat{X}|Y}=p_{X|Y}$ (equivalently, $p_{\hat{X},Y}=p_{X,Y}$), and thus, a deterministic estimator cannot concurrently attain both of these properties (only a stochastic sampler from the posterior can).
Since both of these properties are desired, it is therefore important to ask what would be the consequences if a deterministic estimator still tries to satisfy them.
As \citet{Ohayon2023} empirically demonstrate, a deterministic estimator $\hat{X}$ that satisfies $\smash{p_{\hat{X}}\approx p_{X}}$ (high perceptual quality) and $\smash{p_{Y|\hat{X}}=p_{Y|X}}$ should be erratic\footnote{By erratic we mean that the Lipschitz constant of $\hat{X}$ is high.} and sensitive to input adversarial attacks.

While considering estimators that produce perfectly consistent outputs is important, it is often practically sufficient that the distance between $D(\hat{X})$ and $Y$ would just be very small~\citep{Lugmayr_2021_CVPR,Lugmayr_2022_CVPR}, in which case $p_{Y|X}$ is not precisely equal to $p_{Y|\hat{X}}$.
Indeed, image restoration algorithms with only near-perfect consistency are much more prevalent in practice (e.g.,~\citep{Ledig_2017_CVPR,Wang_2018_ECCV_Workshops,Lugmayr_2021_CVPR,Lugmayr_2022_CVPR}).
Does the aforementioned sensitivity issue holds for deterministic restoration algorithms that strive for high perceptual quality and only \emph{near-perfect} consistency?
Instead of breaking down this question into separate discussions on perceptual quality and consistency, we can more generally explore the behavior of deterministic estimators that satisfy $p_{\hat{X},Y}\approx p_{X,Y}$.
This includes estimators with high perceptual quality producing outputs with either perfect or near-perfect consistency.
As we show in~\cref{theorem:erratic_behavior} below, estimators that satisfy $p_{\hat{X},Y}\approx p_{X,Y}$ must indeed behave erratically.
\subsection{Stochastic Degradations}
While deterministic degradations are prevalent in academic discussions, real-world degradations often involve a stochastic component (noise, random motion blur, etc.).
Our~\cref{theorem:erratic_behavior} holds for any form of $p_{Y|X}$ (including stochastic measurements). Therefore, not only do we confirm the result of~\citet{Ohayon2023} with a rigorous proof, but we also extend it to any type of degradation.

One may wonder why we are interested in deterministic estimators that satisfy \mbox{$p_{\hat{X},Y}\approx p_{X,Y}$} in the case of stochastic degradations. 
Our motivation to explore such estimators stems from a practical observation: Aiming for this objective, whether implicitly or explicitly, is a recurring trend in the literature when using deterministic estimators in stochastic degradation settings, as we describe next.
\subsection{CGAN Image Restoration Algorithms}
CGANs have become highly popular for solving image inverse problems, such as image-to-image translation tasks~\citep{pix2pix2017,wang2018pix2pixHD,park2019SPADE,Shaham_2021_CVPR}, image denoising~\citep{yi2018sharpness,ma2018speckle,joung2020,ohayon2021high,wang2021,chen2021generative,sun2023deep}, motion deblurring~\citep{Kupyn_2018_CVPR,Kupyn_2019_ICCV,Zhang_2020_CVPR}, JPEG decoding~\citep{Man_2023_CVPR,Agustsson_2023_CVPR}, and more.
Their popularity can be partly attributed to the fact that they do not require a problem-specific loss, a model of the prior distribution, nor any assumptions about the relation between $X$ and $Y$. 
To train a CGAN, one only needs a dataset of paired i.i.d.~samples from $p_{X,Y}$.

In the image restoration literature, the term CGAN is commonly used to describe models that are trained with two different types of optimization frameworks.
Some models use this term because the discriminator receives either pairs of natural images and their degraded measurements (``real'' examples), or pairs of estimated outputs and the corresponding degraded inputs (``fake'' examples)~\cite{pix2pix2017,park2019SPADE,Shaham_2021_CVPR,wang2018pix2pixHD,ma2018speckle,ohayon2021high,Man_2023_CVPR,yi2018sharpness,joung2020,sun2023deep,Agustsson_2023_CVPR,islam2019fast}.
This approach follows from the original definition of a CGAN~\cite{MirzaO14}.
Theoretically, the optimal solution of a parametric model trained solely with such an adversarial discriminator is a stochastic estimator $\hat{X}$ which satisfies $p_{\hat{X},Y}=p_{X,Y}$.

Other models use the term CGAN simply because the estimator is regarded as a conditional ``generator'' (conditioned on the degraded measurement), trained together with an adversarial discriminator which does not receive the degraded measurements as an additional input~\cite{Kupyn_2018_CVPR,Kupyn_2019_ICCV,wang2018esrgan,wang2021realesrgan,Zhang_2020_CVPR,CycleGAN2017,liang2021swinir,chen2022femasr,Umer_2020_CVPR_Workshops,zhang2021designing}.
Although these models do not explicitly aim to attain $p_{\hat{X},Y}=p_{X,Y}$, in~\cref{section:real-world-experiments} we demonstrate that they still attain a relatively small statistical distance between $p_{\hat{X},Y}$ and $p_{X,Y}$. Therefore, according to our~\cref{theorem:erratic_behavior} they must possess a high Lipschitz constant, which we indeed demonstrate through experiments. It is important to note that the means by which a small distance between $p_{\hat{X},Y}$ and $p_{X,Y}$ is obtained is irrelevant. Any deterministic estimator with high joint perceptual quality must be erratic.

Lastly, it is important to note that many CGAN-based image restoration algorithms make use of a deterministic estimator (even though the optimal solution is a stochastic one).
One reason is that, in practice, the estimator often becomes nearly deterministic anyway, mapping each input to (almost) a single output ~\citep{Mathieu_2016,pix2pix2017,ohayon2021high,park2019SPADE,dsganICLR2019}.
Thus, many implementations simply omit the use of input random seed, anticipating that the estimator will likely ignore it.
Another reason to omit the use of such a random seed is that in some applications it is not desirable to obtain a stochastic estimator, but rather to provide a point estimate which simply ``corresponds'' to the input measurement.
See~\cref{appendix:desire_for_deterministic_estimators} for the motivation to use deterministic estimators.

We further address other related works in~\cref{appendix:related-work}, e.g., studies that discuss a tradeoff between robustness and distortion for deterministic restoration algorithms.
\section{Problem Formulation}\label{section:problem_setting}
We adopt the Bayesian perspective for solving inverse problems~\citep{davison_2003,kaipio_2005,Blau2018}, where a natural image $x\in\mathbb{R}^{n_{x}}$ is regarded as a realization of a random vector $X$ with probability density function $p_{X}$.
The degraded measurement $y\in\mathbb{R}^{n_{y}}$ is related to $x$ via the conditional density $p_{Y|X}$.
We are dealing with non-invertible degradations, where $x$ cannot be retrieved from $y$ with zero-error.
Formally:
\begin{definition}\label{definition:non-invertible-degradation}
A degradation is invertible if $p_{X|Y}(\cdot|y)$ is a Dirac delta function for almost every $y\in\supp{p_{Y}}$.
\end{definition}
The task of an estimator $\hat{X}$ is to estimate $X$ from $Y$, i.e., $X\rightarrow Y\rightarrow \hat{X}$ is a Markov chain ($X$ and $\hat{X}$ are statistically independent given $Y$).
We say that $\hat{X}$ is a deterministic estimator if it can be written as a function of $Y$, i.e. $\hat{X}=f(Y)$.
This means that $p_{\hat{X}|Y}(\cdot|y)$ is a Dirac delta function for any~$y$.
If this does not hold, then the algorithm is stochastic.

We say that the degradation is deterministic if $p_{Y|X}(\cdot|x)$ is a Dirac delta function for almost any $x$, and otherwise it is a stochastic degradation.
In this paper we consider any form of $p_{Y|X}$, including both deterministic degradations (e.g., JPEG compression, down-sampling) and stochastic degradations (e.g., additive noise, random blur, color-jitter).

\subsection{Lipschitz Continuity}
We define the robustness of a deterministic estimator \mbox{$\hat{X}=f(Y)$} using the well known notion of Lipschitz continuity.
Formally, $\hat{X}$ is said to be Lipschitz continuous \mbox{(or $K$-Lipschitz)} if there exists $K\geq 0$ such that
\begin{gather}
\norm{f\left(y_1\right)-f\left(y_2\right)}\leq K\norm{y_1-y_2}
\end{gather}
for every $y_1,y_2\in\supp{p_{Y}}$, and we denote $\text{Lip}(\hat{X})\coloneqq K$.
When saying that a deterministic estimator $\hat{X}$ is robust, it is meant that $\text{Lip}(\hat{X})$ is small. Similarly, when a deterministic estimator is said to be erratic, it means that $\text{Lip}(\hat{X})$ is very large (it may be $\infty$).

\subsection{Joint Perceptual Quality}
We focus on the case where statistical distances between distributions are measured via the Wasserstein distance defined on a norm-induced metric space.
The Wasserstein distance between $p_{\hat{X},Y}$ and $p_{X,Y}$ is defined by
\begin{align}
    &W_{p}(p_{X,Y},p_{\hat{X},Y}) \nonumber\\
    &=\inf_{p_{X_{1},Y_{1},X_{2},Y_{2}}\in\mathcal{V}}\Bigg\{\left(\mathbb{E}\left[\norm{\begin{pmatrix}X_{1}\\Y_{1}\end{pmatrix}-\begin{pmatrix}X_{2}\\Y_{2}\end{pmatrix}}^{p}\right]\right)^{\frac{1}{p}}\Bigg\},
\end{align}
where\mbox{ $\mathcal{V}=\{p_{X_{1},Y_{1},X_{2},Y_{2}}: p_{X_{1},Y_{1}}\!=\!p_{X,Y}, p_{X_{2},Y_{2}}\!=\!p_{\hat{X},Y}\}$} and the notation 
$\left(\begin{smallmatrix}z_1\\z_2\end{smallmatrix}\right)$ refers to concatenation of the vectors $z_{1},z_{2}$. 
We refer to the statistical distance between $\smash{p_{\hat{X},Y}}$ and $\smash{p_{X,Y}}$ as the \emph{joint perceptual index} of $\smash{\hat{X}}$.
When this index is small, we say that the estimator attains high joint perceptual quality.

\subsection{Consistency}\label{section:consistency}
\citet{Ohayon2023} considered only the case of deterministic degradations, where one can write $Y=D(X)$, i.e., $\smash{p_{Y|X}(y|x)=\delta(y-D(x))}$ for all $x,y$.
In this case, $\hat{X}$ is said to produce perfectly consistent outputs if $\smash{D(\hat{X})=Y}$.
This condition holds if and only if $\smash{p_{Y|\hat{X}}(\cdot|x)=p_{Y|X}(\cdot|x)}$ for almost every $\smash{x\in\supp{p_{\hat{X}}}}$, which we write as $\smash{p_{Y|\hat{X}}=p_{Y|X}}$ in short.
Since here we consider any type of degradation, we generalize the notion of perfect consistency and say that $\hat{X}$ produces perfectly consistent outputs if $p_{Y|\hat{X}}=p_{Y|X}$, regardless of the form of $p_{Y|X}$.
In~\cref{section:residual_noise_toy_example} we discuss the meaning of this definition of consistency for degradations of the form $Y=X+N$, where $X$ and $N$ are statistically independent.

While~\citet{Ohayon2023} discussed estimators that satisfy $p_{Y|\hat{X}}=p_{Y|X}$ (where the degradation is deterministic), here we also consider estimators that do not necessarily satisfy this equality.
Yet, we avoid referring to consistency as a stand-alone property of an estimator, but rather consider perceptual quality and consistency jointly via the joint perceptual index.
When the joint perceptual index is zero ($\smash{p_{\hat{X},Y}=p_{X,Y}}$), from Bayes' rule we have that $\smash{\hat{X}}$ attains perfect perceptual quality and perfect consistency according to the previous notations, $p_{\hat{X}}=p_{X}$ and $p_{Y|\hat{X}}=p_{Y|X}$.
In that sense, minimizing the joint perceptual index means approaching perfect perceptual quality and perfect consistency.
\begin{figure*}
    \centering
\includegraphics[width=\linewidth]{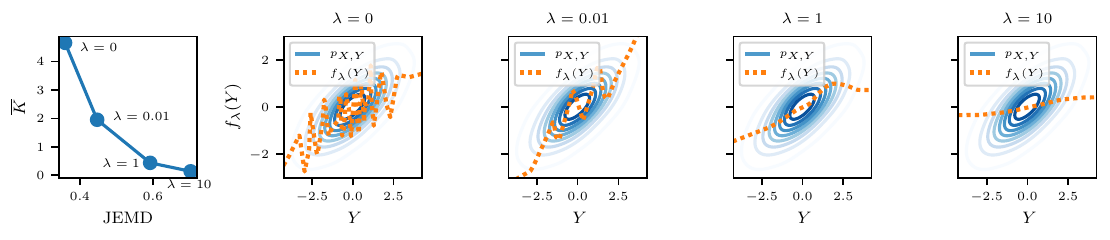}
    \caption{An illustration of~\cref{theorem:erratic_behavior} on the toy example from~\cref{section:toy_example}. \textbf{On the left}, we plot the Lipschitz constant lower bound $\overline{K}$ versus the JEMD (joint perceptual index) of $\hat{X}_{\lambda}=f_{\lambda}(Y)$, for several values of $\lambda$ (the coefficient of the robustness loss). \textbf{On the right}, we present contour plots of the density $p_{X,Y}$ (blue concentric ellipses) and outputs from $\hat{X}_{\lambda}$ as a function of $Y$, for several values of $\lambda$. We clearly see that $\hat{X}_{\lambda}$ is more erratic for smaller values of $\lambda$, as anticipated by~\cref{theorem:erratic_behavior}.
    Refer to~\cref{section:toy_example} for more details.}
    \label{figure:toy-wasserstein-rob-plot}
\end{figure*}
\section{The Perception-Robustness Tradeoff}\label{section:erratic_behavior}
We move on to provide the main result of our paper.
\begin{restatable}{theorem}
{erraticbehavior}\label{theorem:erratic_behavior}
Consider any joint probability density function $p_{X,Y}$ of the random variables $X$ and $Y$, such that the degradation is not invertible (according to~\cref{definition:non-invertible-degradation}).
For every $\gamma>0$, there exist constants $m_{1},m_{2}>0$ such that
\begin{gather}
    \text{Lip}(\hat{X})\geq \frac{m_{1}}{\sqrt{W_{p}(p_{X,Y},p_{\hat{X},Y})}}-m_{2},
\end{gather}
for any deterministic estimator $\hat{X}=f(Y)$ of $X$ from $Y$ with joint perceptual index ${W_{p}(p_{X,Y},p_{\hat{X},Y})\leq\gamma}$.
\end{restatable}
\begin{proof}
    See~\cref{appendix:main_theorem_proof}.
\end{proof}
Note that when the degradation is not invertible, a deterministic estimator $\hat{X}$ cannot satisfy \mbox{$p_{\hat{X},Y}=p_{X,Y}$}~\citep{Ohayon2023}, so we always have \mbox{$W_{p}(p_{X,Y},p_{\hat{X},Y})>0$}.
\cref{theorem:erratic_behavior} shows that the Lipschitz constant of $\hat{X}$ is bounded from below by a function that strictly increases when $W_{p}(p_{X,Y},p_{\hat{X},Y})$ decreases.
Thus, as long as the joint perceptual index $W_{p}(p_{X,Y},p_{\hat{X},Y})$ is small, $\hat{X}$ is not robust, and this is regardless of the nature of the degradation, the model architecture, the distribution of the images, etc. 
We next demonstrate~\cref{theorem:erratic_behavior} on a toy example and on widely recognized image super-resolution models.
\subsection{Toy Example Illustration}\label{section:toy_example}
Consider a simple example where \mbox{$X\sim\mathcal{N}(0,1)$}, \mbox{$N\sim\mathcal{N}(0,1)$}, $X$ and $N$ are statistically independent, and $Y=X+N$. Our goal is to predict $X$ from $Y$. To do so, we train a deterministic denoiser $\hat{X}_{\lambda}=f_{\lambda}(Y)$ (a neural network) by optimizing the objective
\begin{gather}
    \mathcal{L}_{CGAN}+\lambda \mathcal{L}_{R}.
\end{gather}
Here, $\mathcal{L}_{CGAN}$ is a CGAN loss~\cite{MirzaO14}, where the discriminator is conditioned on both $X$ and $Y$, so $\mathcal{L}_{CGAN}$ is theoretically minimized only when $p_{\hat{X}_{\lambda},Y}=p_{X,Y}$ (such an objective promotes low joint perceptual index).
The loss $\mathcal{L}_{R}$ is defined by
\begin{gather}
    \mathcal{L}_{R}=\mathbb{E}\left[\norm{f_{\lambda}(Y)-f_{\lambda}(Y+Z)}_{2}^{2}\right],
\end{gather}
where $Z\sim\mathcal{N}(0,0.2)$ is statistically independent of $Y$.
This loss drives the outputs originating from $Y$ and from (randomly chosen) inputs close to $Y$ to be roughly equal, i.e., such an objective promotes robustness, and the level of robustness is controlled with the coefficient $\lambda\geq 0$.
See~\cref{appendix:toy_example} for more training details.

\paragraph{Lipschitz Constant Lower Bound.}We compute the average of $K(y,y_{adv})=\frac{\norm{f_{\lambda}(y)-f_{\lambda}(y_{adv})}_{2}}{\norm{y-y_{adv}}_{2}}$ over the entire validation set, where $y$ is the original input, $y_{adv}=y+z$, and $z$ is independently sampled from $p_{Z}$.
The result is denoted by $\overline{K}$, and serves as a lower bound on the Lipschitz constant of the estimator.

\paragraph{Joint Perceptual Index.}We measure the Earth Mover's Distance (EMD) between $p_{\hat{X}_{\lambda},Y}$ and $p_{X,Y}$, which is the Wasserstein 1-distance between these distributions, and denote the result by Joint-EMD (JEMD).
To compute this distance, we sample 10,000 points from $p_{\hat{X}_{\lambda},Y}$ and $p_{X,Y}$, compute the pairwise $L_{1}$ distances between each pair of points, and then use the \verb|ot.emd2| function from~\citep{flamary2021pot} to compute the JEMD.
Each sample from $p_{X,Y}$ is obtained by independently sampling a point $x$ from $p_{X}$ and a point $n$ from $p_{N}$, and concatenating $x$ with $y=x+n$.
The samples from $p_{\hat{X}_{\lambda},Y}$ are obtained by concatenating $f_{\lambda}(y)$ to each $y$.

In~\cref{figure:toy-wasserstein-rob-plot} we plot $\overline{K}$ versus the JEMD for several choices of $\lambda$.
As anticipated by~\cref{theorem:erratic_behavior}, we clearly see that the lower the value of JEMD, the higher the value of $\overline{K}$.
We also present the outputs of $\hat{X}_{\lambda}$, and indeed see that $\hat{X}_{\lambda}$ is more erratic for smaller values of $\lambda$.
We further discuss this toy example in~\cref{appendix:toy_example}, revealing additional observations and ideas for future research.

Note that, while $\overline{K}$ is indeed a lower bound of $\text{Lip}(\hat{X}_{\lambda})$, it is not necessarily equal to the lower bound provided in~\cref{theorem:erratic_behavior}.
However, we demonstrate that $\overline{K}$ still exhibits the behavior anticipated by the theorem: It grows inversely proportional to the joint perceptual index.
This inverse proportionality between $\overline{K}$ and the joint perceptual index offers a valuable insight.
A large value of $\overline{K}$ suggests that $\hat{X}_{\lambda}$ is unstable near \emph{many} different inputs, rather than just near a single point (as a large value of $\text{Lip}(\hat{X}_{\lambda})$ implies).
Thus, our experiments suggest that deterministic estimators achieving a low joint perceptual index are likely to be unstable near many inputs.

\begin{figure*}
    \centering
\includegraphics[scale=1]{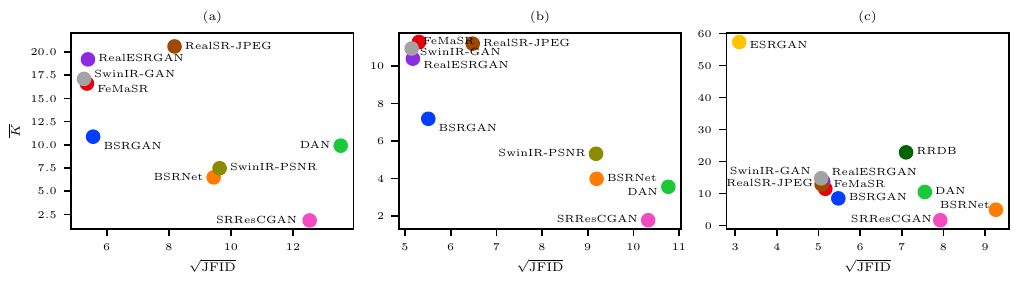}
    \caption{Quantitative demonstration of~\cref{theorem:erratic_behavior}.
    We plot $\overline{K}$ versus $\sqrt{\text{JFID}}$ of several image super-resolution algorithms evaluated on several degradations.
    (a) Results on the Track2 challenge degradation by~\citet{lugmayr2019aim}. (b) Results on the Track1 challenge degradation by~\citet{Lugmayr_2020_CVPR_Workshops}. (c) Results on the standard bicubic $\times 4$ down-sampling degradation. As anticipated by~\cref{theorem:erratic_behavior}, we see a tradeoff between $\overline{K}$ and $\sqrt{\text{JFID}}$ for all three degradations, i.e., the Lipschitz constant is lower bounded by a function that increases as the joint perceptual index decreases (See~\cref{section:real-world-experiments}).}
    \label{fig:real-world-attacks}
\end{figure*}
\begin{figure}
        \centering
        \includegraphics[width=\linewidth]{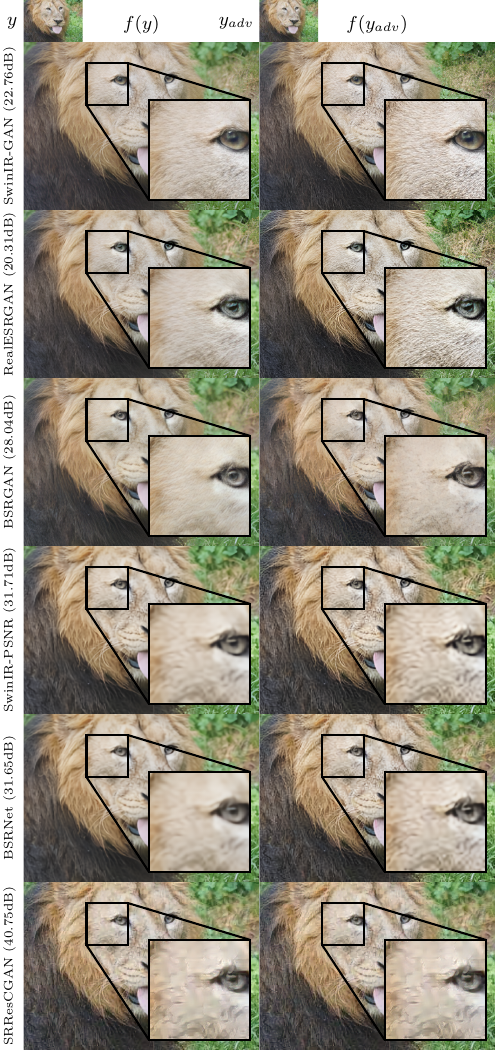}
    \caption{Visual comparison of some of the super-resolution algorithms evaluated in~\cref{section:real-world-experiments} on the Track2 challenge degradation by~\citet{lugmayr2019aim}, sorted from top to bottom by their JFID (increasing). The original and the attacked outputs are denoted by $f(y)$ and $f(y_{adv})$, respectively.
    The PSNR between $y$ and $y_{adv}$ is at least $48.13\text{dB}$ (obtained with $\alpha=1/255$ in I-FGSM), so the difference is visually negligible (see the attacked input $y_{adv}$ of SwinIR-GAN at the top). 
    The PSNR between $f(y)$ and $f(y_{adv})$ is reported next to the name of each algorithm. As can be seen, the better the joint perceptual quality, the higher the sensitivity to adversarial attacks.}
    \label{fig:visual-comparison}
\end{figure}
\subsection{Quantitative Demonstration}\label{section:real-world-experiments}
We support~\cref{theorem:erratic_behavior} via a quantitative evaluation on the task of image super-resolution, by showing that algorithms that attain a lower joint perceptual index tend to behave more erratically.
The algorithms we evaluate are RealESRGAN~\citep{wang2021realesrgan}, BSRGAN and BSRNet~\citep{zhang2021designing}, FeMaSR~\citep{chen2022femasr}, DAN (setting 2)~\citep{luo2020unfolding}, ESRGAN and RRDB~\citep{wang2018esrgan}, RealSR-JPEG and RealSR-DPED~\citep{Ji_2020_CVPR_Workshops}, FSSR-JPEG and FSSR-DPED~\citep{fritsche2019frequency}, SRResCGAN~\citep{Umer_2020_CVPR_Workshops}, SwinIR-GAN and SwinIR-PSNR~\citep{liang2021swinir}, all of which perform $4\times$ up-scaling.
We use the DIV2K~\citep{Agustsson_2017_CVPR_Workshops,Timofte_2017_CVPR_Workshops} test set with the degradations from the Track1 and Track2 challenges in~\citep{lugmayr2019aim}, as well as the common $4\times$ bicubic down-sampling.

\paragraph{Lipschitz Constant Lower Bound.}We conduct the $\ell_\infty$ I-FGSM basic attack~\citep{Choi_2019_ICCV,kurakin2017adversarial} on each algorithm and each input $y$ using $\alpha=1/255$ and compute the ratio \mbox{$K(y,y_{adv})=\frac{\norm{f(y)-f(y_{adv})}_{2}}{\norm{y-y_{adv}}_{2}}$}, where $f(\cdot)$ is the model and $y_{adv}$ is the attacked version of $y$.
We specifically choose a small value of $\alpha$ in order to assess the rate of change in each algorithm's output while using inputs that remain within $\supp{p_{Y}}$ (this constraints the attacked outputs to be in $\supp{p_{\hat{X}}}$).
Indeed, at $\alpha=1/255$, the difference between each pixel in $y$ and $y_{adv}$ is at most 1 gray level, i.e., the PSNR between $y$ and $y_{adv}$ is at least 48.13dB, making this a minor change for images.
We compute the average of $K(y,y_{adv})$ over all pairs $(y,y_{adv})$ and denote the result by $\overline{K}$, which serves as a lower bound on the Lipschitz constant.
Besides, by taking the average, we can better determine if an algorithm exhibits erratic behavior across many inputs rather than being influenced by a single outlier.
Note that $\overline{K}$ has an intuitive scale: Changing the input pixels by an average of $1$ gray levels leads to an average change of $\overline{K}$ gray levels in the output pixels.

\paragraph{Joint Perceptual Index.}We approximate the joint perceptual index with a slightly tweaked version of the Fréchet Inception Distance (FID)~\citep{fid} for joint distributions.
Given pairs $(x,y)$ of source images and their degraded versions, we construct pairs $(\hat{x},y)$ using each algorithm $f(\cdot)$, where $\hat{x}=f(y)$.
We extract all patches of size $32\times 32$ with stride $16$ from each $y$, and all the corresponding patches of size $128\times 128$ with stride $64$ from $x$ and $\hat{x}$.
From the 100 images of the DIV2K test set, this leads to a set of $N=46,500$ pairs of patches $\{(x^{(i)}, y^{(i)})\}_{i=1}^{N}$, and $N$ pairs of patches $\{(\hat{x}^{(i)}, y^{(i)})\}_{i=1}^{N}$.
We then feed each $x^{(i)}$, $y^{(i)}$ and $\hat{x}^{(i)}$ into the Inception-V3 network~\citep{inceptionv3} and obtain their features $x^{(i)}_{F},y^{(i)}_{F},$ and $\hat{x}^{(i)}_{F}$ from the last pooling layer, each of which is a vector of size 2048.
Lastly, we compute 
\begin{gather}
    \text{FID}\left(\left\{\begin{pmatrix}
    x_{F}^{(i)}\\ y^{(i)}_{F}
\end{pmatrix}\right\}_{i=1}^{N},\left\{\begin{pmatrix}
    \hat{x}_{F}^{(i)}\\ y^{(i)}_{F}
\end{pmatrix}\right\}_{i=1}^{N}\right)
\end{gather}
and denote the result by Joint-FID (JFID).

In~\cref{fig:real-world-attacks} we present the plot of $\overline{K}$ versus $\sqrt{\text{JFID}}$ for each degradation.
We take the square-root of JFID so that both axes have the same scale.
The results clearly show that smaller values of $\sqrt{\text{JFID}}$ correspond to larger values of $\overline{K}$.
For a better view of the tradeoff between $\overline{K}$ and $\sqrt{\text{JFID}}$, notice that in each plot we omit the algorithms which are too far-off from the tradeoff.
In~\cref{appendix:real-world-algorithms-additional-details} we provide complementary figures that include all the evaluated algorithms, as well as distortion-perception plots of RMSE versus $\sqrt{\text{FID}}$, where FID is computed between the sets $\{x_{F}^{(i)}\}_{i=1}^{N}$ and $\{\hat{x}_{F}^{(i)}\}_{i=1}^{N}$ (i.e., it is a perceptual quality evaluation of each algorithm).
In~\cref{appendix:real-world-algorithms-additional-details} we present additional plots to support~\cref{theorem:erratic_behavior}, by replacing the JFID with alternative measures related to Kernel Inception Distance (KID)~\cite{kid} and Precision and Recall~\cite{precision_recall}.
In~\cref{fig:visual-comparison} we present visual results of selected methods that appear in favorable locations in~\cref{fig:real-world-attacks}.

\subsection{Adversarial Attacks}\label{section:adv_attacks}
While slightly perturbing the original input $y$ of a deterministic estimator with high joint perceptual quality may lead to a large change in its output, the new output may still look ``valid'', i.e., it may still be in $\supp{p_{X|Y}(\cdot|y)}$ (see, e.g., the outputs of SwinIR-GAN in~\cref{fig:visual-comparison}).
In what sense then can this new output be considered as ``adversarial''?
More generally, what is the practical relevance of~\cref{theorem:erratic_behavior} when the outputs originating from any slightly perturbed input $y_{adv}$ may remain valid, even when the algorithm is highly erratic?
Here we demonstrate a possible drawback of such erratic deterministic estimators that arises when using them in a decision-making pipeline.

Consider an experiment where the goal is to predict the gender in low-resolution noisy face images, by performing gender classification after enhancing the images' quality with a super-resolution algorithm.
We take the first 1000 face images from the CelebA-HQ~\cite{karras2018progressive} data set and degrade each image by resizing it to $32\times 32$ pixels using bilinear interpolation, and then adding isotropic Gaussian noise with zero mean and standard deviation $10/255$.
The super-resolution algorithm we use is \mbox{GFPGAN~\cite{wang2021gfpgan}}, a deterministic face image restoration model with high perceptual quality, which we expect to also attain high joint perceptual quality (we use the v1.3 checkpoint and official code provided by the authors).
This algorithm performs $\times16$ up-scaling, resulting in an output image of size $512\times 512$.
We attack each degraded image using a tweaked version of the I-FGSM basic attack with $\alpha=16/255$ and $T=100$.
Instead of using the $L_{2}$ loss in I-FGSM like in~\cite{Choi_2019_ICCV}, we forward each attacked output through a vision transformer model~\cite{dosovitskiy2021an} trained to classify the gender in face images~\cite{gender_classification}, and then try to maximize the log-probability of the class ``female''.
Put simply, we forward each low-quality image through GFPGAN and then feed the result to the face gender classification model.
We then use the I-FGSM update rule~\cite{Choi_2019_ICCV} to maximize the softmax probability of the category ``female''.

\begin{figure}[t!]
\renewcommand{\arraystretch}{0.0}
    \begin{tabular}
{P{0.2\linewidth}P{0.2\linewidth}P{0.2\linewidth}P{0.2\linewidth}}
    \multicolumn{4}{l}{\hskip3.25cm GFPGAN}\\
   \vspace*{0.1cm} \ & & &\\
    \multicolumn{4}{l}{\hskip-0.19cm\includegraphics[width=0.98\linewidth]{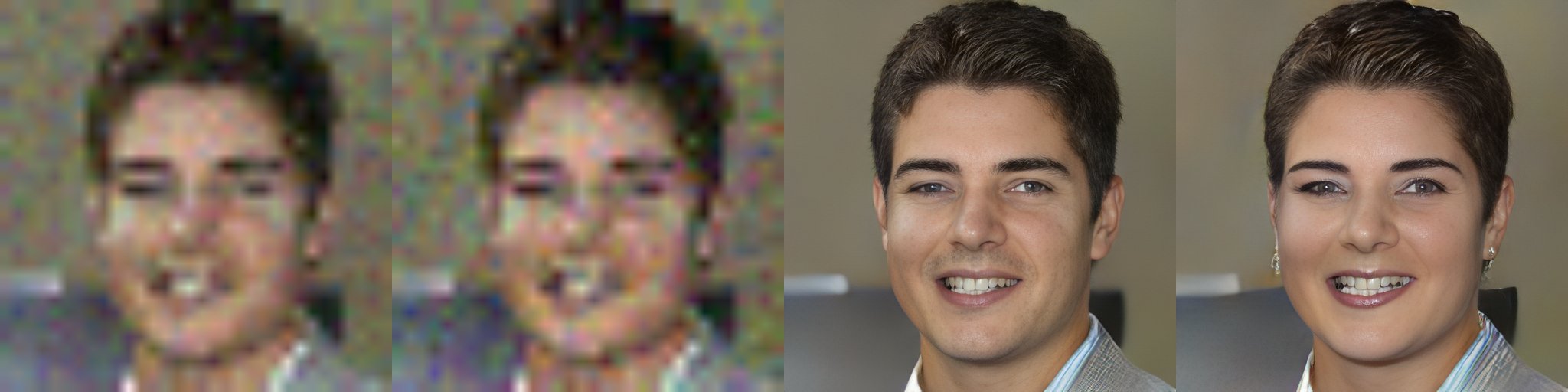}} \\
    \multicolumn{4}{l}{\hskip-0.19cm\includegraphics[width=0.98\linewidth]{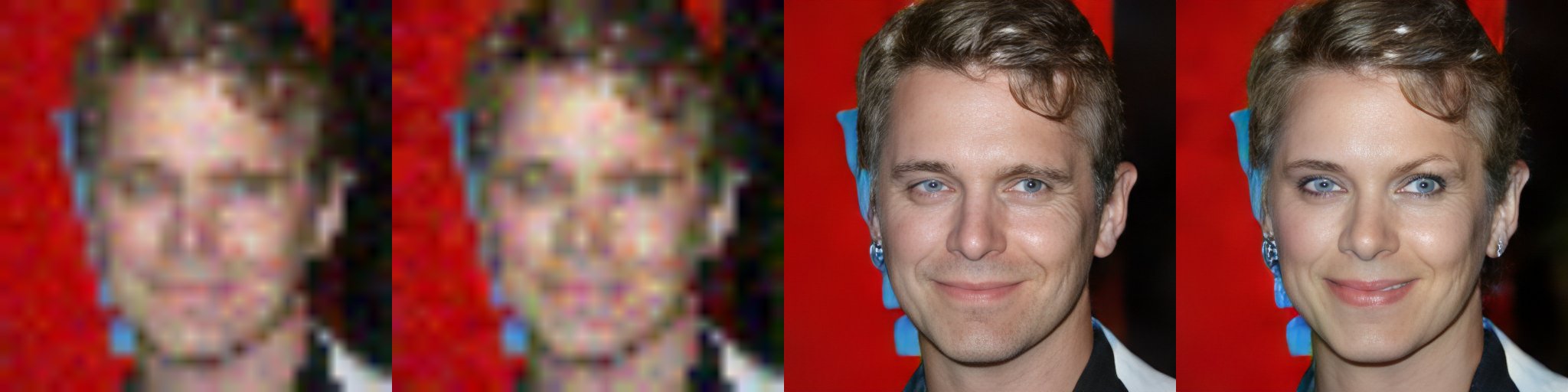}} \\
    \multicolumn{4}{l}{\hskip-0.19cm\includegraphics[width=0.98\linewidth]{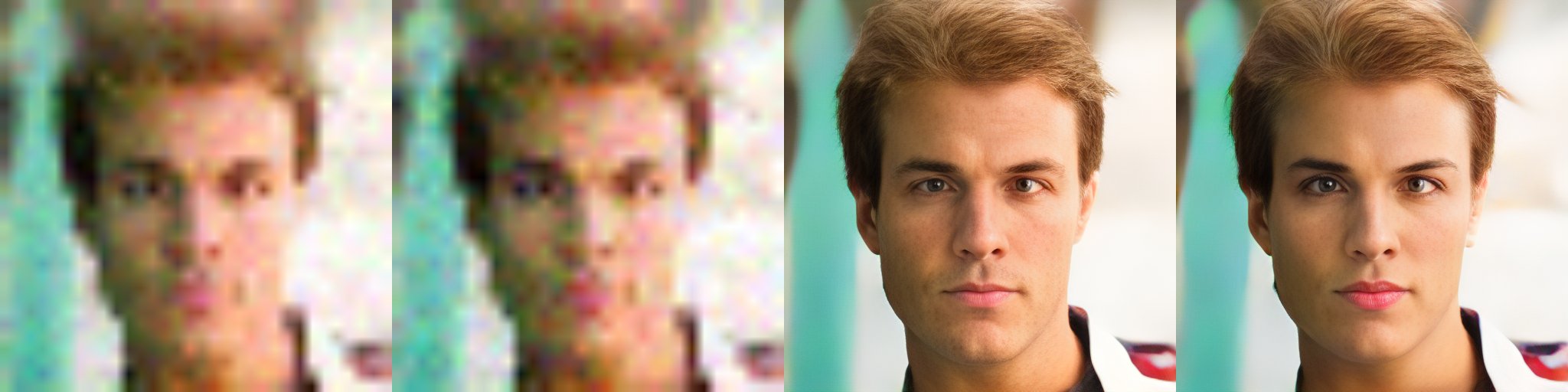}} \\
   \vspace*{0.2cm} \ & & &\\
       \multicolumn{4}{l}{\hskip3.45cm RRDB}\\
   \vspace*{0.1cm} \ & & &\\
    \multicolumn{4}{l}{\hskip-0.19cm\includegraphics[width=0.98\linewidth]{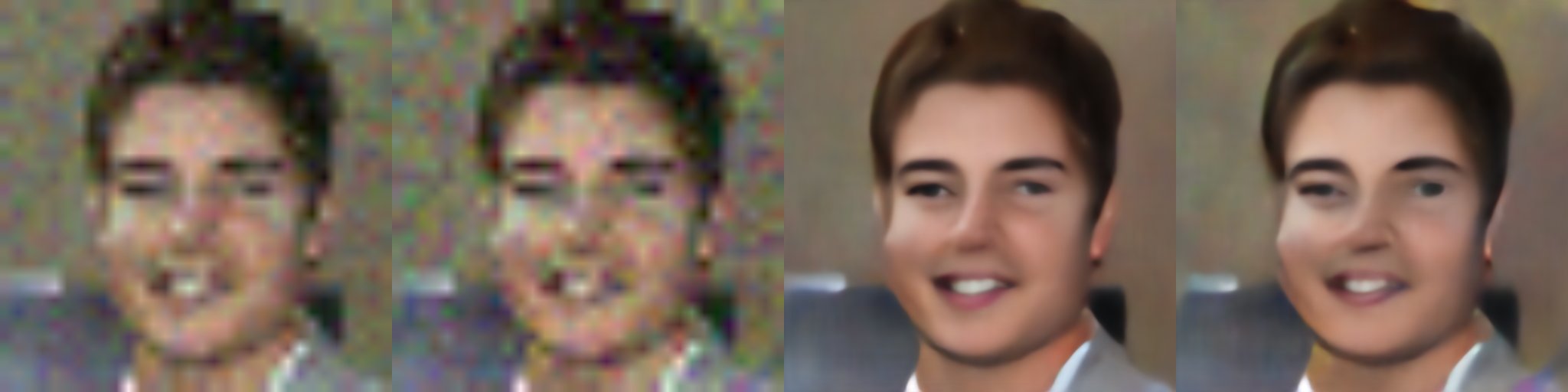}} \\
    \multicolumn{4}{l}{\hskip-0.19cm\includegraphics[width=0.98\linewidth]{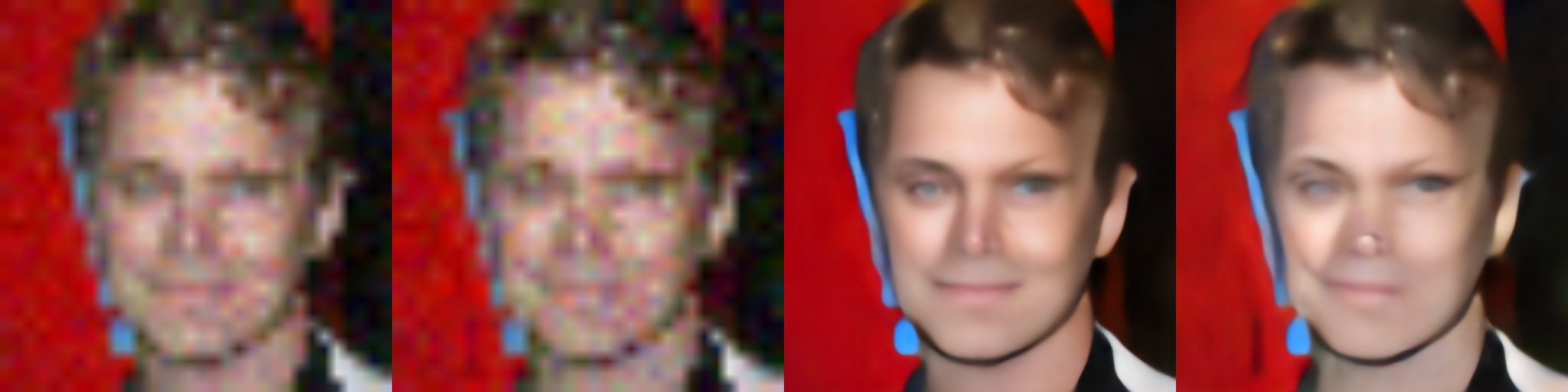}} \\
    \multicolumn{4}{l}{\hskip-0.19cm\includegraphics[width=0.98\linewidth]{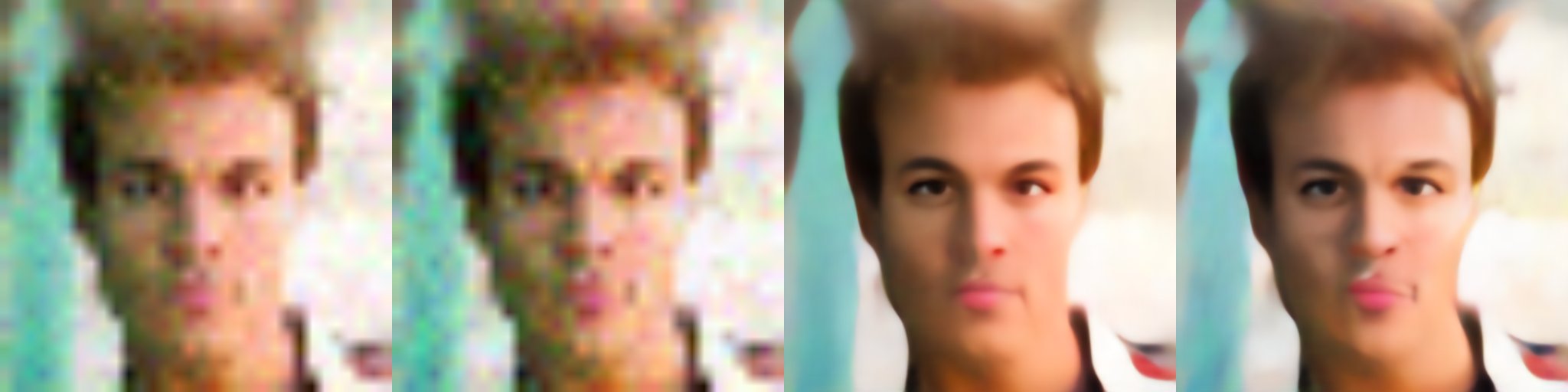}} \\
    \vspace*{0.1cm} \ & & & \\
    {\small $y$} & 
    {\small $y_{adv}$} & 
    {\small $f(y)$} & 
    {\small $f(y_{adv})$} \\
\end{tabular}

    \caption{Adversarial attacks on low-resolution face images intended to alter the outputs of GFPGAN and RRDB to produce a face which is classified as ``female'' rather than ``male''. From left to right: Original input $y$, attacked input $y_{adv}$, original output $f(y)$, output obtained from the attacked input $f(y_{adv})$. 
    The perturbed inputs $y_{adv}$ are obtained with $\alpha=16/255$ in I-FGSM.
    With such a value of $\alpha$ we barely see any visual difference between $y$ and $y_{adv}$.
    As anticipated, $y_{adv}$ indeed leads to outputs with newly generated features when using GFPGAN (e.g., makeup), yet we barely see any significant change for RRDB. An image gender classifier associates these features with the ``female'' category, as the predicted class in the GFPGAN outputs switch from ``male'' when the input is $y$, to ``female'' when the input is $y_{adv}$. Refer to~\cref{section:adv_attacks} for more details.
    }
    \label{fig:emotion_adv_attack}
\end{figure}
\begin{figure}
\renewcommand{\arraystretch}{0.0}
    \hskip-0.19cm\begin{tabular}{l}
        \hspace{0.08\linewidth}{\small $y$}\hspace{0.12\linewidth}{\small Exploring the posterior by slightly perturbing $y$}\\
        \vspace*{0.1cm}\\
        \includegraphics[width=1\linewidth]{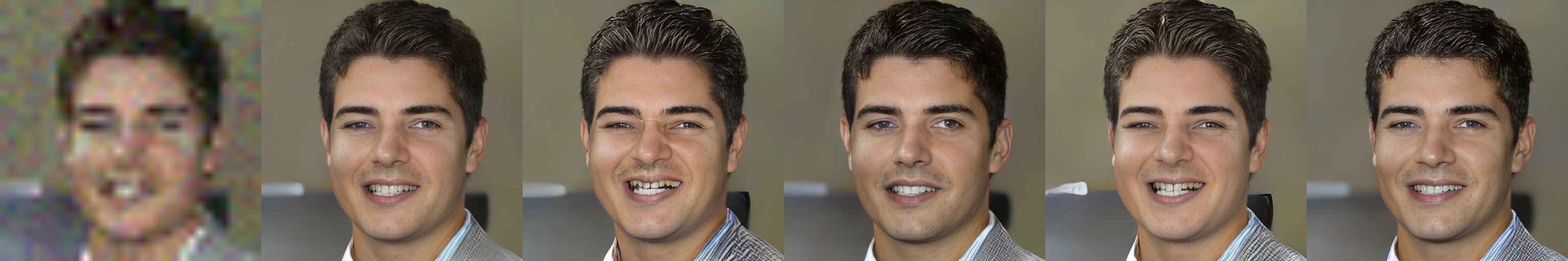}\\
        \includegraphics[width=1\linewidth]{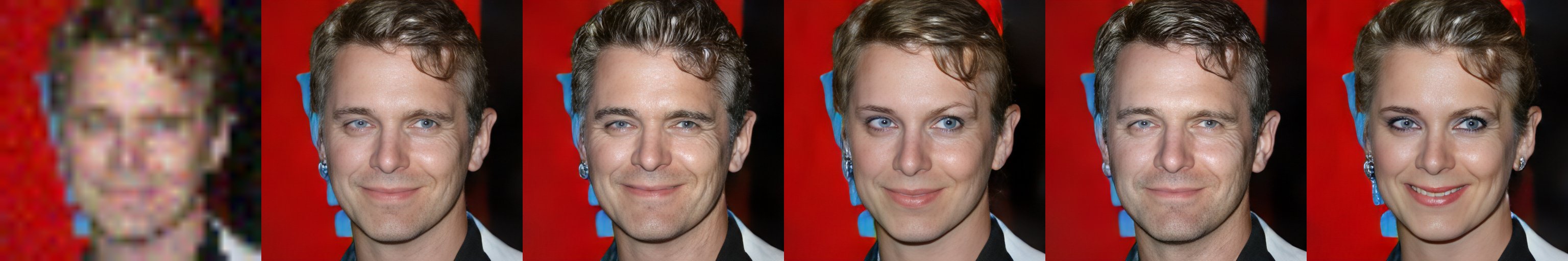}\\
    \includegraphics[width=1\linewidth]{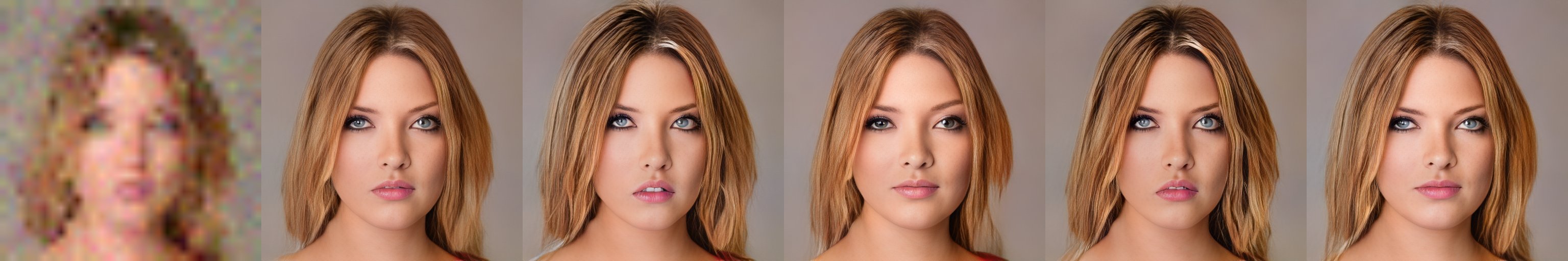}
    \\\includegraphics[width=1\linewidth]{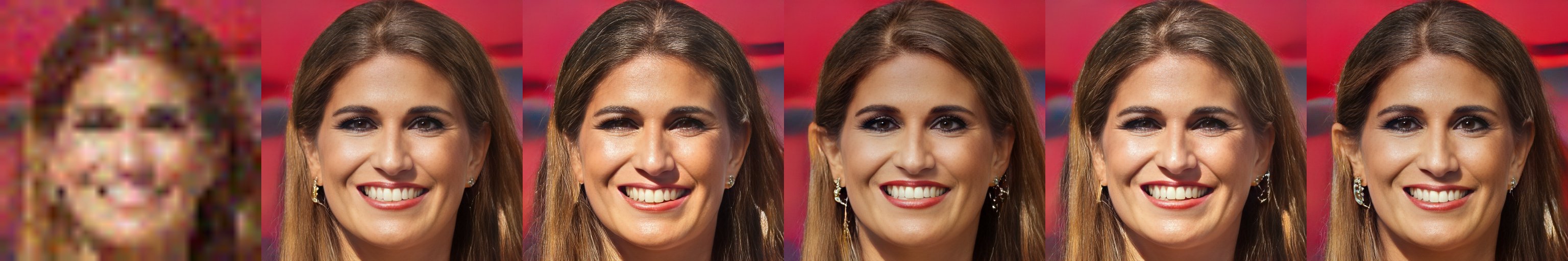}
    \end{tabular}
    \caption{Exploring the posterior distribution with \mbox{GFPGAN}~\cite{wang2021gfpgan}.
    In each row we show the original input $y$ and outputs generated from ``adversarial'' inputs, each of which is a slight perturbation of $y$ (with a minimum of 30dB PSNR).
    By slightly perturbing $y$, we observe meaningful semantic variation in the outputs, such as change of gender and hair textures, all of which are expected to be possible varying features in the posterior.}
    \label{fig:posterior-imitation}
\end{figure}
Observe in~\cref{fig:emotion_adv_attack} that there is barely any visual difference between the original and the attacked inputs.
Yet, compared to the original outputs, the ones resulting from the attacked inputs exhibit newly generated features, such as makeup and jewelry.
As for a quantitative analysis, we use~\cite{serengil2021lightface} to perform face gender classification on all the source images, the outputs resulting from the original inputs, and the outputs resulting from the attacked inputs.
Note that we use a different gender classification model for evaluation to alleviate the possibility of over-fitting.
61.2\% of the source images and 63.3\% of the original reconstructed images are classified as ``female'', so GFPGAN is quite accurate in terms of reconstructing the true gender population in the data.
Yet, the percentage of ``female''-classified attacked images is increased to 72.2\%.

Lastly, to support our claim that such a vulnerability is associated with the joint perceptual quality of the algorithm, we also train the RRDB model~\cite{wang2018esrgan} solely with the $L_{1}$ reconstruction loss, using the exact same training setup as that of GFPGAN (same training data, same degradation, same input normalization, etc.).
Additional details are disclosed in~\cref{appendx:adv_attacks}.
This model is expected to produce blurry images with low perceptual quality~\cite{Blau2018}, so it is also expected to produce lower joint perceptual quality compared to GFPGAN (see~\cref{fig:emotion_adv_attack} for a qualitative comparison).
We evaluate the outputs of the trained model and observe that, unlike GFPGAN, this model is robust to such attacks, as 59.1\% and 56.9\% of its original and attacked outputs, respectively, are classified as ``female''.
Interestingly, even though we attack the model to produce ``female''-like outputs, the fraction of ``female''-classified images is \emph{reduced} for the attacked outputs.
We discuss this phenomenon and provide more visual results in~\cref{appendix:male-to-female}.
In~\cref{appendix:aging_attacks} we present similar experiments where we attack the models to produce faces of older people, and find that, compared to RRDB, GFPGAN is more vulnerable to such attacks as well.

\subsection{Exploring the Posterior Distribution}\label{section:posterior-sampling-imitation}
Exploring the posterior distribution in imaging inverse problems may hold a practical significance in many scenarios (please refer to~\cref{appendix:posterior-imitation-details} for a discussion).
Here, we demonstrate that, as an implication of~\cref{theorem:erratic_behavior}, doing so is possible with a \emph{deterministic} estimator.
Specifically, when $\hat{X}$ is a deterministic estimator and $W_{p}(p_{X,Y},p_{\hat{X},Y})$ is small, we can intuitively say that the outputs produced by $\hat{X}$ given the inputs in a small ball around $y$ should approximately constitute samples from the posterior distribution (this idea inspired our proof of~\cref{theorem:erratic_behavior}).
Let us leverage this insight.

To explore the posterior, for each input $y^{(1)}$ we search for a new input $y^{(2)}$ close to $y^{(1)}$, such that $f(y^{(2)})$ is far from $f(y^{(1)})$.
Then, to obtain another sample, we seek another input $y^{(3)}$ such that $f(y^{(3)})$ is far from both $f(y^{(1)})$ and $f(y^{(2)})$.
We continue in this fashion to obtain more samples.
This is a Farthest Point Sampling (FPS)~\citep{fps} approach to generate an arbitrary number of outputs that correspond to (almost) the same input (refer to~\cref{appendix:posterior-imitation-details} for a formal description).
We use this algorithm only for demonstration purposes and do not claim any optimality.

\cref{fig:posterior-imitation} presents such an experiment. 
We take several images from the CelebA-HQ data set and degrade each image using the same procedure as described in~\cref{section:adv_attacks}.
To explore the posterior, we execute the FPS algorithm on GFPGAN.
In~\cref{fig:posterior-imitation} we present several examples of degraded inputs and corresponding sequential samples that result from the FPS algorithm.
We indeed see that by slightly perturbing the original input, we obtain diverse outputs that all seem to correspond to the measurements.
\section{Summary}\label{section:summary}
We prove that the better the perceptual quality and consistency of a deterministic image restoration algorithm, the higher its Lipschitz constant must be.
We confirm that widely used image super-resolution algorithms indeed adhere to such a tradeoff, and perform experiments that showcase the practical consequences (both positive and negative) of this result.

\section{Limitations}\label{section:limitations}
While~\cref{theorem:erratic_behavior} holds true for any signal restoration task involving a non-invertible degradation, in this work we limit its demonstration to the image super-resolution task. It would be interesting to explore the discovered tradeoff in alternative tasks, such as MRI reconstruction, image-to-image translation, or other signal restoration tasks such as video, audio, or even text restoration.

In~\cref{section:real-world-experiments}, we evaluate deterministic algorithms with high joint perceptual quality, all of which are GAN-based, as currently such methods are the state-of-the-art.
However, other types of deterministic methods that achieve high joint perceptual quality may emerge in the future.
If so, it would be interesting to assess where those methods lie on the perception-robustness plane compared to GAN-based methods.

In~\cref{section:posterior-sampling-imitation}, we illustrate the feasibility of exploring the posterior distribution using a deterministic estimator with high joint perceptual quality.
However, we provide only qualitative results and do not showcase the practical applications of this phenomenon.
For instance, it could prove beneficial in cases where practitioners employ a deterministic algorithm but still desire to perform uncertainty quantification of the restoration task at hand (e.g., to observe the range of possible pixel values in a specific region in the image).

\section*{Impact Statement}
Our work uncovers a fundamental tradeoff between the stability of deterministic restoration algorithms and their performance, as measured by their joint perceptual index. This tradeoff presents a vulnerability that could be exploited by malicious users to introduce bias and manipulate the outcomes of imaging systems (e.g., through adversarial attacks). We encourage practitioners to address this inherent tradeoff by trading performance in order to safeguard against such misuse.

\section*{Acknowledgments}
We thank Yonatan Shadmi for reviewing the proof of our theorem.
This research was partially supported by the Israel Science Foundation (ISF) under Grant 2318/22 and by the Council For Higher Education - Planning \& Budgeting Committee.

\bibliography{example_paper}
\bibliographystyle{icml2024}

\newpage
\appendix
\onecolumn
\section{On the Desire to Use Deterministic Image Restoration Methods}\label{appendix:desire_for_deterministic_estimators}
In many applications users desire a deterministic outcome every time an input image is processed.
This allows practitioners to trust the results and make informed decisions based on them.
For instance, a deterministic image restoration algorithm ensures that doctors and radiologists do not get different interpretations from the same input.
They would typically desire a single ``best'' estimate to diagnose a condition, rather than multiple images from which they may not draw a definite conclusion. 
While in those cases some may suggest to somehow select only one of the outputs produced by a stochastic estimator, why go through the burden of obtaining a sampler if we do not intend to leverage its full potential anyways (e.g., uncertainty evaluation)?

Another reason that deterministic image restoration algorithms are often desirable is that they are simpler to obtain, especially in high dimensional distributions such as that of natural images.
In the era of deep learning, designing a well-performing model is as straightforward as designing a loss function that targets our specific desired objective.
With such a loss function, we can optimize a neural network to obtain the objective.
Hence, it is appealing (and typical) to address image inverse problems simply by training a neural network (deterministic estimator) to meet certain objectives.
Perhaps the most common objectives are high perceptual quality (e.g., obtaining a low Wasserstein distance between $p_{\hat{X}}$ and $p_{X}$) and low average distortion (e.g., obtaining a low MSE), which may be at odds with each other~\citep{Blau2018}.

While there might be situations where leveraging the diverse outcomes from stochastic methods could be beneficial, especially when drawing insights about uncertainties or making probabilistic decisions, these come with their own set of challenges.
Processing multiple outcomes to derive a singular, actionable insight often involves additional computational steps.
This not only increases the processing time but can also introduce complexities in decision-making.
On the other hand, a deterministic estimator streamlines this process by directly predicting the desired result.
It offers a straightforward approach that eliminates the need for sifting through multiple potential outcomes and consolidates the decision-making process.
In many scenarios, especially where timely and clear responses are essential (e.g., self driving cars), the efficiency and directness of deterministic algorithms make them a favorable choice.
Perhaps for the reasons above, among potentially others, we see that deterministic image restoration algorithms are much more commonly used in practice (compared to stochastic ones).

\section{Additional Related Works}
\label{appendix:related-work}
Previous works have discussed that image restoration algorithms with extremely low values of distortion tend to be unstable~\cite{gottschling2023troublesome,1d5bf173073f46c19e7093d05b0b9b01}.
Namely, as~\citet{gottschling2023troublesome} prove, a restoration algorithm must be unstable (with a high Lipschitz constant) if it produces accurate estimates for two \emph{distinct} source images that correspond to almost identical degraded measurements.
This phenomenon stems directly from the definition of Lipschitz continuity.
Indeed, an algorithm that generates two distinct outputs from (almost) indistinguishable inputs, is, by definition, unstable.
Importantly, such a tradeoff is fundamentally different than the tradeoff we reveal in our paper: We are discussing perceptual quality \& consistency, not distortion.
Moreover, the tradeoff we reveal cannot be alleviated.
It has nothing to do with issues such as over-fitting, ``over-performance'', or ``inconsistent-performance'', as discussed in~\cite{gottschling2023troublesome}.

As we show in~\cref{section:adv_attacks}, one can imitate stochastic sampling by slightly perturbing the input of an unstable deterministic restoration algorithm.
This approach allows to explore the posterior distribution if the algorithm attains high joint perceptual quality.
The idea to imitate stochastic sampling by perturbing the input of a deterministic algorithm has been proposed before~\cite{gandikota2023evaluating}, but not fully demonstrated with a concrete algorithm such as~\cref{algorithm:posterior-sampling-imitation}.
\section{Proof of~\cref{theorem:erratic_behavior}}\label{appendix:main_theorem_proof}
Here we prove~\cref{theorem:erratic_behavior}.
We start with a useful definition and several lemmas.
\begin{definition}\label{definition:beta-ill-posed}
A degradation is called $\beta$-ill-posed if there exist $\beta>0$ and $\mathcal{S}_{y}$ with $P(Y\in\mathcal{S}_{y})>0$, such that for all $y\in\mathcal{S}_{y}$ there exist sets $Q_{1}(y)$ and $Q_{2}(y)$, where \mbox{$P(X\in Q_{i}(y)|Y=y)>0$} and $\norm{x^{(1)}-x^{(2)}}\geq\beta$ for every \mbox{$(x^{(1)},x^{(2)})\in Q_{1}(y)\times Q_{2}(y)$}.
\end{definition}
We use~\cref{definition:beta-ill-posed} since it allows us to link $\beta$ to the Lipschitz constant of a deterministic estimator.
Importantly, we show that such a definition does not pose any special requirements on the degradation, i.e., any non-invertible degradation is also $\beta$-ill-posed for some $\beta>0$.
The lower bound of the Lipschitz constant $K$ will then depend on $\beta$.
\begin{lemma}\label{corollary:non-invertible-is-beta-ill-posed}
Any non-invertible degradation is $\beta$-ill-posed for some $\beta>0$.
\end{lemma}
\begin{proof}
The degradation is not invertible, so there exists $\mathcal{S}_{y}$ with $P(Y\in\mathcal{S}_{y})>0$, such that $p_{X|Y}(\cdot|y)$ is not a delta for every $y\in\mathcal{S}_{y}$.
From the definition of a probability density function, for any $y\in\mathcal{S}_{y}$ there exists a point $x_{1}\in\supp{p_{X|Y}}(\cdot|y)$ such that
for every $\lambda>0$ we have
\begin{gather}
    P(X\in B(x_{1},\lambda)|Y=y)>0,
\end{gather}
where
\begin{gather}
    B(x,\lambda)=\{x'\: :\: \norm{x-x'}<\lambda\}.
\end{gather}
This is true since otherwise $p_{X|Y}(\cdot|y)$ integrates to zero.
It is well known that
\begin{gather}
    \{x_{1}\}=\cap_{n\geq 1}B(x_{1},\frac{1}{n}),\, n\in\mathbb{N}.
\end{gather}
Suppose by contradiction that for every $n\in\mathbb{N}$ it holds that 
\begin{gather}
    P(X\in B(x_{1},\frac{1}{n})|Y=y)=1,
\end{gather}
so
\begin{gather}
    P(X\notin B(x_{1},\frac{1}{n})|Y=y)=P(X\in B^{c}(x_{1},\frac{1}{n})|Y=y)=0.
\end{gather}
Thus, it holds that
\begin{align}
    P(X\in \{x_{1}\}|Y=y)&=P(X\in \cap_{n\geq 1}B(x_{1},\frac{1}{n})|Y=y)\\
    &=1-P(X\in\cup_{n\geq 1}B^{c}(x_{1},\frac{1}{n})|Y=y)\\
    &\geq 1-\sum_{n\geq 1}P(X\in B^{c}(x_{1},\frac{1}{n})|Y=y)=1,
\end{align}
so $P(X\in \{x_{1}\}|Y=y)=1$, which is a contradiction to the fact that $p_{X|Y}(\cdot|y)$ is not a delta.
Hence, there exists $n^{*}\in\mathbb{N}$ such that
\begin{gather}
    P(X\in B(x_{1},\frac{1}{n^{*}})|Y=y)<1,
\end{gather}
and it holds that
\begin{gather}
    P(X\in B^{c}(x_{1},\frac{1}{n^{*}})|Y=y)>0.
\end{gather}
Again, from the definition of a density, there exists $x_{2}\in B^{c}(x_{1},\frac{1}{n^{*}})$, $x_{2}\neq x_{1}$, such that
\begin{gather}
     P(X\in B(x_{2},\lambda)|Y=y)>0,
\end{gather}
and this is true for every $\lambda>0$.
In particular, for $\lambda^{*}=\norm{x_{1}-x_{2}}>0$ we have
\begin{gather}
P(X\in B(x_{1},\frac{\lambda^{*}}{4})|Y=y)>0,\\
P(X\in B(x_{2},\frac{\lambda^{*}}{4})|Y=y)>0,
\end{gather}
and thus, for every $(x^{(1)},x^{(2)})\in B(x_{1},\frac{\lambda^{*}}{4})\times B(x_{2},\frac{\lambda^{*}}{4})$ we have $\norm{x^{(1)}-x^{(2)}}\geq \frac{\lambda^{*}}{2}$.

To summarize the proof thus far, for every $y\in\mathcal{S}_{y}$ we have found two sets $Q_{1}(y)=B(x_{1},\frac{\lambda^{*}}{4})$ and $Q_{2}(y)=B(x_{2},\frac{\lambda^{*}}{4})$ (notice that $\lambda^{*}$ depends on $y$), such that for every $(x^{(1)},x^{(2)})\in Q_{1}(y)\times Q_{2}(y)$ we have $\norm{x^{(1)}-x^{(2)}}\geq \frac{\lambda^{*}}{2}$.
Yet, we are not finished with the proof since we need to find a ``shared'' $\beta>0$ over a set of $y$'s with positive probability.

Let
\begin{gather}
    \mathcal{S}_{y}(k)=\left\{y\in\mathcal{S}_{y}\: :\: \frac{\lambda^{*}}{2}\geq\frac{1}{k}\right\},\, k\in\mathbb{N}.
\end{gather}
For every $y\in\mathcal{S}_{y}$ 
it holds that $\lambda^{*}>0$, so it is clear that there exists $k\geq 1$ such that $\frac{\lambda^{*}}{2}\geq\frac{1}{k}$.
Hence, we have
\begin{gather}
    \mathcal{S}_{y}=\cup_{k\geq 1}\mathcal{S}_{y}(k).
\end{gather}
Now, there exists $k^{*}\geq1$ such that $P(Y\in\mathcal{S}_{y}(k^*))>0$, since otherwise $\mathcal{S}_{y}$ is a countable union of sets with zero probability, which is a contradiction to the fact that $P(Y\in\mathcal{S}_{y})>0$.
Hence, for every $y\in\mathcal{S}_{y}(k^*)$ there exists sets $Q_{1}(y),Q_{2}(y)$, such that for every $(x^{(1)},x^{(2)})\in Q_{1}(y)\times Q_{2}(y)$ we have $\norm{x^{(1)}-x^{(2)}}\geq \frac{\lambda^{*}}{2}\geq\frac{1}{k^{*}}$.
Choosing $\beta=\frac{1}{k^{*}}$, the set $\mathcal{S}_{y}(k^*)$, and the sets $Q_{1}(y),Q_{2}(y)$ (for every $y\in\mathcal{S}_{y}(k^*)$) concludes the proof.
\end{proof}
\begin{lemma}\label{corollary:subset-with-required-probability}
    Assume a $\beta$-ill-posed degradation with the sets $\mathcal{S}_{y}, Q_{1}(y), Q_{2}(y)$.
    There exist $k>0$ and a subset $\tilde{S}_{y}\subseteq\mathcal{S}_{y}$ with $P(Y\in \tilde{S}_{y})>0$, such that for every $y\in\tilde{S}_{y}$ it holds that $P(X\in Q_{i}(y)|Y=y)\geq k$.
\end{lemma}
\begin{proof}
Let
\begin{gather}
    \mathcal{S}_{y}(n)=\left\{y\in\mathcal{S}_{y}\: :\: P(X\in Q_{1}(y)|Y=y)\geq\frac{1}{n},\, P(X\in Q_{2}(y)|Y=y)\geq\frac{1}{n}\right\},
\end{gather}
and notice that
\begin{gather}
    \mathcal{S}_{y}=\cup_{n\geq 1}\mathcal{S}_{y}(n).
\end{gather}
There exists $n^{*}\geq 1$ such that 
\begin{gather}
    P(Y\in\mathcal{S}_{y}(n^*))>0,
\end{gather}
since otherwise $\mathcal{S}_{y}$ is a countable union of sets with zero probability, implying that $P(Y\in\mathcal{S}_{y})=0$.
By the definition of $\mathcal{S}_{y}(n^*)$, for every $y\in\mathcal{S}_{y}(n^{*})$ we have
\begin{align}
    &P(X\in Q_{1}(y)|Y=y)\geq\frac{1}{n^*},\\
    &P(X\in Q_{2}(y)|Y=y)\geq\frac{1}{n^*}.
\end{align}
Choosing $k=\frac{1}{n^{*}}$ and $\tilde{\mathcal{S}}_{y}=\mathcal{S}_{y}(n^*)$ concludes the proof.
\end{proof}
\begin{lemma}\label{corollary:l2_norm_bound}
There exists $C>0$ such that, for any vectors $x_{1},y_{1},x_{2},y_{2}$ in finite dimensional space,
if \begin{gather}
    \norm{\begin{pmatrix}
        x_{1}\\y_{1}
    \end{pmatrix}-\begin{pmatrix}
        x_{2}\\y_{2}
    \end{pmatrix}}\leq\epsilon,
\end{gather}
then 
\begin{align}
    &\norm{x_{1}-x_{2}}_{2}\leq C\epsilon,\\
    &\norm{y_{1}-y_{2}}_{2}\leq C\epsilon,
\end{align}
where $\norm{\cdot}_{2}$ is the standard Euclidean norm.
\end{lemma}
\begin{proof}
From the equivalence of norms, for any norm $\norm{\cdot}$ there exists $C>0$ such that for all $z$ in finite dimensional space we have $C\norm{z}\geq\norm{z}_{2}$.
    Hence, we have
    \begin{gather}
        C\epsilon\geq C\norm{\begin{pmatrix}
        x_{1}\\y_{1}
    \end{pmatrix}-\begin{pmatrix}
        x_{2}\\y_{2}
    \end{pmatrix}}\geq \norm{\begin{pmatrix}
        x_{1}\\y_{1}
    \end{pmatrix}-\begin{pmatrix}
        x_{2}\\y_{2}
    \end{pmatrix}}_{2}.
    \end{gather}
    By the definition of the $L_{2}$ norm we have
    \begin{gather}
        \norm{\begin{pmatrix}
        x_{1}\\y_{1}
    \end{pmatrix}-\begin{pmatrix}
        x_{2}\\y_{2}
    \end{pmatrix}}^{2}_{2}=\norm{x_{1}-x_{2}}_{2}^{2}+\norm{y_{1}-y_{2}}_{2}^{2}\leq C^{2}\epsilon^{2},
    \end{gather}
    so
    \begin{gather}
        \norm{x_{1}-x_{2}}_{2}^{2}\leq C^{2}\epsilon^{2},\\
        \norm{y_{1}-y_{2}}_{2}^{2}\leq C^{2}\epsilon^{2}.
    \end{gather}
    Taking square root on both sides concludes the proof.
\end{proof}
\erraticbehavior*
\begin{proof}
Let $\gamma\geq\epsilon$, suppose that $\hat{X}$ satisfies
\begin{gather}\label{equation:wasserstein-p-epsilon}
    W_{p}(p_{X,Y},p_{\hat{X},Y})=\inf_{p_{X_{1},Y_{1},X_{2},Y_{2}}}\left\{\left(\mathbb{E}\left[\norm{\begin{pmatrix}
        X_{1}\\Y_{1}
    \end{pmatrix}-\begin{pmatrix}
        X_{2}\\Y_{2}
    \end{pmatrix}}^{p}\right]\right)^{\frac{1}{p}}\: :\: p_{X_{1},Y_{1}}=p_{X,Y},\: p_{X_{2},Y_{2}}=p_{\hat{X},Y}\right\}=\epsilon,
\end{gather}
and let $p_{X_{1}^{*},Y_{1}^{*},X_{2}^{*},Y_{2}^{*}}$ be the solution attaining the infimum in~\cref{equation:wasserstein-p-epsilon}.
Denote
\begin{gather}
  D=\norm{\begin{pmatrix}
    X_{1}^{*}\\Y_{1}^{*}
\end{pmatrix}-\begin{pmatrix}
    X_{2}^{*}\\Y_{2}^{*}
\end{pmatrix}}^{p},
\end{gather}
so by the definition of $D$ we have
\begin{gather}\mathbb{E}[D]=\epsilon^{p}.
\end{gather}
Since $D\geq 0$ almost surely, from Markov's inequality it holds that, for any $a>0$,
\begin{gather}
    P(D\geq a)\leq \frac{\mathbb{E}[D]}{a}=\frac{\epsilon^{p}}{a}.
\end{gather}
Let us choose $a=t^{p}\epsilon^{0.5p}$ for some $t>0$, so we have
\begin{gather}
    P(D\geq t^{p}\epsilon^{0.5p})\leq\frac{\epsilon^p}{t^{p}\epsilon^{0.5p}}=\frac{1}{t^{p}}\epsilon^{0.5p}.
\end{gather}

The degradation is not invertible, so from~\cref{corollary:non-invertible-is-beta-ill-posed} the degradation is $\beta$-ill-posed for some $\beta>0$, and from~\cref{corollary:subset-with-required-probability} there exist $k>0$ and a set $\mathcal{S}_{y}$ with $P(Y\in\mathcal{S}_{y})>0$ (denoted as $\tilde{\mathcal{S}}_{y}$ in~\cref{corollary:subset-with-required-probability}), where for every $y\in\mathcal{S}_{y}$ there exist sets $Q_{1}(y),Q_{2}(y)$ such that
\begin{align}
    &P(X\in Q_{1}(y)|Y=y)\geq k,\\
    &P(X\in Q_{2}(y)|Y=y)\geq k,
\end{align}
and for every $(x^{(1)},x^{(2)})\in Q_{1}(y)\times Q_{2}(y)$ we have $\norm{x^{(1)}-x^{(2)}}\geq\beta$.

Let
\begin{align}
    & A_{i}=\left\{(x_{1},y_{1},x_{2},y_{2})\::\: \norm{\begin{pmatrix}
        x_{1}\\y_{1}
    \end{pmatrix}-\begin{pmatrix}
        x_{2}\\y_{2}
    \end{pmatrix}}^{p}<t^{p}\epsilon^{0.5p},\,y_{1}\in\mathcal{S}_{y},\,x_{1}\in Q_{i}(y_1),(x_{2},y_{2})\in\supp{p_{\hat{X},Y}}\right\},\\
    & B_{i}=\left\{y_{1}\: :\: (\_,y_{1},\_,\_)\in A_{i}\right\}.
\end{align}
We will show that for a large enough $t>0$ we have
\begin{align}
P(Y\in B_{1}\cap B_{2})>0,
\end{align}
and this will hold true as long as $\epsilon\leq\gamma$.

By the definition of $ A_{i}$ and $ B_{i}$ we have
\begin{align}
P(X\in Q_{i}(Y),Y\in B_{i})&=P(X_{1}^{*}\in Q_{i}(Y_{1}^{*}),Y_{1}^{*}\in B_{i})\\&\geq P((X_{1}^{*},Y_{1}^{*},X_{2}^{*},Y_{2}^{*})\in A_{i})\\&=P(D<t^{p}\epsilon^{0.5p}, X_{1}^{*}\in Q_{i}(Y_{1}^{*}),Y_{1}^{*}\in\mathcal{S}_{y})\\&=P(X_{1}^{*}\in Q_{i}(Y_{1}^{*}),Y_{1}^{*}\in\mathcal{S}_{y})-P(D\geq t^{p}\epsilon^{0.5p}, X_{1}^{*}\in Q_{i}(Y_{1}^{*}),Y_{1}^{*}\in\mathcal{S}_{y})\\
    &=P(X\in Q_{i}(Y),Y\in\mathcal{S}_{y})-P(D\geq t^{p}\epsilon^{0.5p}, X_{1}^{*}\in Q_{i}(Y_{1}^{*}),Y_{1}^{*}\in\mathcal{S}_{y})\\
    &\geq P(X\in Q_{i}(Y),Y\in\mathcal{S}_{y})-P(D\geq t^{p}\epsilon^{0.5p})\\
    &\geq P(X\in Q_{i}(Y),Y\in\mathcal{S}_{y})-\frac{1}{t^{p}}\epsilon^{0.5p},
\end{align}
so $P(X\in Q_{i}(Y),Y\in B_{i})\geq P(X\in Q_{i}(Y),Y\in\mathcal{S}_{y})-\frac{1}{t^{p}}\epsilon^{0.5p}$.
Since $ B_{i}\subseteq\mathcal{S}_{y}$, by rearranging the inequality we have
\begin{gather}
\frac{1}{t^{p}}\epsilon^{0.5p}\geq P(X\in Q_{i}(Y),Y\in\mathcal{S}_{y}\setminus B_{i}),
\end{gather}
so it holds that
\begin{align}
    \frac{1}{t^{p}}\epsilon^{0.5p}&\geq P(X\in Q_{i}(Y),Y\in\mathcal{S}_{y}\setminus B_{i})\\
    &=\int_{\mathcal{S}_{y}\setminus B_{i}}p_{Y}(y)\int_{Q_{i}(y)}p_{X|Y}(x|y)dxdy\\
    &=\int_{\mathcal{S}_{y}\setminus B_{i}}p_{Y}(y)P(X\in Q_{i}(y)|Y=y)dy\\
    &\geq k\int_{\mathcal{S}_{y}\setminus B_{i}}p_{Y}(y)dy\\
    &=kP(Y\in\mathcal{S}_{y}\setminus B_{i})\\
    \Rightarrow \frac{1}{kt^{p}}\gamma^{0.5p}&\geq \frac{1}{kt^{p}}\epsilon^{0.5p}\geq P(Y\in\mathcal{S}_{y}\setminus B_{i}).
\end{align}
Since $P(Y\in\mathcal{S}_{y})>0$, for a large enough $t>0$ it holds that $\frac{1}{kt^{p}}\gamma^{0.5p}<\frac{1}{2}P(Y\in\mathcal{S}_{y})$ (this choice does not depend on $\epsilon$), and so
\begin{gather}
    P(Y\in\mathcal{S}_{y}\setminus B_{i})<\frac{1}{2}P(Y\in\mathcal{S}_{y}).
\end{gather}
Therefore, we have
\begin{align}
P(Y\in B_{i})&=P(Y\in\mathcal{S}_{y})-P(Y\in\mathcal{S}_{y}\setminus B_{i})\\
&> P(Y\in\mathcal{S}_{y})-\frac{1}{2}P(Y\in\mathcal{S}_{y})\\
&=\frac{1}{2}P(Y\in\mathcal{S}_{y}),
\end{align}
and finally we get
\begin{align}
P(Y\in B_{1}\cap B_{2})&=P(Y\in B_{1})+P(Y\in B_{2})-P(Y\in B_{1}\cup B_{2})\\
&> 2\cdot \frac{1}{2}P(Y\in\mathcal{S}_{y})-P(Y\in\mathcal{S}_{y})=0,
\end{align}
so indeed $P(Y\in B_{1}\cap B_{2})>0$.

Let us now construct a lower bound on the Lipschitz constant of $\hat{X}$.
From the definition of $B_{i}$, for any $y_{1}\in B_{i}$ there exists $x_{1}^{(i)}\in Q_{i}(y_{1})$ and $(x_{2}^{(i)},y_{2}^{(i)})\in\supp{p_{\hat{X},Y}}$ such that $(x_{1}^{(i)},y_{1},x_{2}^{(i)},y_{2}^{(i)})\in A_{i}$.
In particular, since $P(Y\in B_{1}\cap  B_{2})>0$ and $ B_{1}\cap  B_{2}\subseteq B_{i}$, we can take $y_{1}\in B_{1}\cap  B_{2}$ so that $(x_{1}^{(i)},y_{1},x_{2}^{(i)},y_{2}^{(i)})\in A_{i}$ are both with the same $y_{1}$ for $i=1,2$.
From the definition of $A_{i}$ we have
\begin{gather}
    \norm{\begin{pmatrix}
        x_{1}^{(i)}\\y_{1}
    \end{pmatrix}-\begin{pmatrix}
        x_{2}^{(i)}\\y_{2}^{(i)}
    \end{pmatrix}}\leq t\sqrt{\epsilon},
\end{gather}
and thus from~\cref{corollary:l2_norm_bound} we have
\begin{gather}
    \norm{x_{1}^{(i)}-x_{2}^{(i)}}_{2}\leq Ct\sqrt{\epsilon},\\
    \norm{y_{1}-y_{2}^{(i)}}_{2}\leq Ct\sqrt{\epsilon}.
\end{gather}
Hence, from the triangle inequality we have
\begin{align}
    \norm{y_{2}^{(1)}-y_{2}^{(2)}}_{2}&\leq \norm{y_{2}^{(1)}-y_{1}}_{2}+\norm{y_{1}-y_{2}^{(2)}}_{2}\\
    &\leq 2Ct\sqrt{\epsilon},
\end{align}
and
\begin{align}
    \norm{x_{1}^{(1)}-x_{1}^{(2)}}_{2}&\leq \norm{x_{1}^{(1)}-x_{2}^{(1)}}_{2}+\norm{x_{2}^{(1)}-x_{1}^{(2)}}_{2}\\
    &\leq \norm{x_{1}^{(1)}-x_{2}^{(1)}}_{2}+\norm{x_{2}^{(1)}-x_{2}^{(2)}}_{2}+\norm{x_{2}^{(2)}-x_{1}^{(2)}}_{2}.\label{eq:inequality_of_x}
\end{align}
We are interested in $\norm{x_{2}^{(1)}-x_{2}^{(2)}}_{2}$. 
Since $(x_{1}^{(1)},x_{1}^{(2)})\in Q_{1}(y_{1})\times Q_{2}(y_{1})$ we have
\begin{gather}
    \norm{x_{1}^{(1)}-x_{1}^{(2)}}_{2}\geq\beta.
\end{gather}
Thus, by rearranging~\cref{eq:inequality_of_x} we get
\begin{align}
    \norm{x_{2}^{(1)}-x_{2}^{(2)}}_{2}&\geq \norm{x_{1}^{(1)}-x_{1}^{(2)}}_{2}-\norm{x_{2}^{(2)}-x_{1}^{(2)}}_{2}-\norm{x_{1}^{(1)}-x_{2}^{(1)}}_{2}\\
    &\geq \beta-2Ct\sqrt{\epsilon},
\end{align}
so
\begin{gather}
    \frac{\norm{x_{2}^{(1)}-x_{2}^{(2)}}_{2}}{\norm{y_{2}^{(1)}-y_{2}^{(2)}}_{2}}\geq\frac{\beta-2Ct\sqrt{\epsilon}}{2Ct\sqrt{\epsilon}}.
\end{gather}
From the equivalence of norms in finite dimensional space, there exist constants $c_{1},c_{2}>0$, such that for any $x_{2}^{(1)},x_{2}^{(2)},y_{2}^{(1)},y_{2}^{(2)}$ it holds that $\norm{x_{2}^{(1)}-x_{2}^{(2)}}\geq c_{1}\norm{x_{2}^{(1)}-x_{2}^{(2)}}_{2}$ and $\norm{y_{2}^{(1)}-y_{2}^{(2)}}\leq c_{2}\norm{x_{2}^{(1)}-x_{2}^{(2)}}_{2}$.
Hence,
\begin{align}
    &\frac{\norm{x_{2}^{(1)}-x_{2}^{(2)}}}{\norm{y_{2}^{(1)}-y_{2}^{(2)}}}\geq\frac{c_{1}\norm{x_{2}^{(1)}-x_{2}^{(2)}}_{2}}{c_{2}\norm{y_{2}^{(1)}-y_{2}^{(2)}}_{2}}\\
    &\geq\frac{c_{1}}{c_{2}}\frac{\beta-2Ct\sqrt{\epsilon}}{2Ct\sqrt{\epsilon}}\\
    &\geq\frac{c_{1}}{c_{2}}\frac{\beta}{2Ct\sqrt{\epsilon}}-\frac{c_{1}}{c_{2}}\\
    &=\frac{m_{1}}{\sqrt{\epsilon}}-m_{2}\label{eq:lip-const-bound},
\end{align}
where $m_{1}=\frac{c_{1}}{c_{2}}\frac{\beta}{2Ct}$ and $m_{2}=\frac{c_{1}}{c_{2}}$.
To connect this result with the Lipschitz constant of $\hat{X}$, notice that from the definition of $A_{i}$ we have $(x_{2}^{(i)},y_{2}^{(i)})\in\supp{p_{\hat{X},Y}}$, so $x_{2}^{(i)}=f(y_{2}^{(i)})$.
Thus, from~\cref{eq:lip-const-bound} we have
\begin{gather}
    \frac{\norm{f\left(y_{2}^{(1)}\right)-f\left(y_{2}^{(2)}\right)}}{\norm{y_{2}^{(1)}-y_{2}^{(2)}}}\geq \frac{m_{1}}{\sqrt{\epsilon}}-m_{2}.\label{eq:lip_final}
\end{gather}
To conclude, for every $0<\epsilon\leq\gamma$, we have found two inputs $y_{2}^{(1)},y_{2}^{(2)}\in\supp{p_{Y}}$ for which~\cref{eq:lip_final} holds, i.e.,
\begin{gather}
   \text{Lip}(\hat{X})\geq\frac{m_{1}}{\sqrt{\epsilon}}-m_{2}.
\end{gather}
Note that $C>0$ depends only on the norm $\norm{\cdot}$, and is constant for any $\epsilon$.
Lastly, while this result holds for any $0<\epsilon\leq\gamma$, one can always choose an arbitrarily large value of $\gamma$ (and accordingly, an appropriate value of $t$) to obtain a lower bound of $K$ for any range of values of $0<\epsilon\leq\gamma$, but this lower bound becomes less tight for larger values of $\gamma$.
\end{proof}
\section{Additional Details of~\cref{section:toy_example} (Toy Example Illustration)}\label{appendix:toy_example}
\subsection{Full Training Details of the CGAN-Based Denoisers}
Both the denoiser $\hat{X}_{\lambda}$ and the discriminator consist of 7 fully-connected layers with ReLU activations.
For both the denoiser and the discriminator, the ReLU activation is applied on all layers except for the last one.
All the layers have an input and output size of 16, except for the first and last layers.
The input size of the last layer is 16 and the output size is 1 for both the denoiser and the discriminator.
The input size of the first layer is 1 for the generator and 2 for the discriminator, and the output size is 16 for both.
To clarify, as in~\citep{MirzaO14}, the discriminator is fed one time with $(X,Y)$ (``real'' samples) and one time with $(\hat{X},Y)$ (``fake'' samples) to compute the loss.
The training set consists of $100,000$ independently drawn samples from $p_{X,Y}$, which is a Gaussian distribution with mean $0$ and covariance $\begin{pmatrix}
    1&1\\1&2
\end{pmatrix}$.
We train the networks with a non-saturating generative adversarial training loss~\citep{gan}, and the discriminator is regularized using the $R_{1}$ gradient penalty~\citep{Mescheder2018} with coefficient $1.0$.
We optimize both of the networks using the Adam optimizer~\citep{adam} with $\beta_{1}=0.5$, $\beta_{2}=0.9$, a learning rate of $10^{-4}$, a batch size of 128, and for a total of $100,000$ training steps for each network (we perform one training step at a time for each network).
We multiply the learning rate by $0.5$ every 5,000 training steps, starting after 50,000 steps, i.e., we perform multi-step learning rate scheduling.

\subsection{Discussion and Ideas for Future Work}\label{appendix:additional_toy_discussion}
\begin{figure}
    \centering
\includegraphics{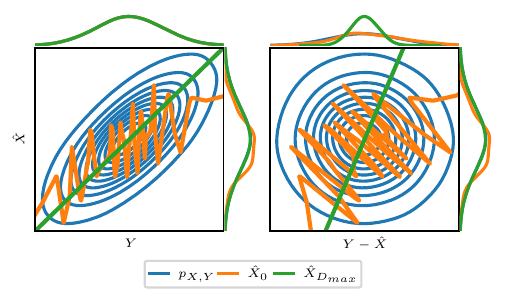}
    \caption{A visual illustration of $\hat{X}_{0}$ and $\hat{X}_{D_{max}}$ from the toy example in~\cref{appendix:additional_toy_discussion} (with $\sigma_{N}=1$). \textbf{Left}: Contour plot of the density $p_{X,Y}$ (blue concentric ellipses), outputs from $\hat{X}_{0}$ (orange ``zigzag'' line), and outputs from $\hat{X}_{D_{max}}$ (green straight line).
    \textbf{Right}: the same as the left plot, but instead of $Y$ we consider $Y-\hat{X}$.
    We also present the marginal density functions of each axis, i.e., the densities $p_{\hat{X}},p_{Y}$ and $p_{Y-\hat{X}}$. These marginal densities are computed using KDE.}
    \label{figure:toy-example}
\end{figure}

Consider the same toy example as in~\cref{section:toy_example}, with \mbox{$N\sim\mathcal{N}(0,\sigma_{N}^{2})$}.
It is well known that
\begin{align}
    &\hat{X}_{\text{MMSE}}=\frac{1}{1+\sigma_{N}^{2}}Y,\\
    &\hat{X}_{p}=\frac{1}{1+\sigma_{N}^{2}}Y+W,
\end{align}
where $\hat{X}_{\text{MMSE}}$ attains the lowest possible Mean-Squared-Error (MSE), $\hat{X}_{p}$ is a sampler from the posterior distribution $p_{X|Y}$, and $W\sim \mathcal{N}(0,1-\frac{1}{1+\sigma_{N}^{2}})$ is independent of $X$, $Y$ and $N$.
Moreover, consider the estimator which attains the lowest possible MSE among all estimators with perfect perceptual quality, $\hat{X}_{D_{max}}$.
In our specific example $X$ and $Y$ are jointly Gaussian, so from~\citep{Freirich2021,Blau2018} we have
\begin{gather}
    \hat{X}_{D_{max}}=\frac{1}{\sqrt{1+\sigma_{N}^{2}}}Y.\label{formula:dmax}
\end{gather}
Now, let us compare $\hat{X}_{p}$ and $\hat{X}_{D_{max}}$ with $\hat{X}_{0}$, the estimator $\hat{X}_{\lambda}$ from~\cref{section:toy_example} with $\lambda=0$, i.e., an estimator that is trained solely with a CGAN loss.
In~\cref{figure:toy-example} we show a contour plot of the density $p_{X,Y}$ ($p_{\hat{X}_{p},Y}=p_{X,Y}$), as well as samples from $p_{\hat{X}_{0},Y}$ and $p_{\hat{X}_{D_{max}},Y}$.
The samples from $p_{\hat{X}_{0},Y}$ and $p_{\hat{X}_{D_{max}},Y}$ are obtained by independently drawing samples $y\in p_{Y}$ and passing each $y$ through $\hat{X}_{0}$ (the trained neural network) and $\hat{X}_{D_{max}}$ (\cref{formula:dmax}).
Observe that $\hat{X}_{0}$, a deterministic estimator that seeks to satisfy $p_{\hat{X},Y}\approx p_{X,Y}$, indeed behaves erratically, as anticipated by~\cref{theorem:erratic_behavior}.

It appears, though, that $\hat{X}_{D_{max}}$ is a robust estimator.
So why would one choose to use $\hat{X}_{0}$ over an estimator such as $\hat{X}_{D_{max}}$, which, amongst all the estimators with perfect perceptual quality, attains the lowest possible MSE?
Namely, instead of using an erratic estimator which attains $p_{\hat{X},Y}\approx p_{X,Y}$, why not choose a robust estimator which attains very high perceptual quality and produces outputs that are sufficiently close to the source image?
If such an estimator exists, we describe next two potential issues.
\subsubsection{Conditional MSE}\label{section:conditional_mse}
\begin{figure}\centering\includegraphics{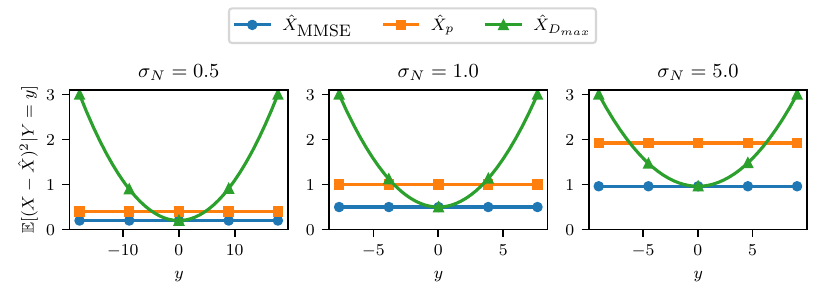}
    \caption{The conditional MSE $\mathbb{E}[(X-\hat{X})|Y=y]$ versus $y$ obtained by each of the estimators $\hat{X}_{\text{MMSE}}$, $\hat{X}_{p}$ and $\hat{X}_{D_{max}}$ which are described in~\cref{appendix:additional_toy_discussion}. Each subplot corresponds to a different value of $\sigma_{N}$ (mentioned in the title).
    The $y$ values in each subplot are taken such that the conditional MSE is below or equal to $3$.}
    \label{fig:conditional-mse-gaussian}
\end{figure}

Let us evaluate the conditional MSE, defined by \mbox{$\mathbb{E}[(X-\hat{X})^{2}|Y=y]$}.
It holds that
\begin{align}
    \mathbb{E}&[(X-\hat{X}_{\text{MMSE}})^{2}|Y=y]=1-\frac{1}{1+\sigma_{N}^{2}},\\
    \mathbb{E}&[(X-\hat{X}_{p})^{2}|Y=y]=2\left(1-\frac{1}{1+\sigma_{N}^{2}}\right),\\
    \mathbb{E}&[(X-\hat{X}_{D_{max}})^{2}|Y=y]=1-\frac{1}{1+\sigma_{N}^{2}}+\left(\frac{\sqrt{1+\sigma_{N}^{2}}-1}{1+\sigma_{N}^{2}}\right)^{2}y^{2},
\end{align}
(see visual illustration in~\cref{fig:conditional-mse-gaussian} and proof below).
Notice that both the conditional MSE of $\hat{X}_{\text{MMSE}}$ and $\hat{X}_{p}$ are constant in $y$, while that of $\hat{X}_{D_{max}}$ is proportional to $y^{2}$ and bounded from below by the conditional MSE of $\hat{X}_{\text{MMSE}}$.
We find it interesting that even though $\hat{X}_{D_{max}}$ attains a lower MSE compared to $\hat{X}_{p}$ when averaging over all $y$'s, it produces worse conditional MSE when the absolute value of $y$ is large.
Hence, $\hat{X}_{D_{max}}$ does not simply attain better MSE ``for free'', as it sacrifices the conditional MSE originating from less likely values of $y$ to serve better more likely values of $y$.
Perhaps this phenomenon can be extended to more general settings, but we leave this for future work.

To see why the expressions above hold, first note that
\begin{align}
    \mathbb{E}[(X-\hat{X}_{\text{MMSE}})^{2}|Y=y]=Var(X|Y=y)=1-\frac{1}{1+\sigma_{N}^{2}}
\end{align}
is a well known result.
Now, for any estimator $\hat{X}$ we have
\begin{align}
    \mathbb{E}[(X-\hat{X})^{2}|Y=y]&=\mathbb{E}[X^{2}|Y=y]+\mathbb{E}[\hat{X}^{2}|Y=y]-2\mathbb{E}[X\hat{X}|Y=y]\\
    &=\mathbb{E}[X^{2}|Y=y]+\mathbb{E}[\hat{X}^{2}|Y=y]-2\mathbb{E}[X|Y=y]\mathbb{E}[\hat{X}|Y=y]\label{eq:independent_given_y},
\end{align}
where~\cref{eq:independent_given_y} is true since $\hat{X}$ and $X$ are independent given $Y$.
Hence, we have
\begin{align}
    \mathbb{E}[(X-\hat{X}_{p})^{2}|Y=y]=&\mathbb{E}[X^{2}|Y=y]+\mathbb{E}[\hat{X}_{p}^{2}|Y=y]-2\mathbb{E}[X|Y=y]\mathbb{E}[\hat{X}_{p}|Y=y]\\
    =&2\mathbb{E}[X^{2}|Y=y]-2\left(\mathbb{E}[X|Y=y]\right)^{2}\\
    =&2Var(X|Y=y)\\
    =&2\left(1-\frac{1}{1+\sigma_{N}^{2}}\right).
\end{align}
We are left with obtaining $\mathbb{E}[(X-\hat{X}_{D_{max}})^{2}|Y=y]$.
It holds that
\begin{gather}
    \mathbb{E}[X^{2}|Y=y]=Var(X|Y=y)+(\mathbb{E}[X|Y=y])^{2}=1-\frac{1}{1+\sigma_{N}^{2}}+\frac{1}{(1+\sigma_{N}^{2})^{2}}y^{2}.
\end{gather}
Moreover, for $\hat{X}_{D_{max}}$ we have
\begin{align}
    &\mathbb{E}[\hat{X}_{D_{max}}|Y=y]=\mathbb{E}\left[\frac{1}{\sqrt{1+\sigma_{N}^{2}}}Y\middle|Y=y\right]=\frac{1}{\sqrt{1+\sigma_{N}^{2}}}y,\\
    &\mathbb{E}[\hat{X}_{D_{max}}^{2}|Y=y]=\mathbb{E}\left[\frac{1}{1+\sigma_{N}^{2}}Y^{2}\middle|Y=y\right]=\frac{1}{1+\sigma_{N}^{2}}y^{2},
\end{align}
and thus,
\begin{align}
    \mathbb{E}[(X-\hat{X}_{D_{max}})^{2}|Y=y]&=1-\frac{1}{1+\sigma_{N}^{2}}+\frac{1}{(1+\sigma_{N}^{2})^{2}}y^{2}+\frac{1}{1+\sigma_{N}^{2}}y^{2}-\frac{2}{(1+\sigma_{N}^{2})\sqrt{1+\sigma_{N}^{2}}}y^{2}\\
    &=1-\frac{1}{1+\sigma_{N}^{2}}+\left(\frac{1+(1+\sigma_{N}^{2})-2(1+\sigma_{N}^{2})^{0.5}}{(1+\sigma_{N}^{2})^{2}}\right)y^{2}\\
    &=1-\frac{1}{1+\sigma_{N}^{2}}+\left(\frac{\sqrt{1+\sigma_{N}^{2}}-1}{1+\sigma_{N}^{2}}\right)^{2}y^{2}.
\end{align}

\subsubsection{Consistency in the additive noise setting}\label{section:residual_noise_toy_example}
\cref{figure:toy-example} provides a visual intuition for another apparent issue with $\hat{X}_{D_{max}}$, which clarifies the distinction between perceptual quality and joint perceptual quality.
While $\hat{X}_{D_{max}}$ attains perfect perceptual quality, observe that it produces each source $x\in\supp{p_{X}}$ for only one particular measurement $y\in\supp{p_{Y}}$, i.e., it is an invertible function.
Is it practically acceptable that an estimator could produce a given natural image only for one particular noisy measurement?
The issue with this phenomenon is clearer when looking at the right plot of $\hat{X}$ versus $Y-\hat{X}$.
There, we see that $\hat{X}_{D_{max}}$ associates each output $\hat{X}$ with one possible residual noise $\hat{N}_{D_{max}}=Y-\hat{X}_{D_{max}}$, so $\hat{N}_{D_{max}}$ and $\hat{X}_{D_{max}}$ are statistically dependent, whereas $N$ and $X$ are not.
Moreover, from the Kernel Density Estimation (KDE) of the marginal distributions in~\cref{figure:toy-example}, we also see that the distribution of $\hat{N}_{D_{max}}$ is quite different from that $N$.
These phenomena reveal that $\hat{X}_{D_{max}}$ does not fully ``separate'' $X$ from $N$, leaving irrelevant portions from the noise in the estimated output.

Let us elaborate on this phenomenon for general distributions of $X$ and $N$.
Suppose that $Y=X+N$, where $N$ is some random noise independent of $X$.
Let $\hat{X}$ be an image denoiser, and let $\hat{N}=Y-\hat{X}$ be the residual noise.
A common, generally desired property of $\hat{X}$ is that $\hat{N}$ would follow the same distribution as $N$ (e.g., isotropic Gaussian) and be statistically uncorrelated with $\hat{X}$.
It turns out that $p_{\hat{X},Y}=p_{X,Y}$ if and only if  $p_{\hat{X}}=p_{X}$ and $\smash{\forall (x,n)\::\:p_{\hat{N}|\hat{X}}(n|x)=p_{N}(n)}$ (which we write as $\smash{p_{\hat{N}|\hat{X}}=p_{N}}$ in short).
Namely, $\smash{\hat{X}}$ is a posterior sampler if and only if it attains perfect perceptual quality, and $\hat{N}$ is statistically \emph{independent} of $\hat{X}$ and follows the same distribution as that of $N$.
To show that this is true, we simply show that $p_{Y|\hat{X}}=p_{Y|X}$ if and only if $p_{\hat{N}|\hat{X}}=p_{N}$.
Assume first that $p_{Y|\hat{X}}=p_{Y|X}$.
We have
\begin{align}
    p_{\hat{N}|\hat{X}}(n|x)&=\int p_{\hat{N},Y|\hat{X}}(n,y|x)dy\\
    &=\int p_{\hat{N}|Y,\hat{X}}(n|y,x)p_{Y|\hat{X}}(y|x)dy\\
    &=\int \delta(n-(y-x))p_{Y|\hat{X}}(y|x)dy\\
    &=\int \delta(n-(y-x))p_{Y|X}(y|x)dy\\
    &=p_{Y|X}(n+x|x)\\
    &=p_{N}(n).
\end{align}
From the other direction, assume that $p_{\hat{N}|\hat{X}}=p_{N}$.
In that case we have
\begin{align}
    p_{Y|\hat{X}}(y|x)&=\int p_{Y,\hat{N}|\hat{X}}(y,n|x)dn\\
    &=\int p_{Y|\hat{X},\hat{N}}(y|x,n)p_{\hat{N}|\hat{X}}(n|x)dn\\
    &=\int \delta(y-(x+n))p_{N}(n)dn\\
    &=p_{N}(y-x)\\
    &=p_{Y|X}(y|x).
\end{align}

This is quite an interesting result.
First, it gives a comprehensible meaning to the condition $p_{Y|\hat{X}}=p_{Y|X}$ in the case of additive noise degradations, a meaning which is already discussed in the literature.
Second, when $\hat{X}$ attains high perceptual quality, and if one desires that the estimator would (almost) perfectly separate the signal from the noise in the sense that $\hat{N}$ is independent of $\hat{X}$ and $p_{\hat{N}}=p_{N}$, then $\hat{X}$ must be close to be a posterior sampler.
Hence, when $\hat{X}$ is a deterministic estimator, to approach perfect perceptual quality and perfect noise separation, $\hat{X}$ must possess a high Lipschitz constant (\cref{theorem:erratic_behavior}).
This also means that, whenever $\hat{X}_{D_{max}}$ is a robust deterministic estimator, it cannot possibly lead to near-perfect noise separation, and therefore cannot lead to near-perfect consistency;
such an estimator cannot satisfy $p_{\hat{X}_{D_{max}},Y}\approx p_{X,Y}$.

\section{Additional Details of~\cref{section:real-world-experiments} (Quantitative Demonstration)}\label{appendix:real-world-algorithms-additional-details}
All our experiments rely on the official codes and checkpoints of the evaluated methods.
\subsection{Complementary Plots of $\overline{K}$ Versus $\sqrt{\text{JFID}}$ with All the Evaluated Methods}
In~\cref{fig:real_world_with_fid} we provide the same plots as in~\cref{fig:real-world-attacks} ($\overline{K}$ versus $\sqrt{\text{JFID}}$) but this time include all the algorithms that are evaluated in~\cref{section:real-world-experiments} which we omitted from~\cref{fig:real-world-attacks}.

\subsection{Plots of Additional Joint Perceptual Quality Measures}
In~\cref{fig:real_world_with_kid,fig:real_world_with_precision,fig:real_world_with_recall} we provide the same plots as in~\cref{fig:real_world_with_fid}, but instead of using FID to compute statistical discrepancy, we use KID~\cite{kid}, Precision~\cite{precision_recall}, and Recall~\cite{precision_recall}.
Similarly to JFID, we denote by Joint-KID (JKID), Joint-Precision (JP), and Joint-Recall (JR), the KID, precision, and recall (respectively) between the sets 
\begin{align}
    \left\{\begin{pmatrix}
    x_{F}^{(i)}\\ y^{(i)}_{F}
\end{pmatrix}\right\}_{i=1}^{N}\text{ and }\left\{\begin{pmatrix}
    \hat{x}_{F}^{(i)}\\ y^{(i)}_{F}
\end{pmatrix}\right\}_{i=1}^{N}.
\end{align}
Moreover, similar to the computation of the perceptual quality via FID, we compute the KID, Precision (P), and Recall (R) between the sets $\{x_{F}^{(i)}\}_{i=1}^{N}$ and $\{\hat{x}_{F}^{(i)}\}_{i=1}^{N}$.
Interestingly, we observe an empirical tradeoff between $\overline{K}$ and each of these additional joint perceptual quality measures as well, i.e., larger values of $\overline{K}$ correspond to lower values of $\sqrt{\text{JKID}}$ (lower is better), higher values of JP (higher is better), and higher values of JR (higher is better).
These results show that the tradeoff between $\overline{K}$ and $\sqrt{\text{JFID}}$ is not merely a byproduct of using FID as a statistical discrepancy measure, since we observe the tradeoff when using other types of statistical discrepancy measures as well.
\subsection{Assessing the Effectiveness of the Joint Perceptual Quality Measures}
\begin{figure}
    \centering
    \includegraphics{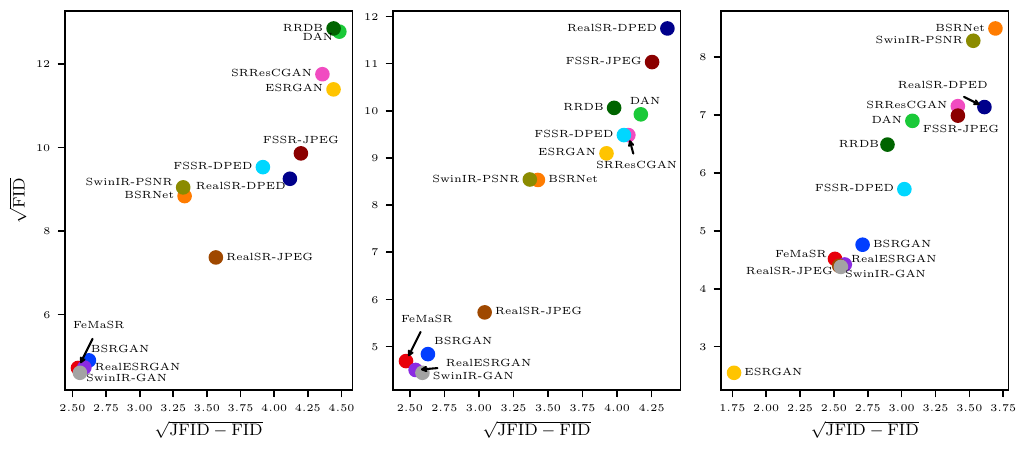}
    \caption{Plots of $\sqrt{\text{FID}}$ versus $\sqrt{\text{JFID-FID}}$. The left plot corresponds to the degradation from the Track2 challenge in~\citep{lugmayr2019aim}, the center plot corresponds to the degradation from the Track1 challenge in~\citep{Lugmayr_2020_CVPR_Workshops}, and the right plot corresponds to the common bicubic degradation.
    We see that two algorithms with a similar FID value (similar perceptual quality) may attain different JFID values (different joint perceptual quality). This shows that JFID is not only influenced by perceptual quality, but also by consistency.
    }
    \label{fig:jfid-relevance}
\end{figure}

\begin{figure}
    \centering
    \includegraphics{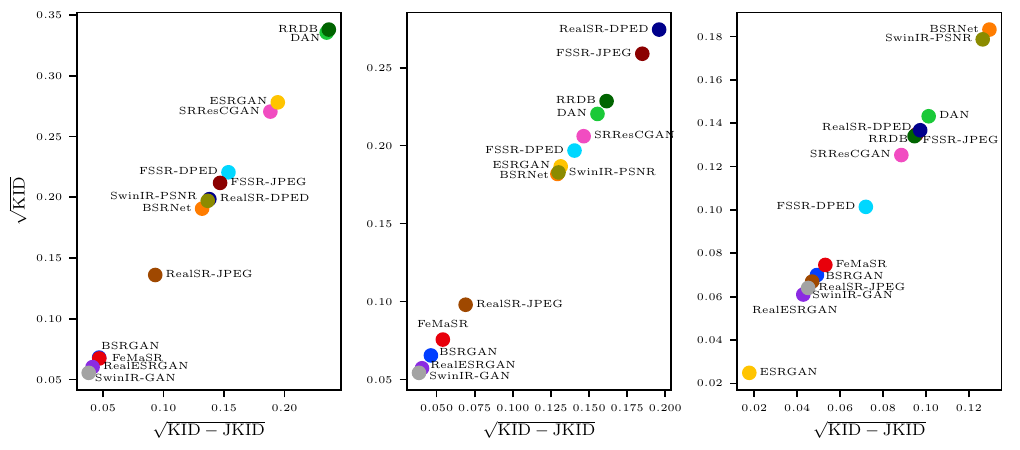}
    \caption{Plots of $\sqrt{\text{KID}}$ versus $\sqrt{\text{KID-JKID}}$. The left plot corresponds to the degradation from the Track2 challenge in~\citep{lugmayr2019aim}, the center plot corresponds to the degradation from the Track1 challenge in~\citep{Lugmayr_2020_CVPR_Workshops}, and the right plot corresponds to the common bicubic degradation.
    We observe a linear relationship between $\sqrt{\text{KID}}$ and $\sqrt{\text{KID-JKID}}$, so JKID is influenced solely by perceptual quality and completely ignores consistency.
    Thus, JKID is an ineffective measure of the joint perceptual quality.}
    \label{fig:jkid-relevance}
\end{figure}

\begin{figure}
    \centering
    \includegraphics{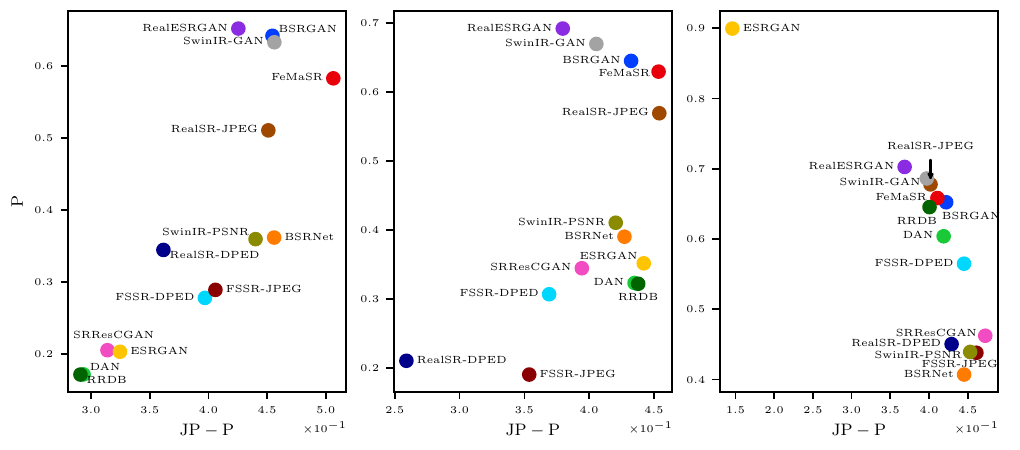}
    \caption{Plots of P versus $\text{JP}-\text{P}$. The left plot corresponds to the degradation from the Track2 challenge in~\citep{lugmayr2019aim}, the center plot corresponds to the degradation from the Track1 challenge in~\citep{Lugmayr_2020_CVPR_Workshops}, and the right plot corresponds to the common bicubic degradation.
    We see that two algorithms with a similar P value may attain different JP values. This shows that JP is not only influenced by perceptual quality, but also by consistency.}
    \label{fig:jp-relevance}
\end{figure}

\begin{figure}
    \centering
    \includegraphics{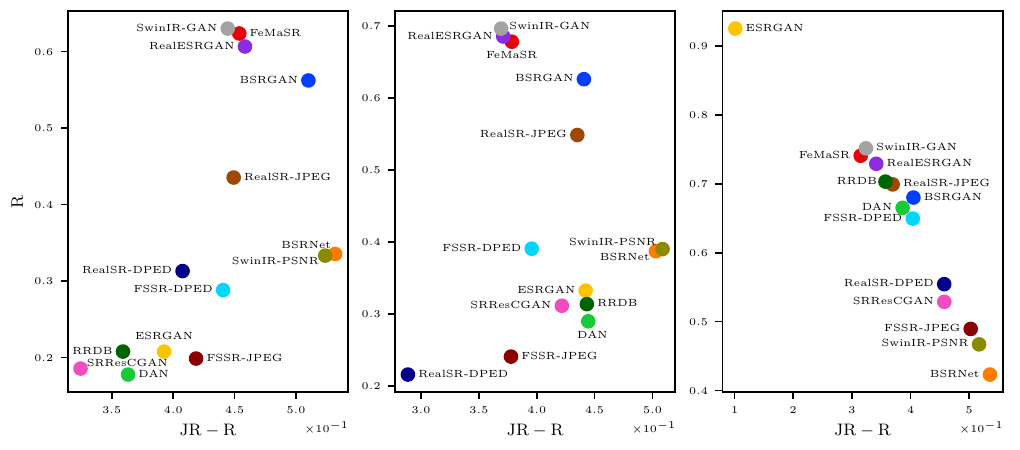}
    \caption{Plots of R versus $\text{JR}-\text{R}$. The left plot corresponds to the degradation from the Track2 challenge in~\citep{lugmayr2019aim}, the center plot corresponds to the degradation from the Track1 challenge in~\citep{Lugmayr_2020_CVPR_Workshops}, and the right plot corresponds to the common bicubic degradation.
    We see that two algorithms with a similar R value may attain different JR values. This shows that JR is not only influenced by perceptual quality, but also by consistency.}
    \label{fig:jr-relevance}
\end{figure}

We question whether our approach to measure joint perceptual quality is predominantly affected by perceptual quality or is it also influenced by consistency.
In~\cref{fig:jfid-relevance,fig:jkid-relevance,fig:jp-relevance,fig:jr-relevance} we plot $\sqrt{\text{FID}}$, $\sqrt{\text{KID}}$, P and R versus $\sqrt{\text{JFID-FID}}$, $\sqrt{\text{KID-JKID}}$, $\text{JP}-\text{P}$ and $\text{JR}-\text{R}$, respectively.
Observe that a particular value of $\sqrt{\text{FID}}$ may correspond to different values of $\sqrt{\text{JFID-FID}}$.
This means that JFID is not simply a scaled version of the FID, i.e., two different algorithms with the same FID may attain a different value of JFID.
This shows that JFID is affected by consistency and not only by perceptual quality.

The same result holds for the JP and JR as well, but it does not hold for the JKID.
Indeed, we observe a linear relationship between $\sqrt{\text{KID}}$ and $\sqrt{\text{KID-JKID}}$, i.e., the observed empirical tradeoff between $\overline{K}$ and $\sqrt{\text{JKID}}$ is actually a tradeoff between $\overline{K}$ and $\sqrt{\text{KID}}$.
Thorough analysis of these results and the search for better ways to measure joint perceptual quality are left for future work.
\subsection{Perception-Distortion Tradeoff}
\begin{figure*}
    \centering
    \includegraphics[scale=0.93]{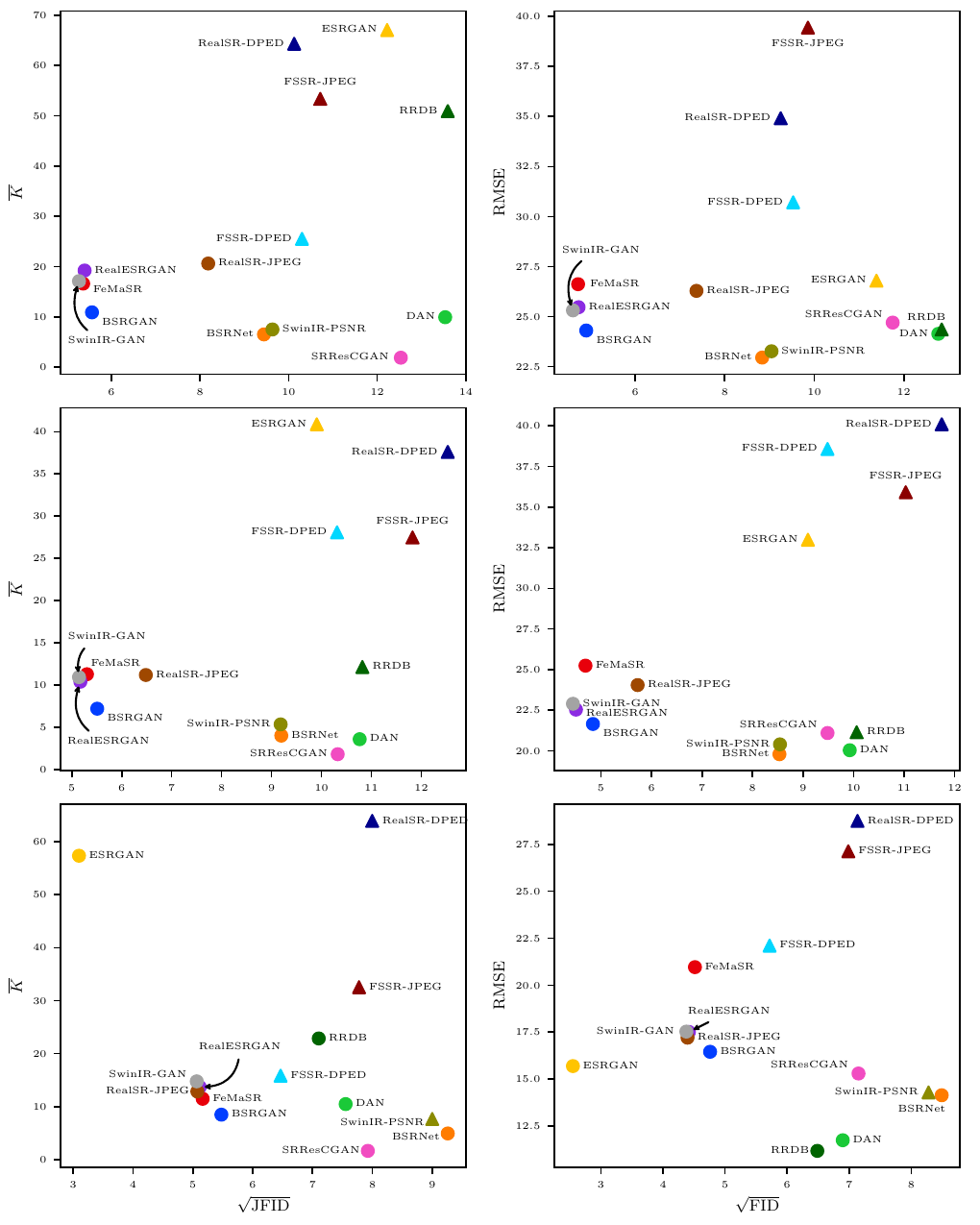}
    \caption{Plots of $\overline{K}$ versus $\sqrt{\text{JFID}}$ and RMSE versus $\sqrt{\text{FID}}$ (distortion-perception plane) of the SISR algorithms evaluated in~\cref{section:real-world-experiments}. These plots are complementary and contain the algorithms which are omitted from~\cref{fig:real-world-attacks}, which are marked here as triangles. The top figures correspond to the degradation from the Track2 challenge in~\citep{lugmayr2019aim}, the center figures correspond to the degradation from the Track1 challenge in~\citep{Lugmayr_2020_CVPR_Workshops}, and the bottom figures correspond to the common bicubic degradation.}
    \label{fig:real_world_with_fid}
\end{figure*}

\begin{figure*}
    \centering
    \includegraphics[scale=0.93]{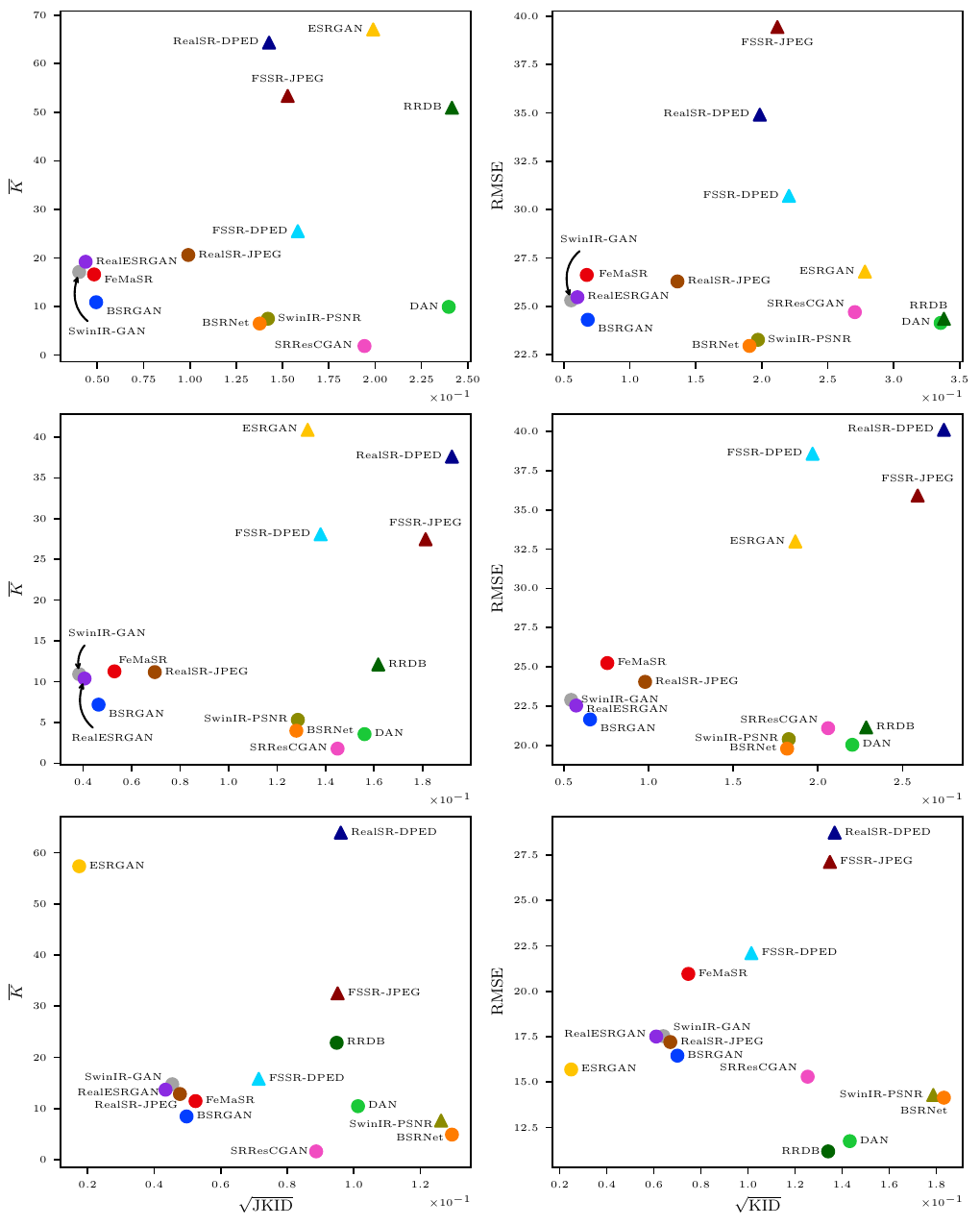}
    \caption{Plots of $\overline{K}$ versus $\sqrt{\text{JKID}}$ and RMSE versus $\sqrt{\text{KID}}$ (distortion-perception plane) of the SISR algorithms evaluated in~\cref{section:real-world-experiments}. The top figures correspond to the degradation from the Track2 challenge in~\citep{lugmayr2019aim}, the center figures correspond to the degradation from the Track1 challenge in~\citep{Lugmayr_2020_CVPR_Workshops}, and the bottom figures correspond to the common bicubic degradation.}
    \label{fig:real_world_with_kid}
\end{figure*}

\begin{figure*}
    \centering
    \includegraphics[scale=0.93]{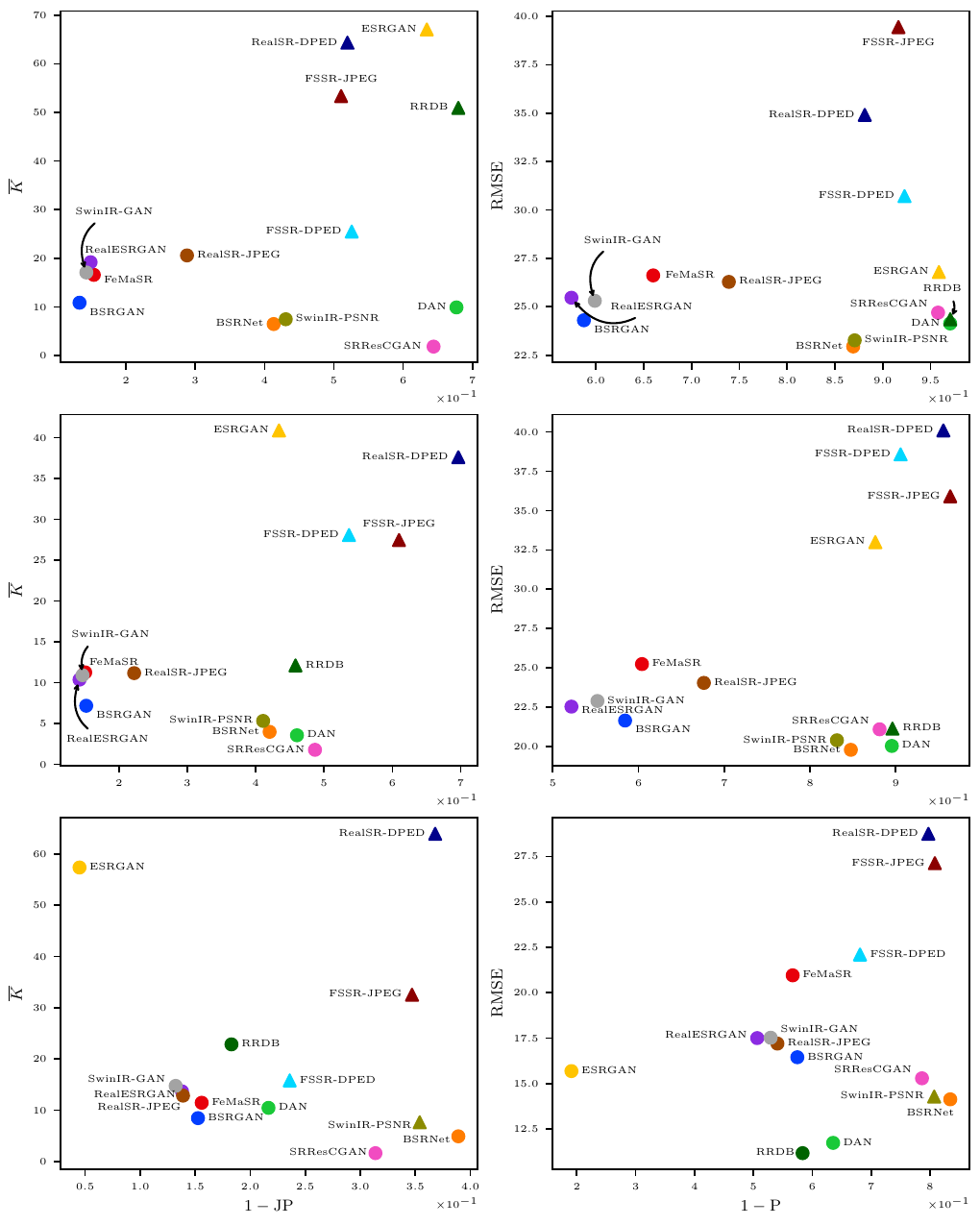}
    \caption{Plots of $\overline{K}$ versus $1-\text{JP}$ and RMSE versus $1-\text{P}$ (distortion-perception plane) of the SISR algorithms evaluated in~\cref{section:real-world-experiments}. The top figures correspond to the degradation from the Track2 challenge in~\citep{lugmayr2019aim}, the center figures correspond to the degradation from the Track1 challenge in~\citep{Lugmayr_2020_CVPR_Workshops}, and the bottom figures correspond to the common bicubic degradation.}
    \label{fig:real_world_with_precision}
\end{figure*}

\begin{figure*}
    \centering
    \includegraphics[scale=0.93]{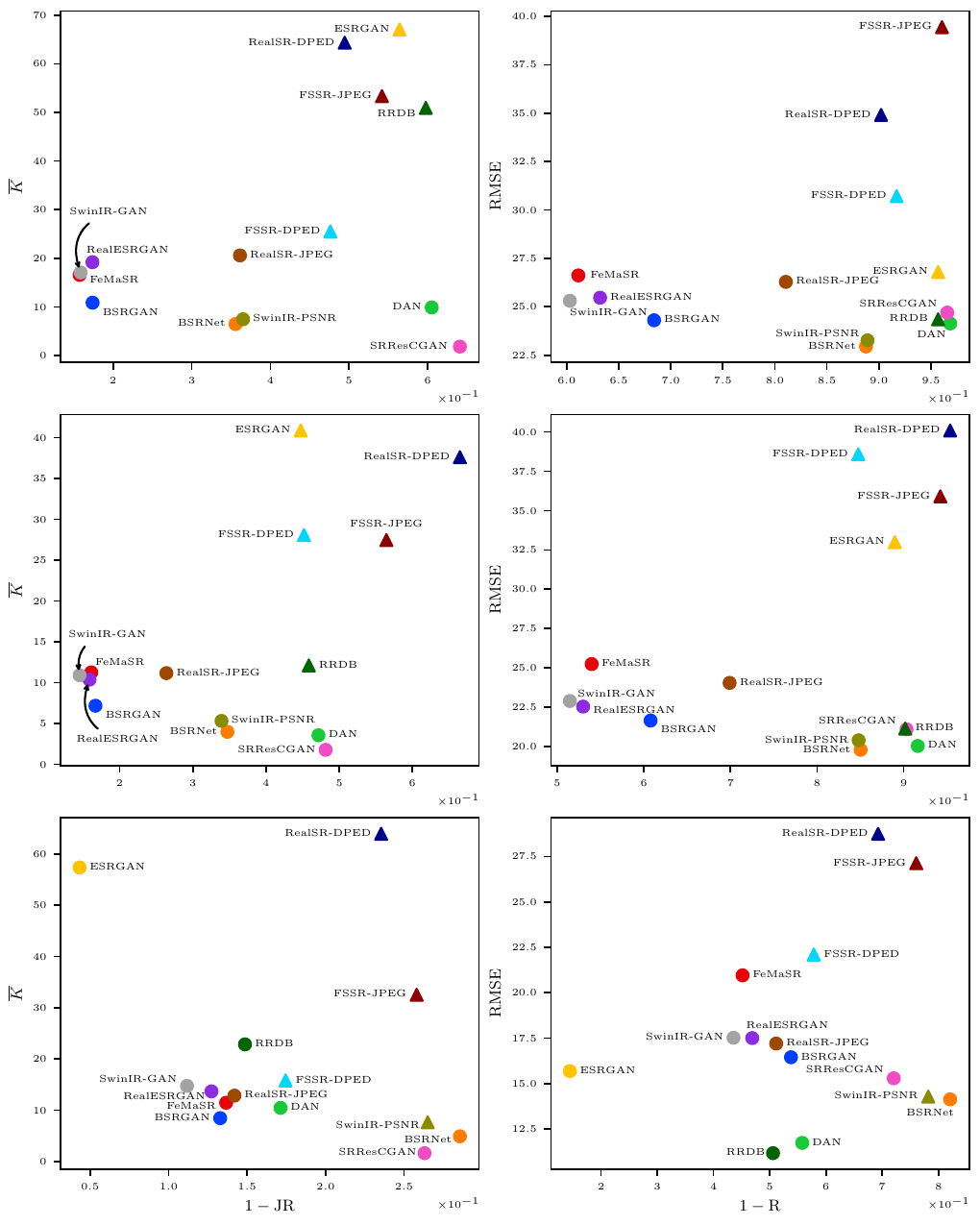}
    \caption{Plots of $\overline{K}$ versus $1-\text{JR}$ and RMSE versus $1-\text{R}$ (distortion-perception plane) of the SISR algorithms evaluated in~\cref{section:real-world-experiments}. The top figures correspond to the degradation from the Track2 challenge in~\citep{lugmayr2019aim}, the center figures correspond to the degradation from the Track1 challenge in~\citep{Lugmayr_2020_CVPR_Workshops}, and the bottom figures correspond to the common bicubic degradation.}
    \label{fig:real_world_with_recall}
\end{figure*}

\begin{figure*}
    \centering
    \includegraphics[scale=0.93]{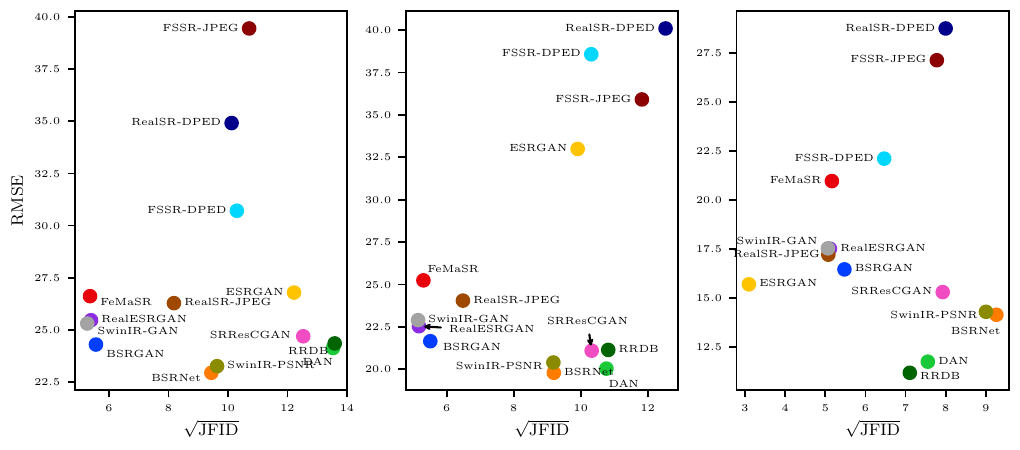}
    \caption{Plots of RMSE versus $\sqrt{\text{JFID}}$ of the SISR algorithms evaluated in~\cref{section:real-world-experiments}. The left plot corresponds to the degradation from the Track2 challenge in~\citep{lugmayr2019aim}, the center plot corresponds to the degradation from the Track1 challenge in~\citep{Lugmayr_2020_CVPR_Workshops}, and the right plot corresponds to the common bicubic degradation.}
    \label{fig:jfid_vs_rmse}
\end{figure*}

\begin{figure*}
    \centering
    \includegraphics[scale=0.93]{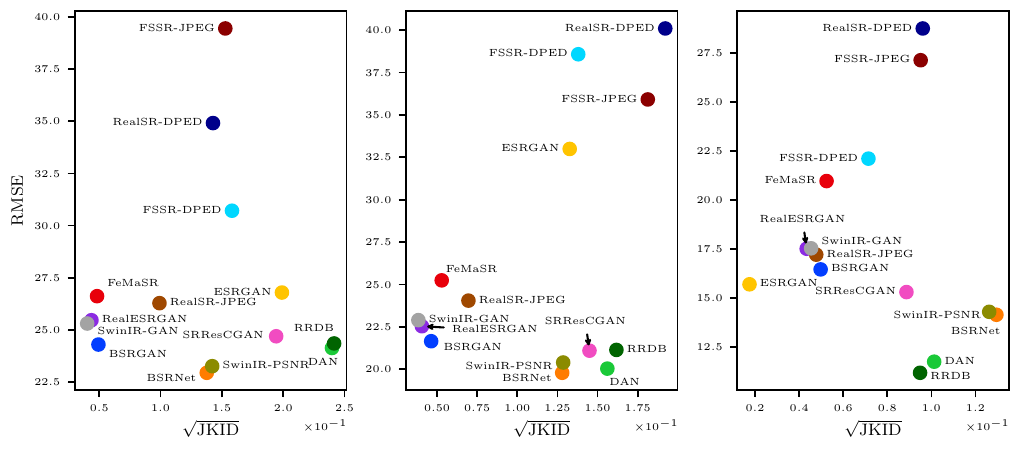}
    \caption{Plots of RMSE versus $\sqrt{\text{JKID}}$ of the SISR algorithms evaluated in~\cref{section:real-world-experiments}. The left plot corresponds to the degradation from the Track2 challenge in~\citep{lugmayr2019aim}, the center plot corresponds to the degradation from the Track1 challenge in~\citep{Lugmayr_2020_CVPR_Workshops}, and the right plot corresponds to the common bicubic degradation.}
    \label{fig:jkid_vs_rmse}
\end{figure*}

\begin{figure*}
    \centering
    \includegraphics[scale=0.93]{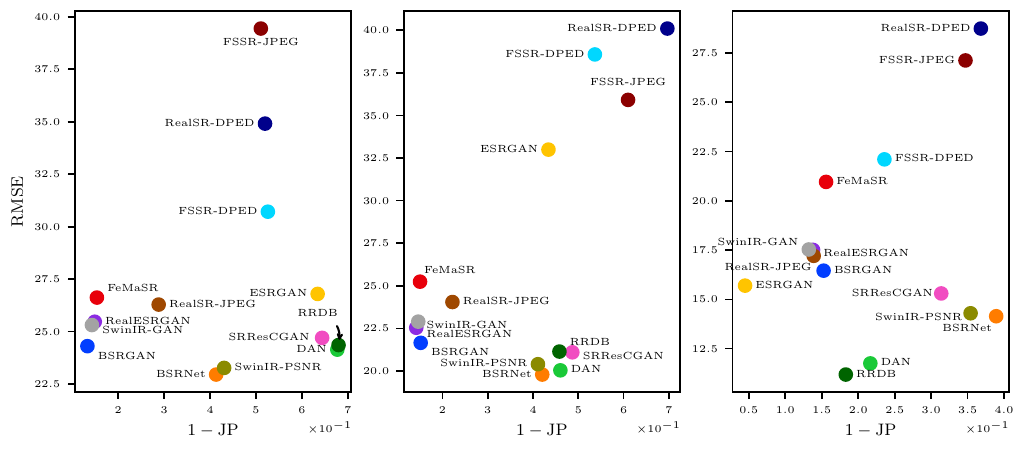}
    \caption{Plots of RMSE versus $1-\text{JP}$ of the SISR algorithms evaluated in~\cref{section:real-world-experiments}. The left plot corresponds to the degradation from the Track2 challenge in~\citep{lugmayr2019aim}, the center plot corresponds to the degradation from the Track1 challenge in~\citep{Lugmayr_2020_CVPR_Workshops}, and the right plot corresponds to the common bicubic degradation.}
    \label{fig:jp_vs_rmse}
\end{figure*}

\begin{figure*}
    \centering
    \includegraphics[scale=0.93]{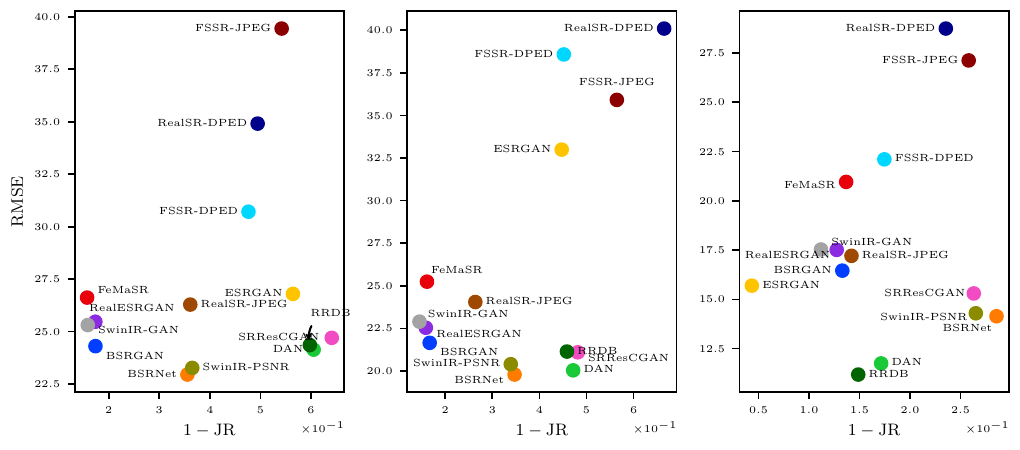}
    \caption{Plots of RMSE versus $1-\text{JR}$ of the SISR algorithms evaluated in~\cref{section:real-world-experiments}. The left plot corresponds to the degradation from the Track2 challenge in~\citep{lugmayr2019aim}, the center plot corresponds to the degradation from the Track1 challenge in~\citep{Lugmayr_2020_CVPR_Workshops}, and the right plot corresponds to the common bicubic degradation.}
    \label{fig:jr_vs_rmse}
\end{figure*}

As a complementary, we provide in~\cref{fig:real_world_with_fid,fig:real_world_with_kid,fig:real_world_with_precision,fig:real_world_with_recall} plots of the perception-distortion tradeoff~\cite{Blau2018} (e.g., $\sqrt{\text{FID}}$ or $\sqrt{\text{KID}}$ versus the RMSE), where each statistical distance (FID, KID, etc.) is computed between the sets $\{x_{F}^{(i)}\}_{i=1}^{N}$ and $\{\hat{x}_{F}^{(i)}\}_{i=1}^{N}$.
Namely, each statistical distance measures the perceptual quality of each algorithm.
Furthermore, in~\cref{fig:jfid_vs_rmse,fig:jkid_vs_rmse,fig:jp_vs_rmse,fig:jr_vs_rmse} we also plot the \emph{joint} perceptual quality (computed via JFID, JKID, etc.) versus the RMSE.
Due to the perception-distortion tradeoff, we anticipate a tradeoff between the joint perceptual quality and distortion as well, which we can indeed observe in these figures.
\section{Additional Details of~\cref{section:adv_attacks} (Adversarial Attacks)}\label{appendx:adv_attacks}
We train the RRDB model~\cite{wang2018esrgan} using the official model's code provided by the authors, deployed into the official training code of GFPGAN~\cite{wang2021gfpgan}.
We omit all the losses of GFPGAN except for the $L_{1}$ reconstruction loss, and use the same training data (FFHQ~\cite{stylegan} face images resized to $512\times 512$) and degradation model as that of GFPGAN v1.
To traing the model, we use the Adam optimizer with $\beta_{1}=0.9,\beta_{2}=0.99$ and a learning rate of $2\cdot 10^{-4}$, and perform the same learning rate scheduling as in the official training configuration of RRDB.
The model is trained with a batch side of 16 for 1,000,000 iterations.
We record a model checkpoint every 5000 steps, and for evaluation we use the checkpoint that attains the highest PSNR (which is 24.7dB).

\subsection{Adversarial Attacks to Produce ``Female''-Classified Faces}\label{appendix:male-to-female}
\begin{figure}
\centering
\begin{subfigure}{.37\textwidth}
  \centering
\renewcommand{\arraystretch}{0.0}
    \begin{tabular}{P{0.185\linewidth}P{0.185\linewidth}P{0.185\linewidth}P{0.185\linewidth}}
    {\small $y$} & 
    {\small $y_{adv}$} & 
    {\small $f(y)$} & 
    {\small $f(y_{adv})$} \\
    \vspace*{0.1cm} \ & & & \\
    \multicolumn{4}{l}{\hskip-0.19cm\includegraphics[width=\linewidth]{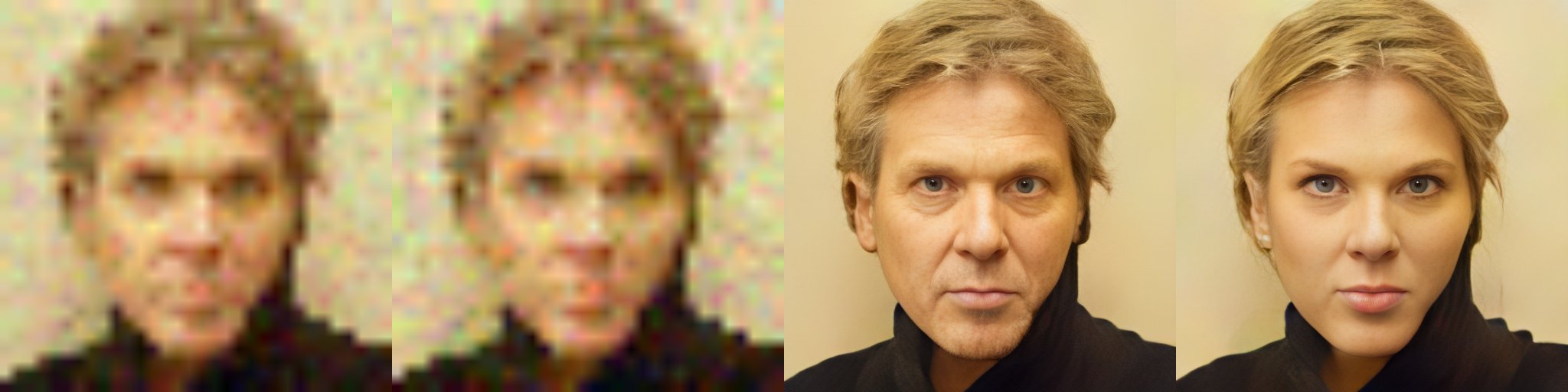}} \\
    \multicolumn{4}{l}{\hskip-0.19cm\includegraphics[width=\linewidth]{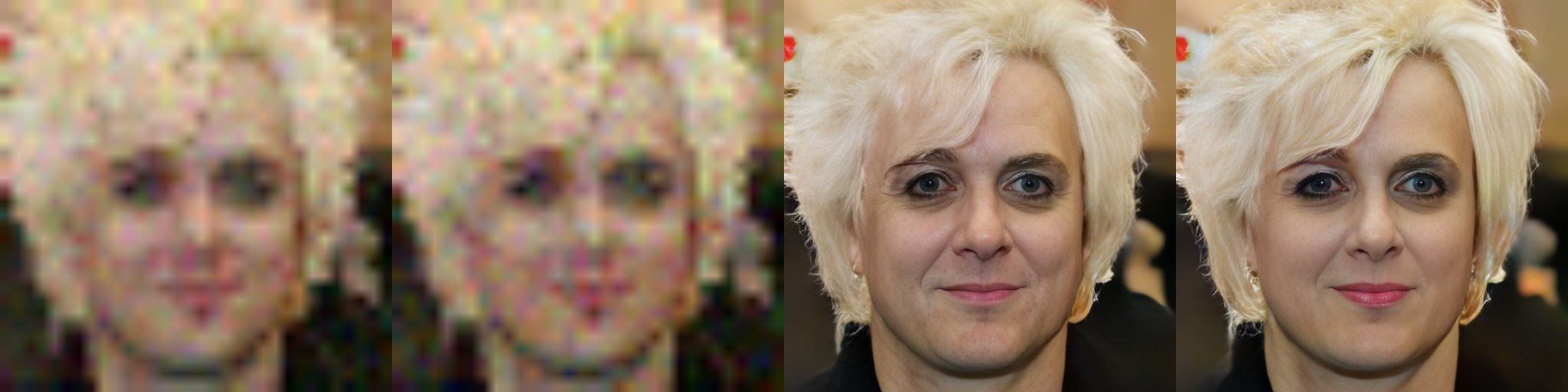}} \\
    \multicolumn{4}{l}{\hskip-0.19cm\includegraphics[width=\linewidth]{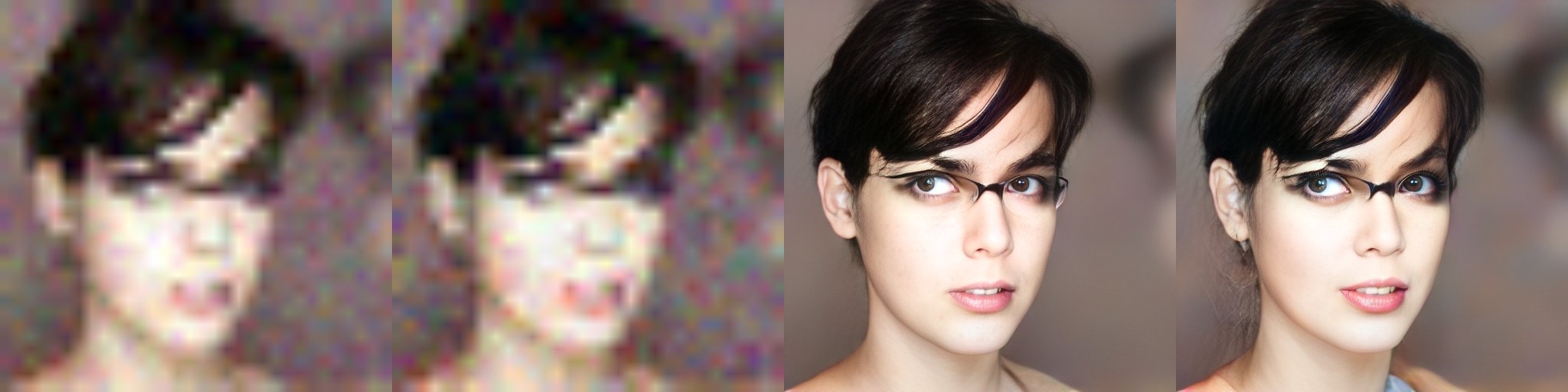}} \\
    \multicolumn{4}{l}{\hskip-0.19cm\includegraphics[width=\linewidth]{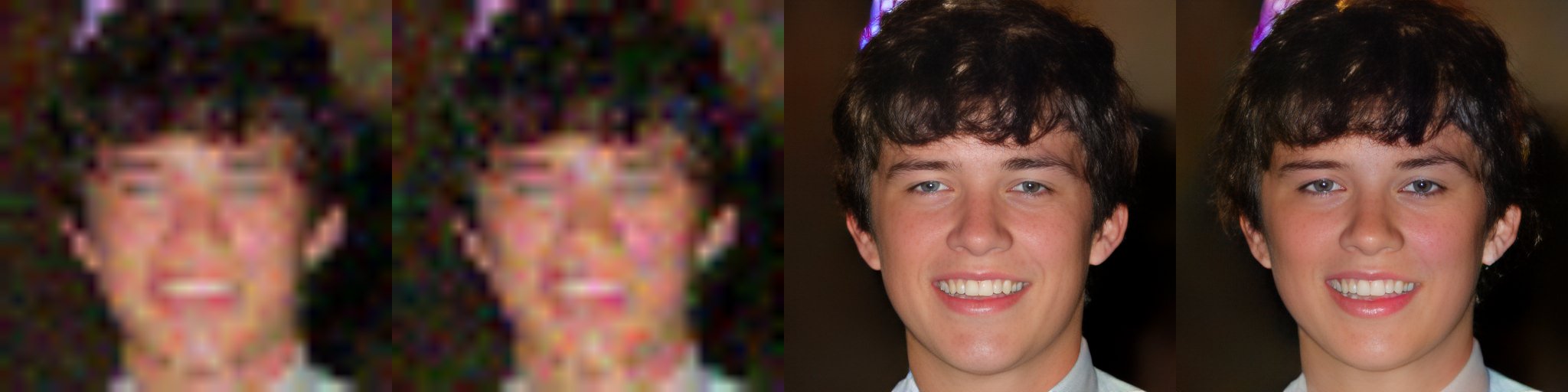}} \\
    \multicolumn{4}{l}{\hskip-0.19cm\includegraphics[width=\linewidth]{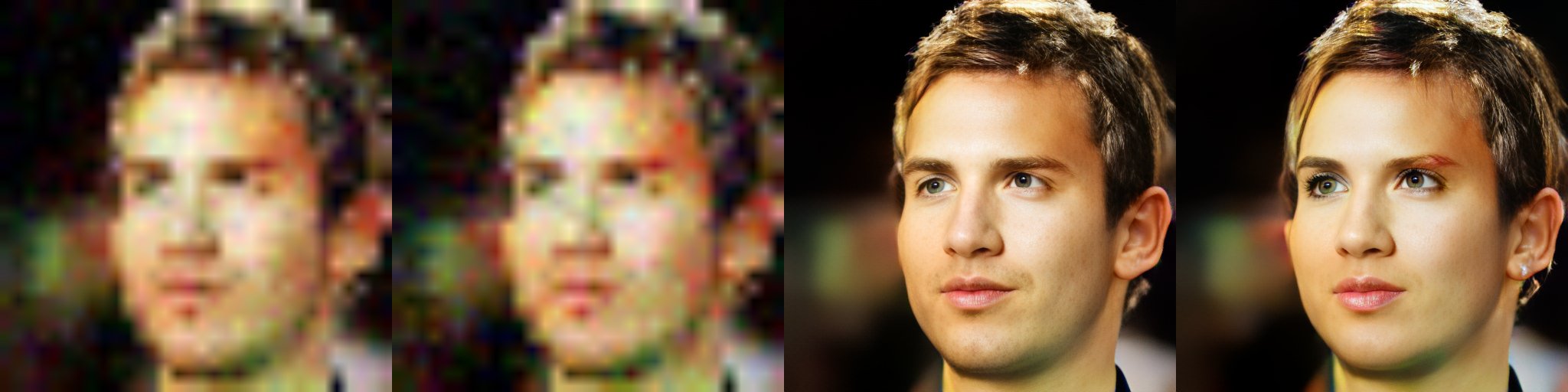}} \\
    \multicolumn{4}{l}{\hskip-0.19cm\includegraphics[width=\linewidth]{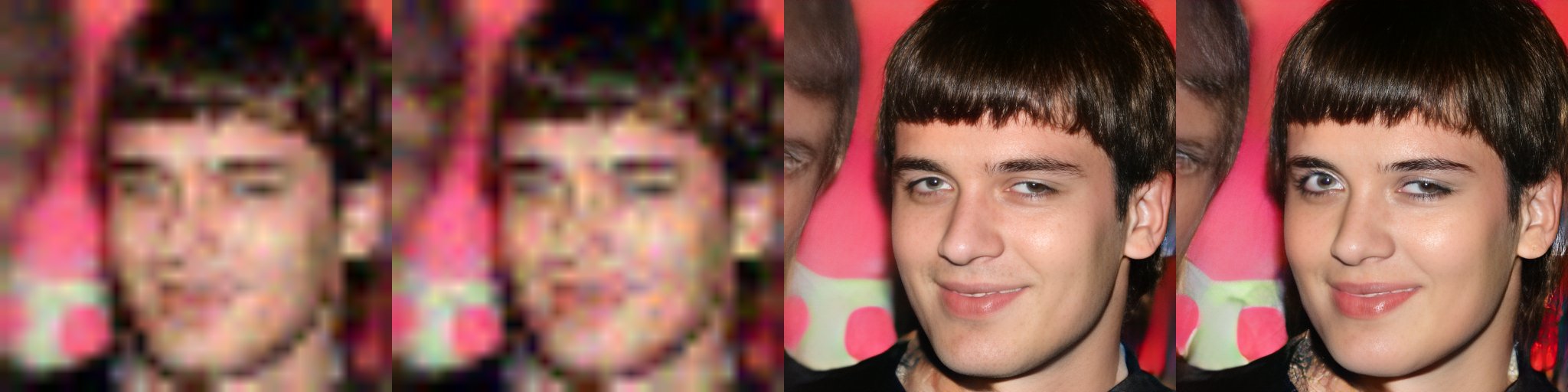}} \\
    \multicolumn{4}{l}{\hskip-0.19cm\includegraphics[width=\linewidth]{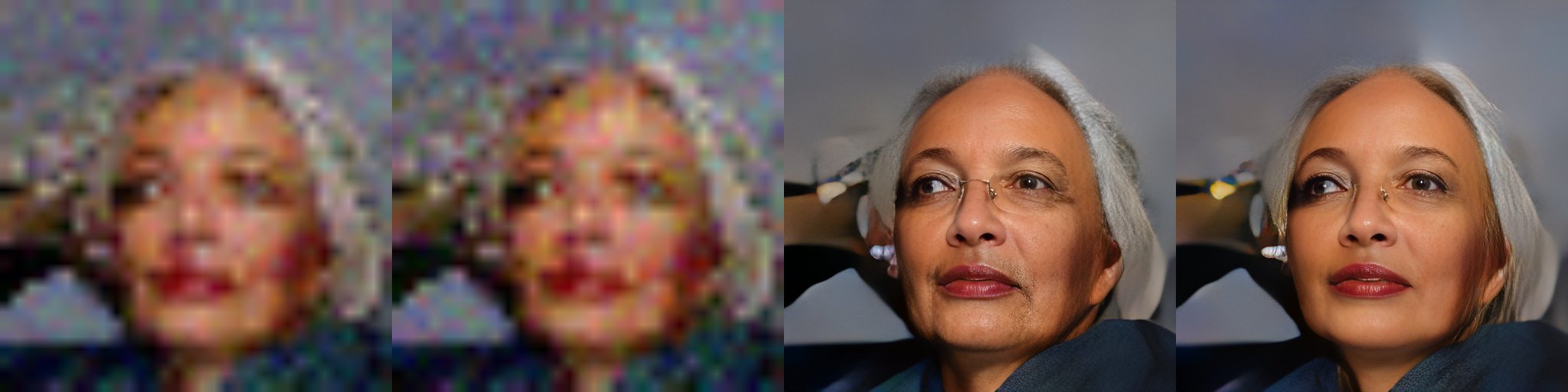}} \\
    \multicolumn{4}{l}{\hskip-0.19cm\includegraphics[width=\linewidth]{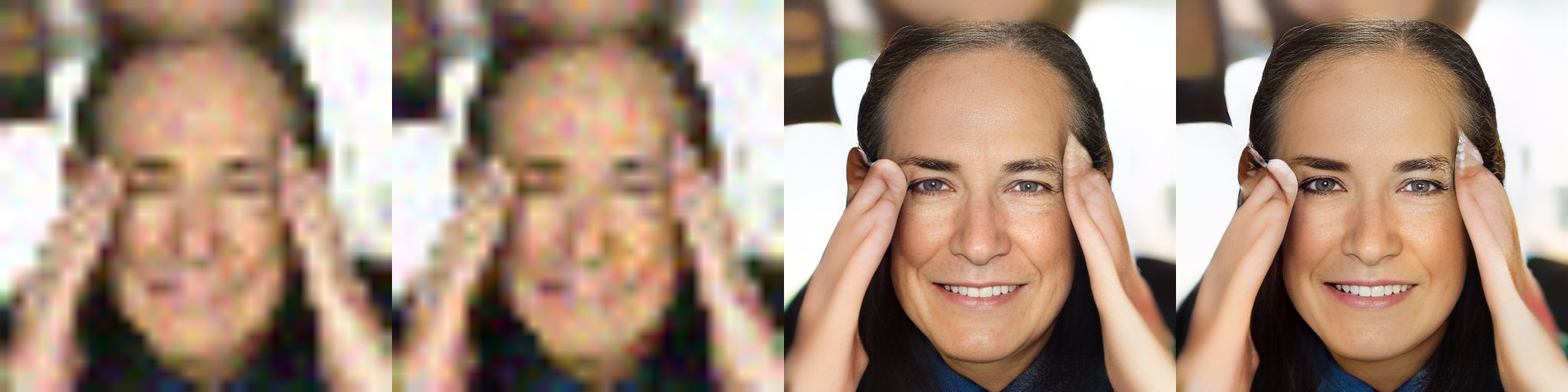}} \\
    \multicolumn{4}{l}{\hskip-0.19cm\includegraphics[width=\linewidth]{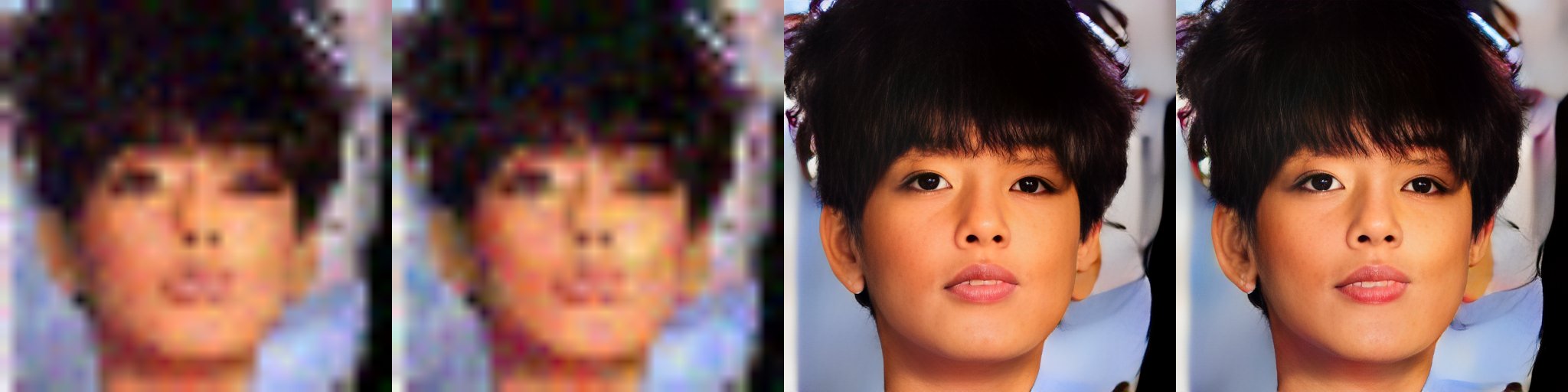}} \\
    \multicolumn{4}{l}{\hskip-0.19cm\includegraphics[width=\linewidth]{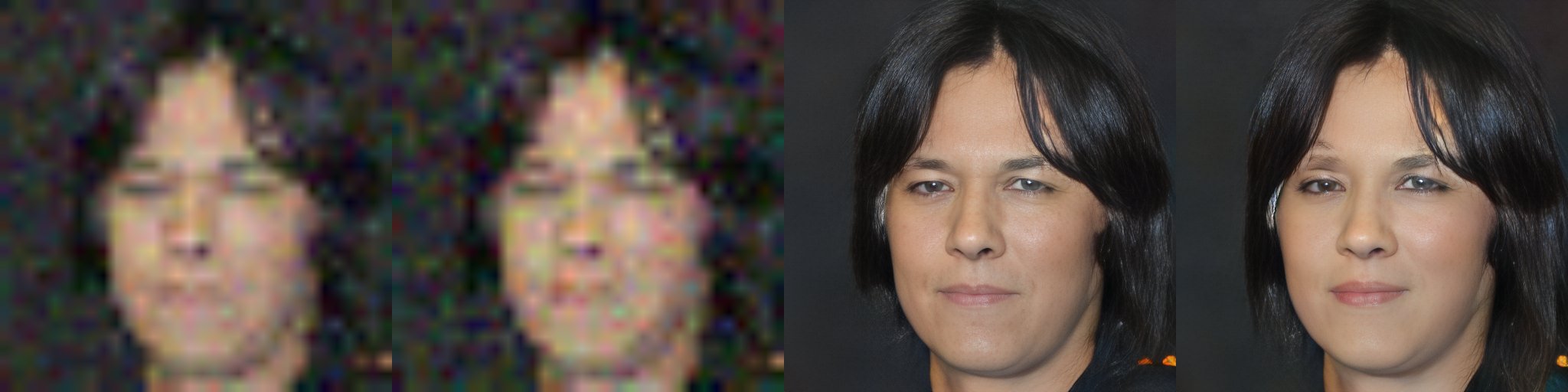}} \\
    \multicolumn{4}{l}{\hskip-0.19cm\includegraphics[width=\linewidth]{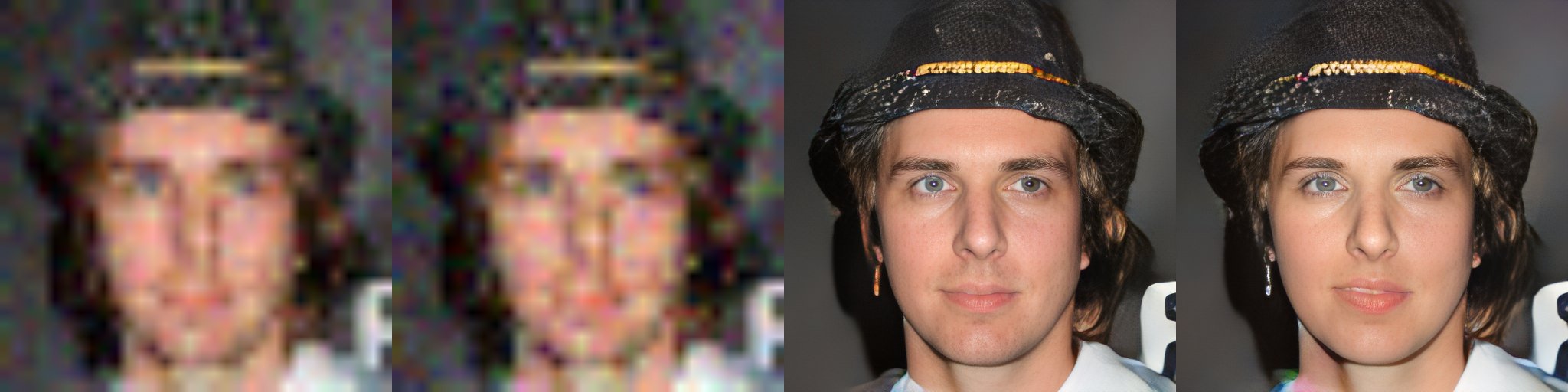}} \\
    \multicolumn{4}{l}{\hskip-0.19cm\includegraphics[width=\linewidth]{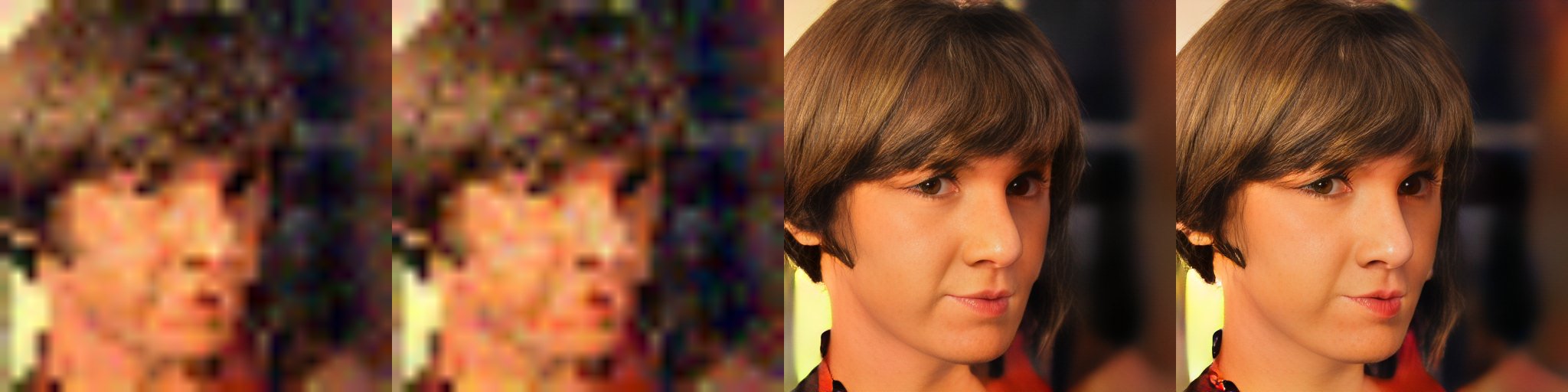}} 
\end{tabular}
\end{subfigure}\hspace*{2cm}
\begin{subfigure}{.37\textwidth}
  \centering
  \renewcommand{\arraystretch}{0.0}
    \begin{tabular}{P{0.185\linewidth}P{0.185\linewidth}P{0.185\linewidth}P{0.185\linewidth}}
    {\small $y$} & 
    {\small $y_{adv}$} & 
    {\small $f(y)$} & 
    {\small $f(y_{adv})$} \\
    \vspace*{0.1cm} \ & & & \\
    \multicolumn{4}{l}{\hskip-0.19cm\includegraphics[width=\linewidth]{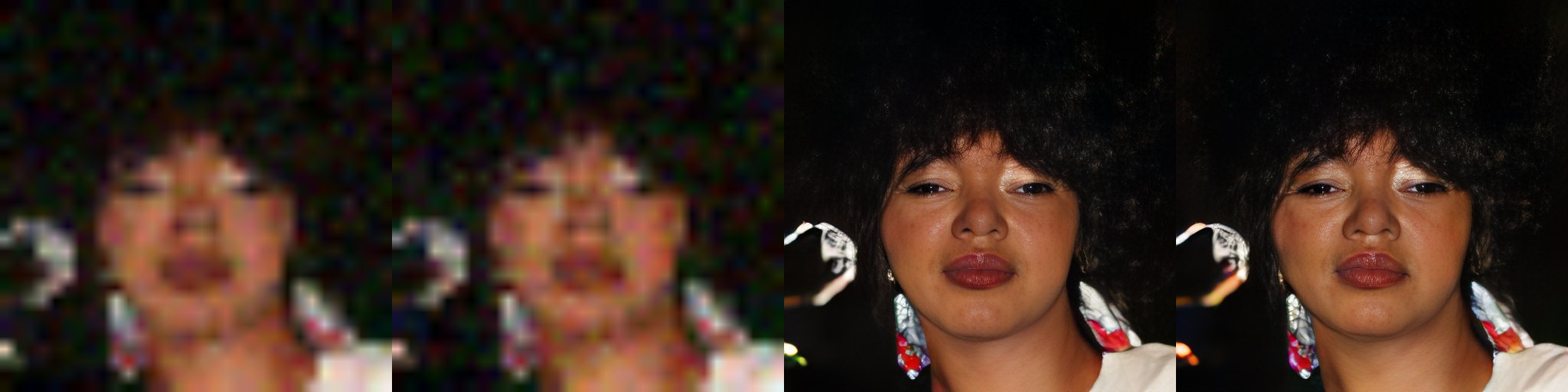}} \\
    \multicolumn{4}{l}{\hskip-0.19cm\includegraphics[width=\linewidth]{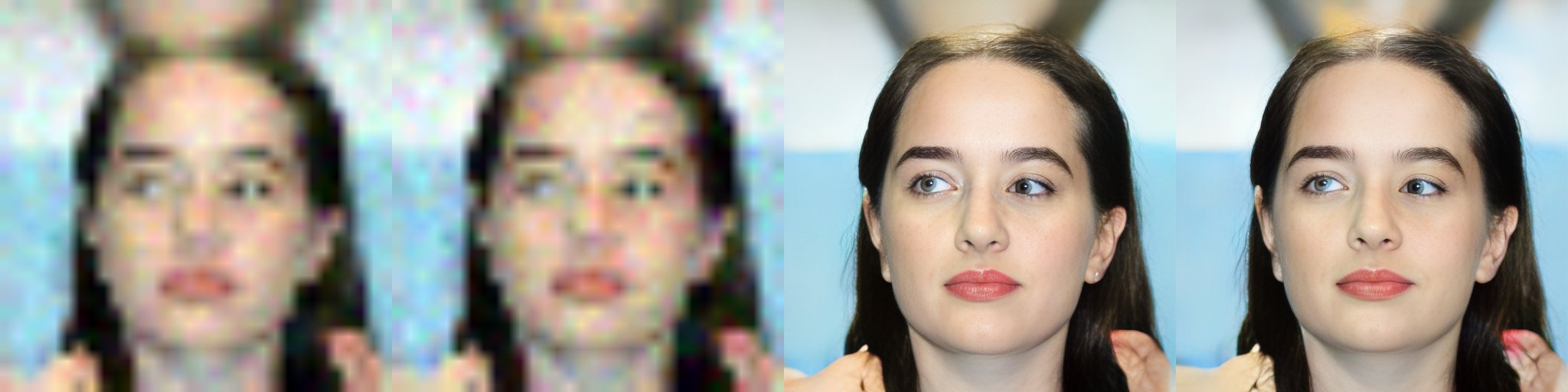}} \\
    \multicolumn{4}{l}{\hskip-0.19cm\includegraphics[width=\linewidth]{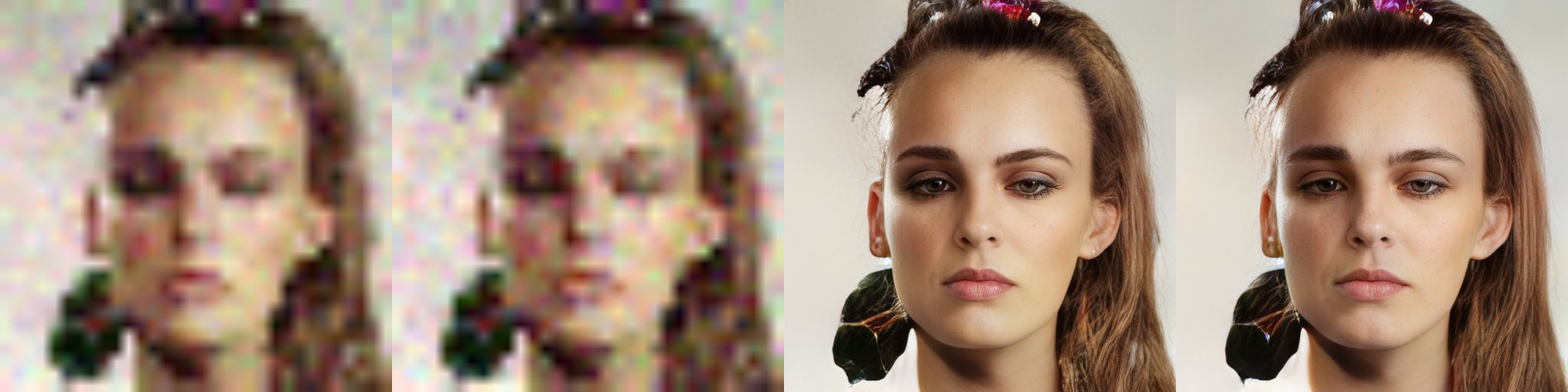}} \\
    \multicolumn{4}{l}{\hskip-0.19cm\includegraphics[width=\linewidth]{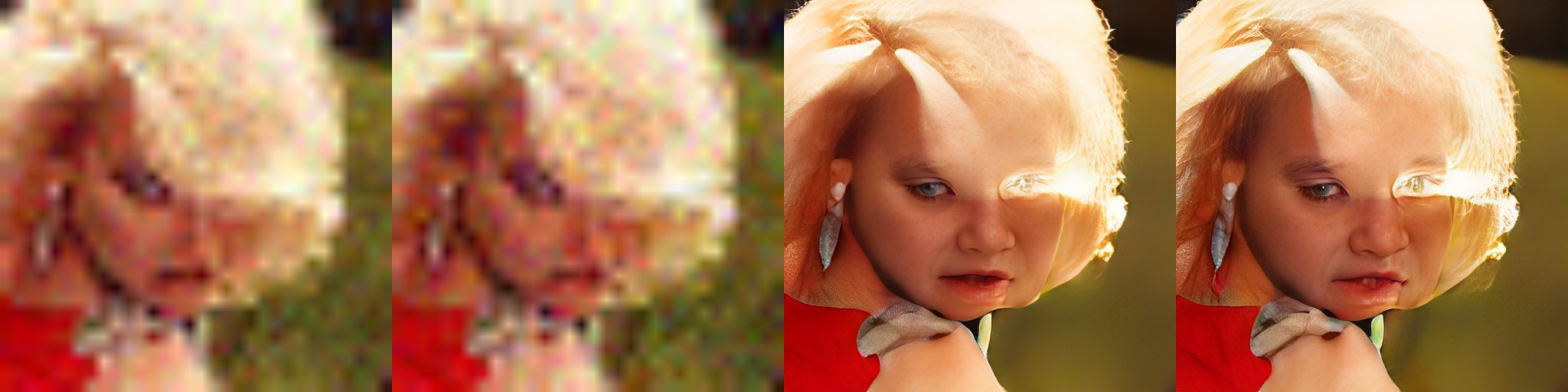}}
    \end{tabular}
\end{subfigure}
    \caption{Adversarial attacks on low-resolution face images intended to alter the output of GFPGAN~\cite{wang2021gfpgan} to produce a female face.
    \textbf{Left}: Successful results, where $f(y)$ is classified as ``male'' and $f(y_{adv})$ is classified as ``female'' by the gender classification model we use for evaluation. The attacks successfully change the classified gender of 9.3\% of the images from ``male'' to ``female''.
    \textbf{Right}: Unsuccessful results, where $f(y)$ is classified as ``female'' and $f(y_{adv})$ is classified as ``male''. 
    For GFPGAN, there are only 4 such unsuccessful examples from the 1000 images we use for evaluation (0.4\% of the test data set).
    }
    \label{fig:gfpgan_gender_switch}
\end{figure}

\begin{figure}
\centering
\begin{subfigure}{.37\textwidth}
  \centering
\renewcommand{\arraystretch}{0.0}
    \begin{tabular}{P{0.185\linewidth}P{0.185\linewidth}P{0.185\linewidth}P{0.185\linewidth}}
    {\small $y$} & 
    {\small $y_{adv}$} & 
    {\small $f(y)$} & 
    {\small $f(y_{adv})$} \\
    \vspace*{0.1cm} \ & & & \\
    \multicolumn{4}{l}{\hskip-0.19cm\includegraphics[width=\linewidth]{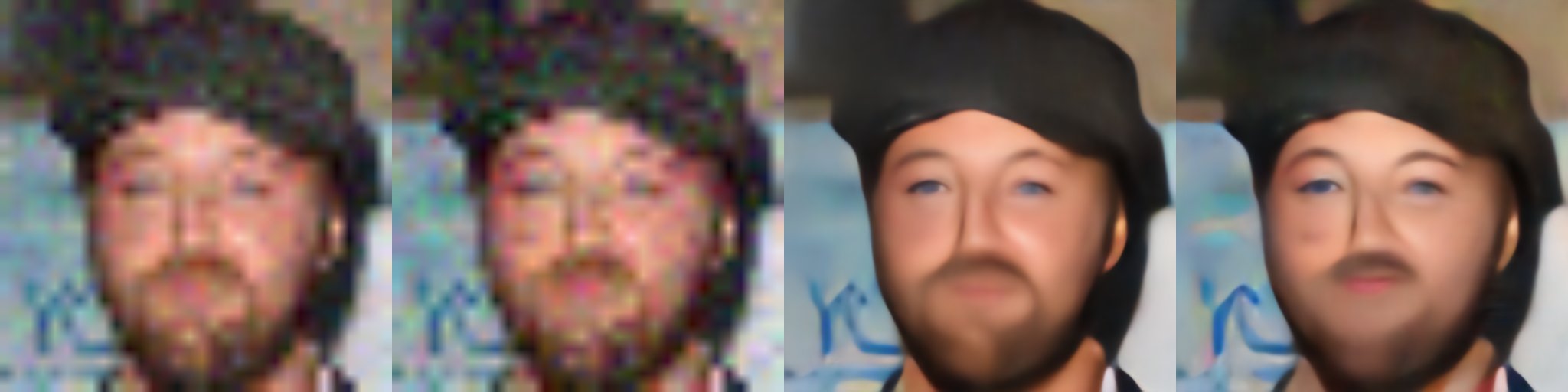}} \\
    \multicolumn{4}{l}{\hskip-0.19cm\includegraphics[width=\linewidth]{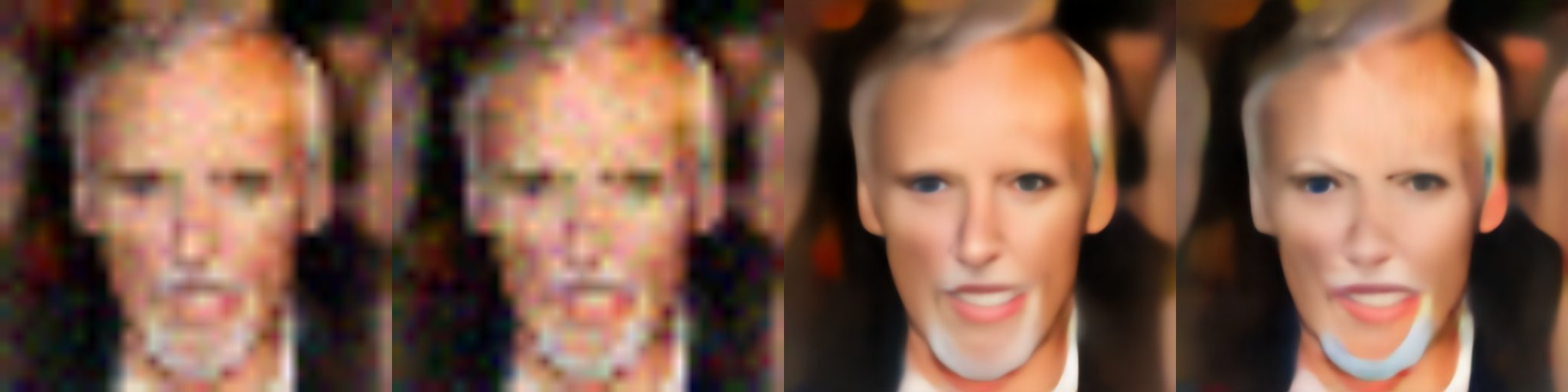}} \\
    \multicolumn{4}{l}{\hskip-0.19cm\includegraphics[width=\linewidth]{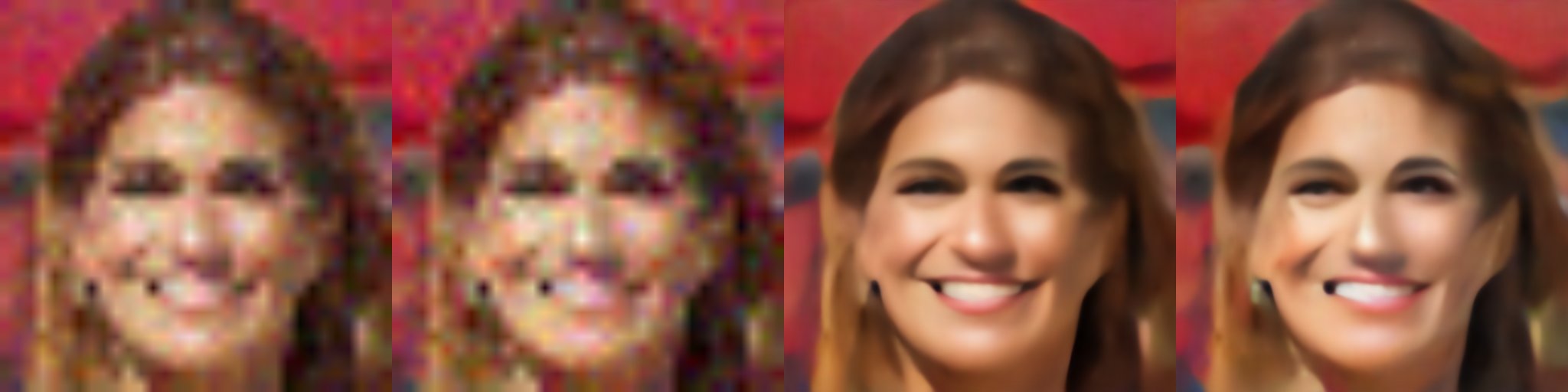}} \\
    \multicolumn{4}{l}{\hskip-0.19cm\includegraphics[width=\linewidth]{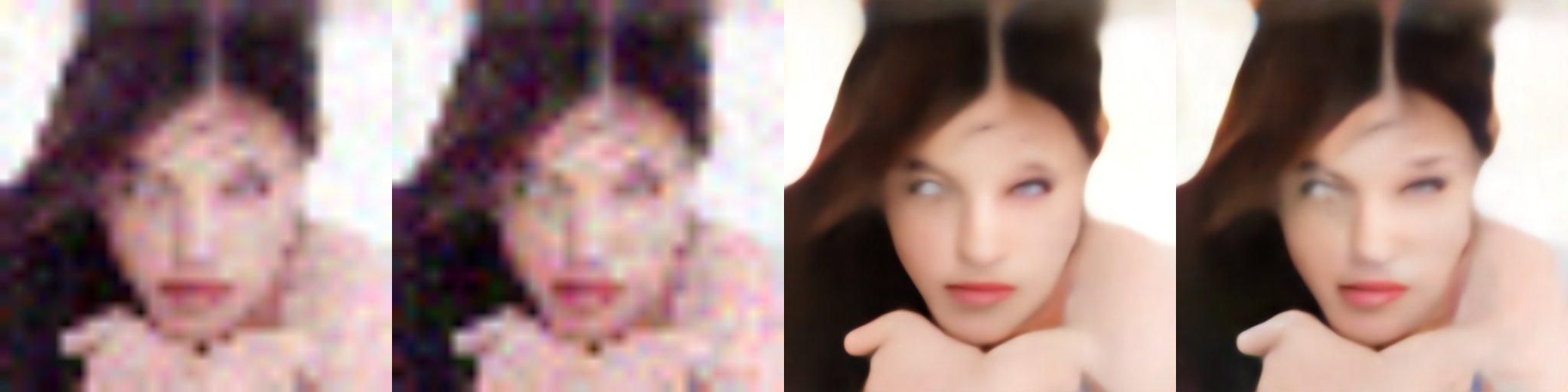}} \\
    \multicolumn{4}{l}{\hskip-0.19cm\includegraphics[width=\linewidth]{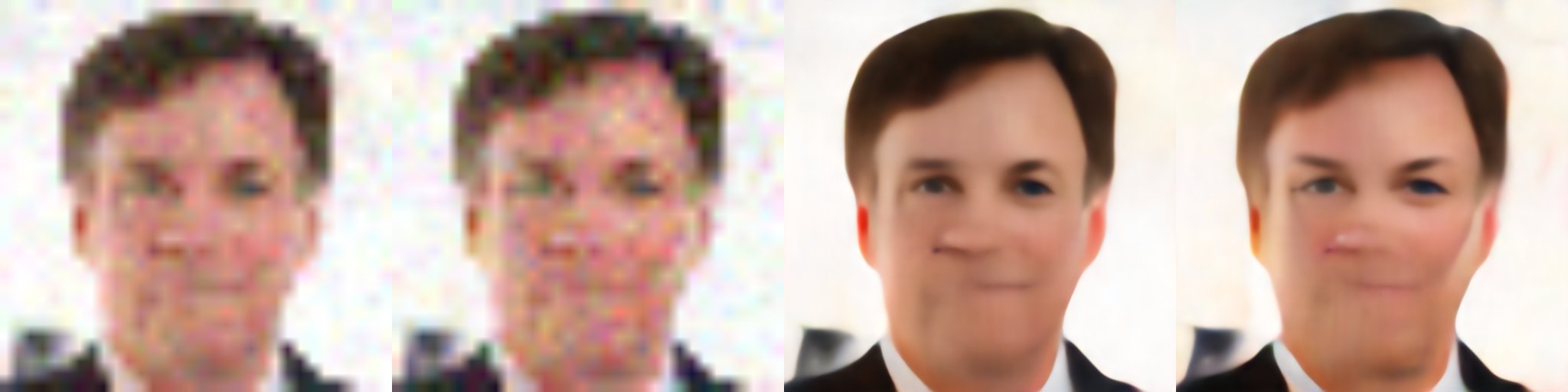}} \\
    \multicolumn{4}{l}{\hskip-0.19cm\includegraphics[width=\linewidth]{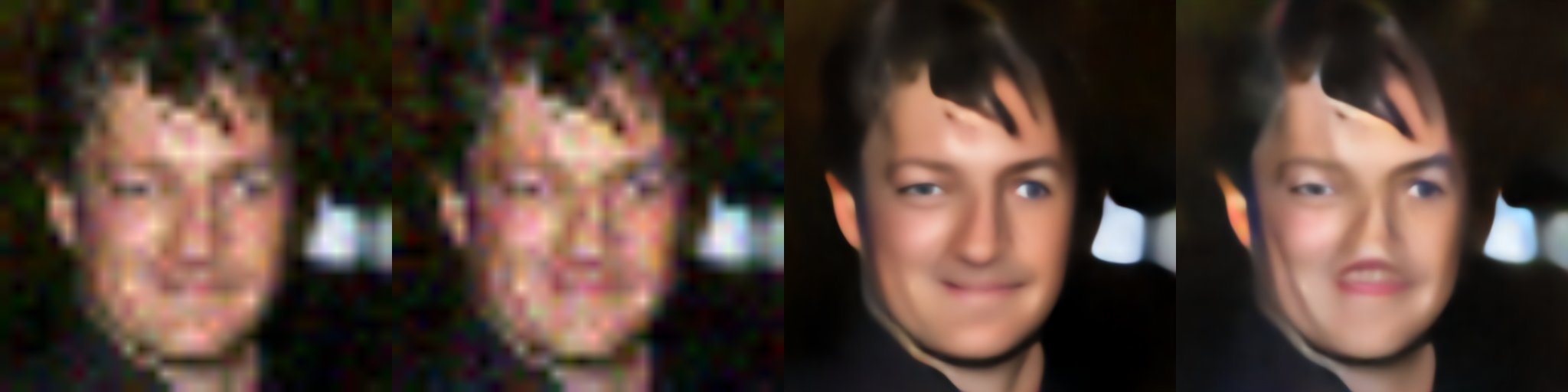}} \\
    \multicolumn{4}{l}{\hskip-0.19cm\includegraphics[width=\linewidth]{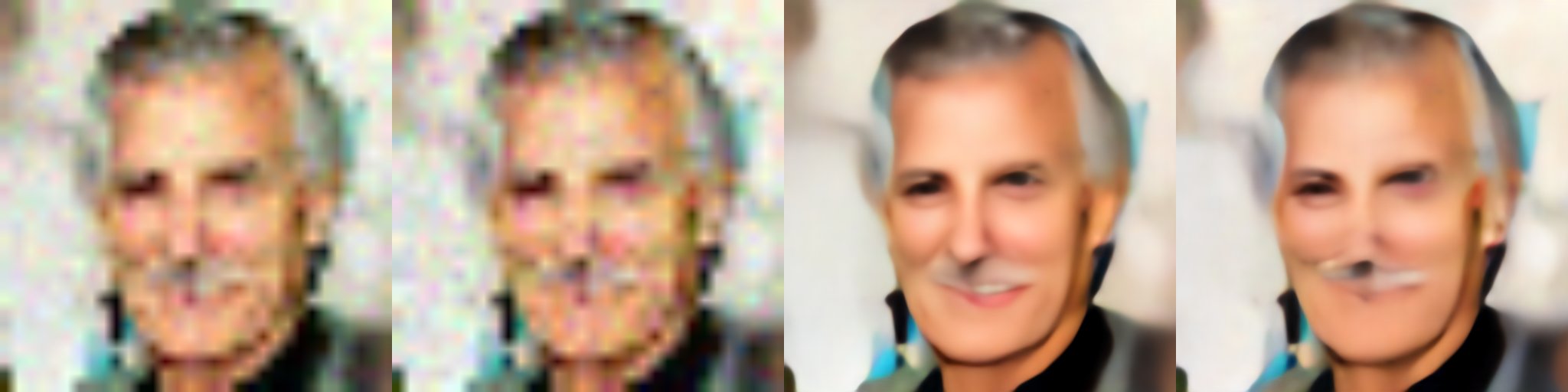}} \\
    \multicolumn{4}{l}{\hskip-0.19cm\includegraphics[width=\linewidth]{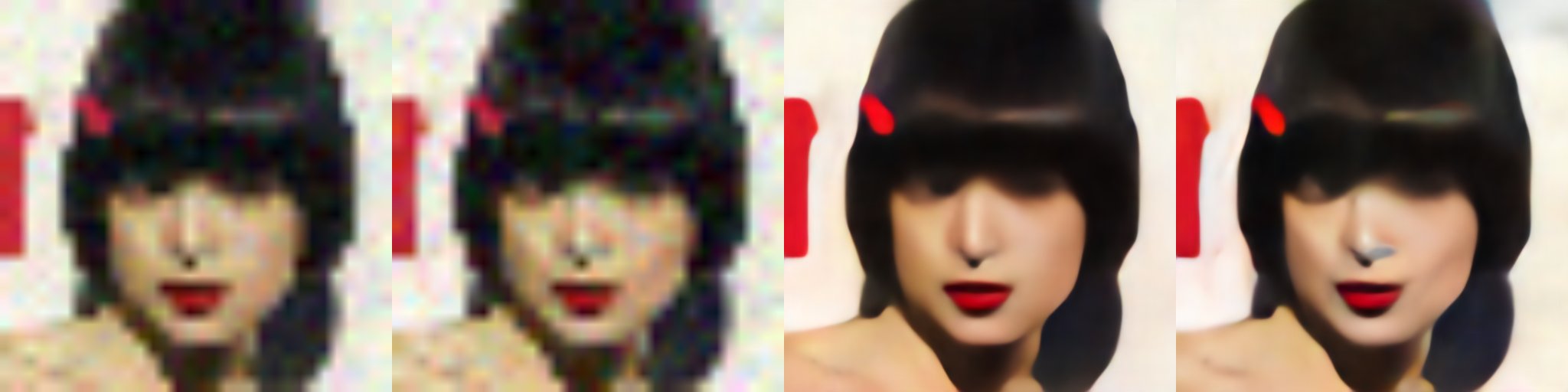}} \\
    \multicolumn{4}{l}{\hskip-0.19cm\includegraphics[width=\linewidth]{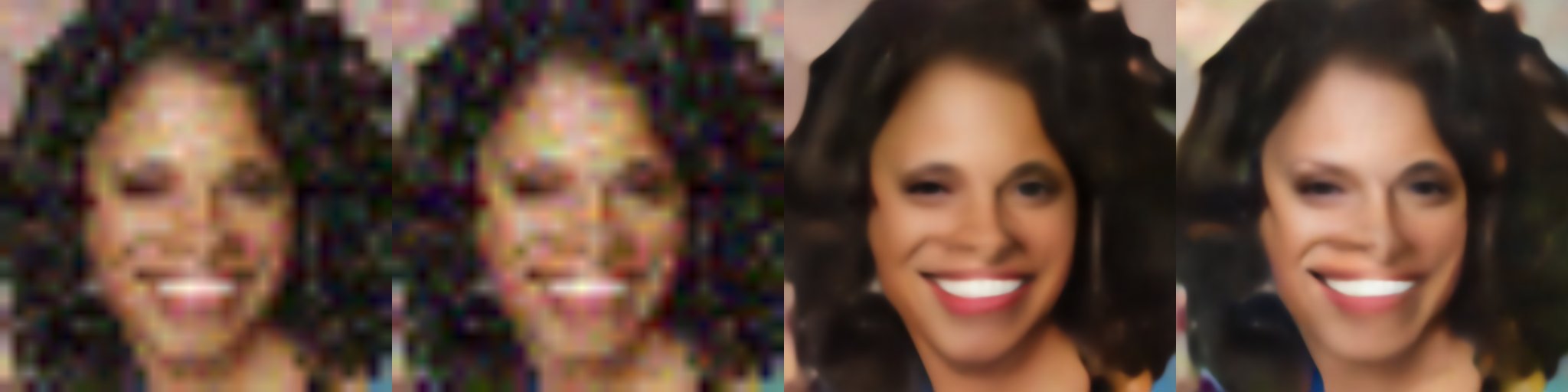}} \\
    \multicolumn{4}{l}{\hskip-0.19cm\includegraphics[width=\linewidth]{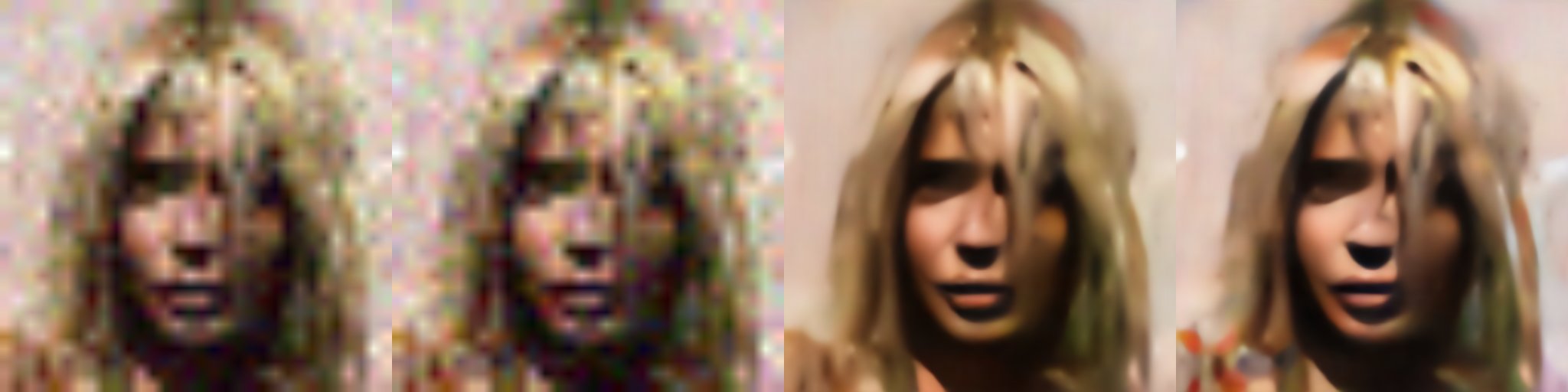}} \\
    \multicolumn{4}{l}{\hskip-0.19cm\includegraphics[width=\linewidth]{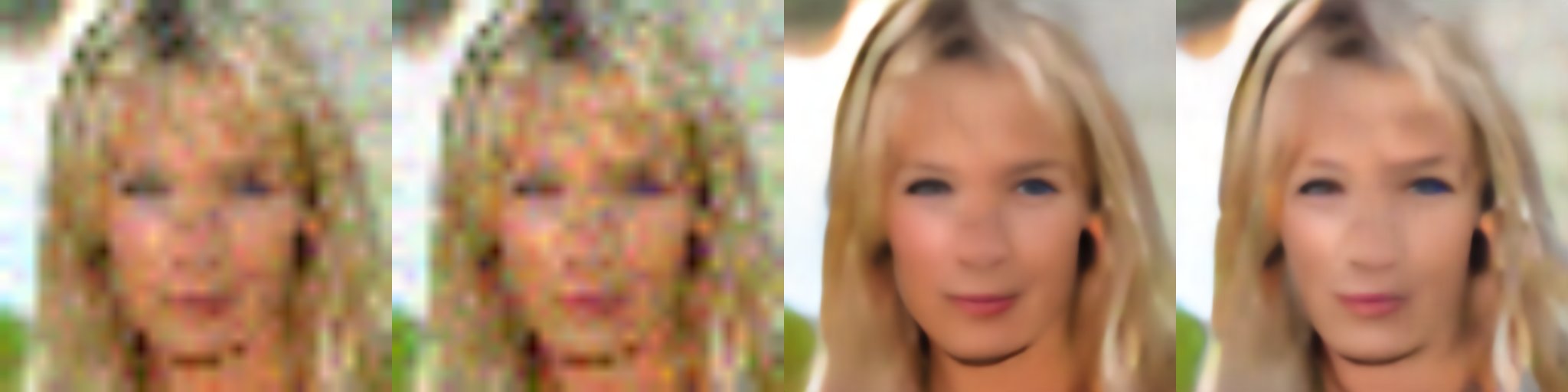}} \\
    \multicolumn{4}{l}{\hskip-0.19cm\includegraphics[width=\linewidth]{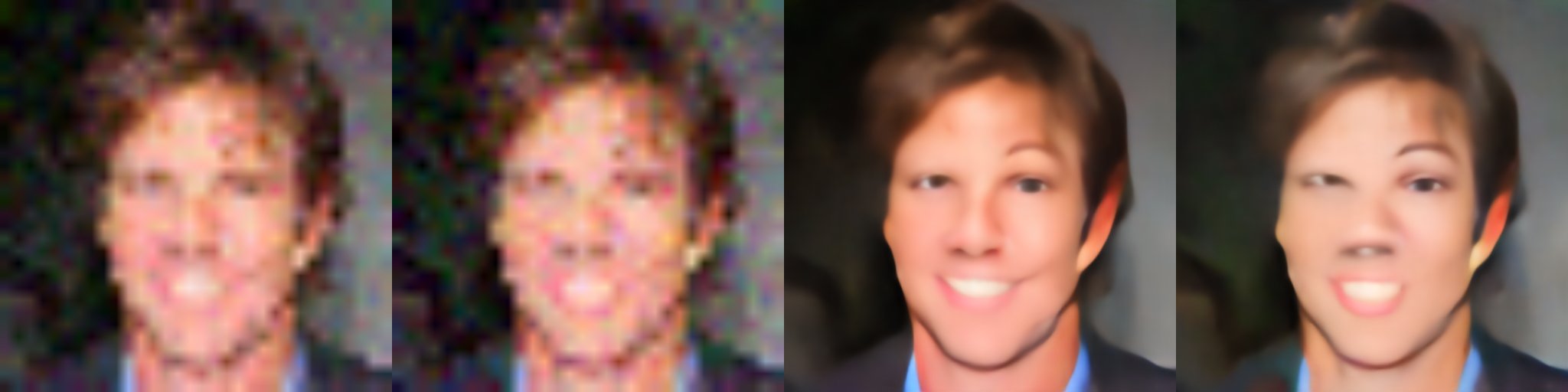}} \\
\end{tabular}
\end{subfigure}\hspace*{2cm}
\begin{subfigure}{.37\textwidth}
  \centering
  \renewcommand{\arraystretch}{0.0}
    \begin{tabular}{P{0.185\linewidth}P{0.185\linewidth}P{0.185\linewidth}P{0.185\linewidth}}
    {\small $y$} & 
    {\small $y_{adv}$} & 
    {\small $f(y)$} & 
    {\small $f(y_{adv})$} \\
    \vspace*{0.1cm} \ & & & \\
    \multicolumn{4}{l}{\hskip-0.19cm\includegraphics[width=\linewidth]{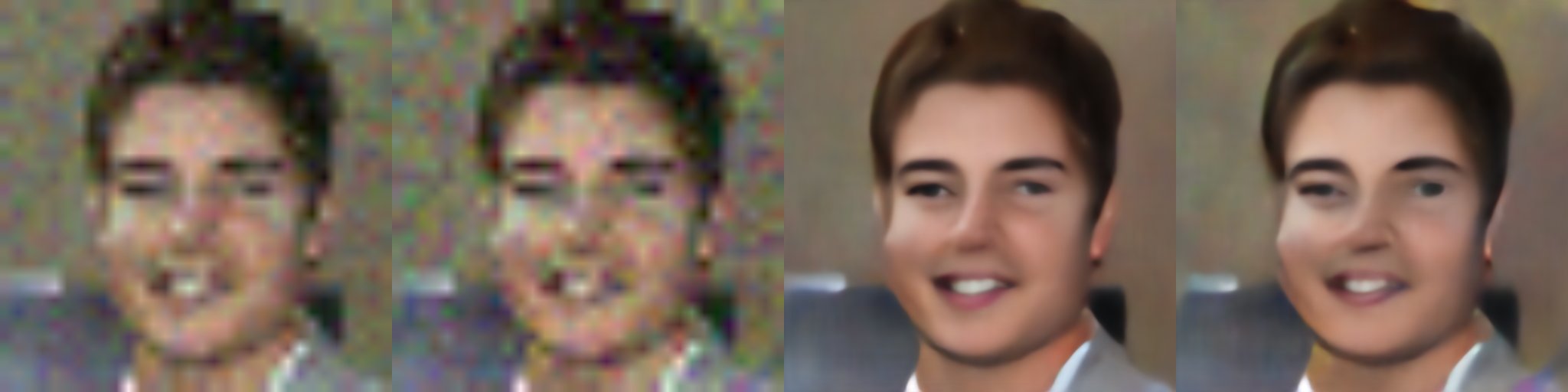}} \\
    \multicolumn{4}{l}{\hskip-0.19cm\includegraphics[width=\linewidth]{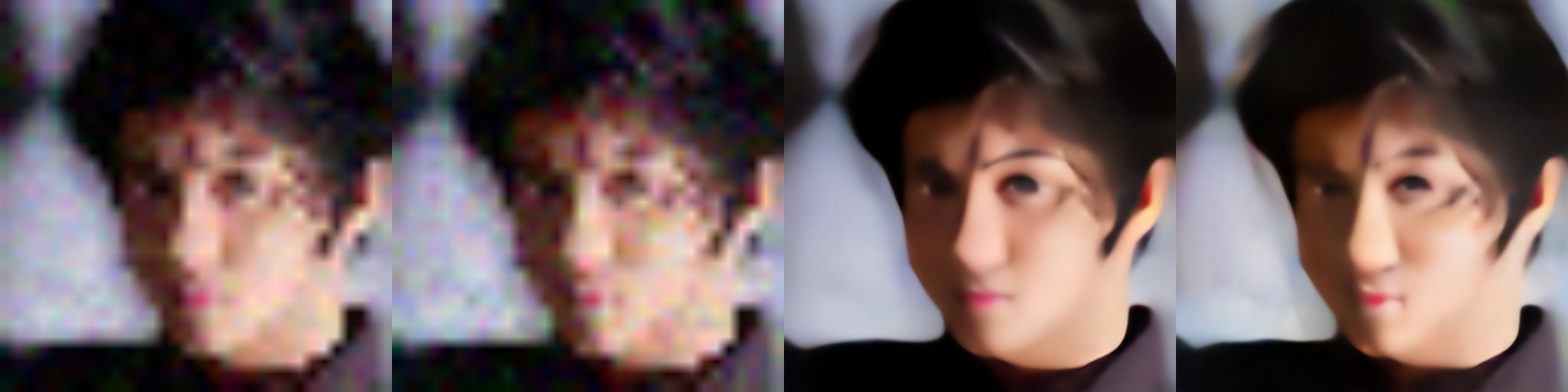}} \\
    \multicolumn{4}{l}{\hskip-0.19cm\includegraphics[width=\linewidth]{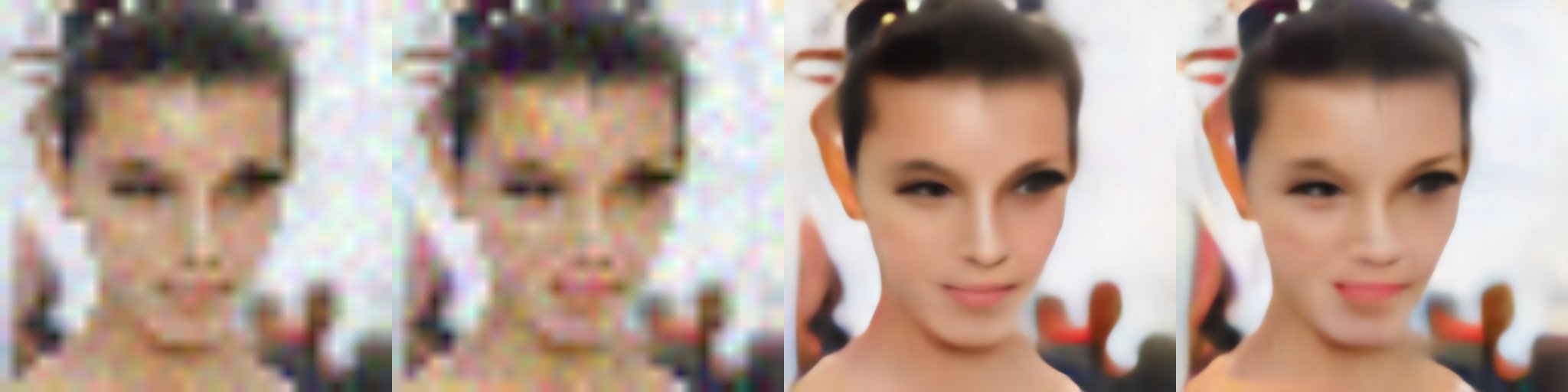}} \\
    \multicolumn{4}{l}{\hskip-0.19cm\includegraphics[width=\linewidth]{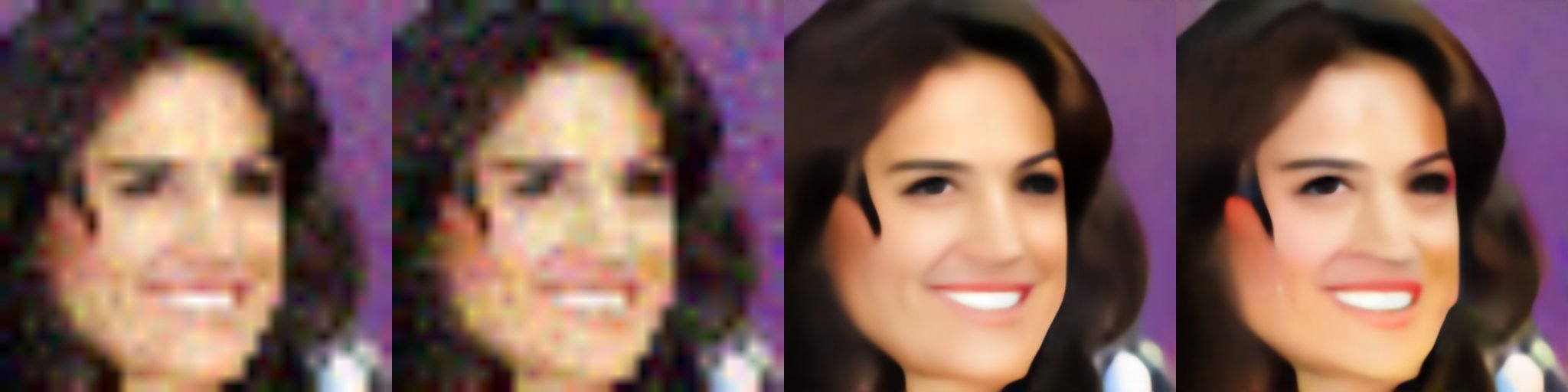}} \\
    \multicolumn{4}{l}{\hskip-0.19cm\includegraphics[width=\linewidth]{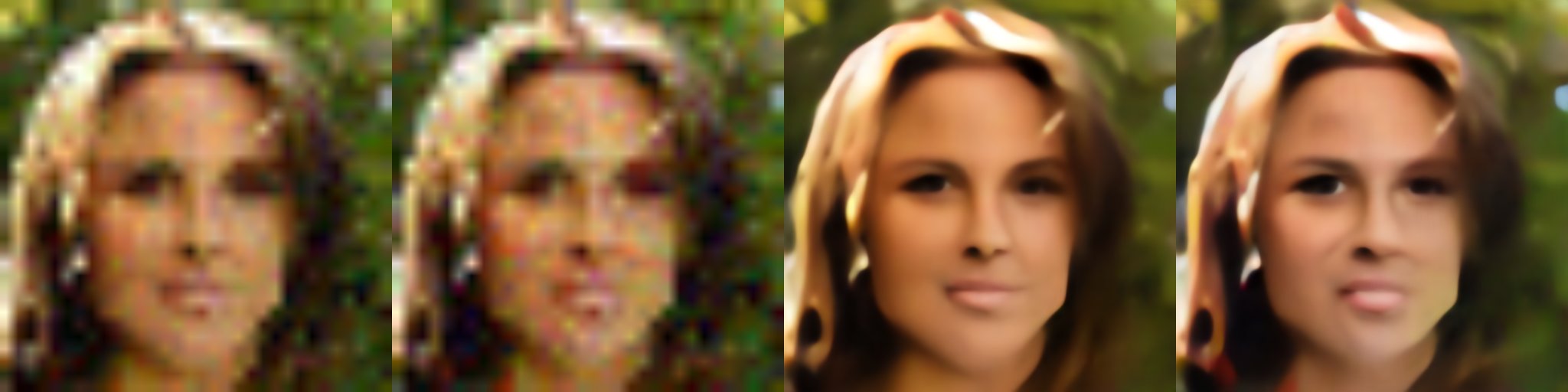}} \\
    \multicolumn{4}{l}{\hskip-0.19cm\includegraphics[width=\linewidth]{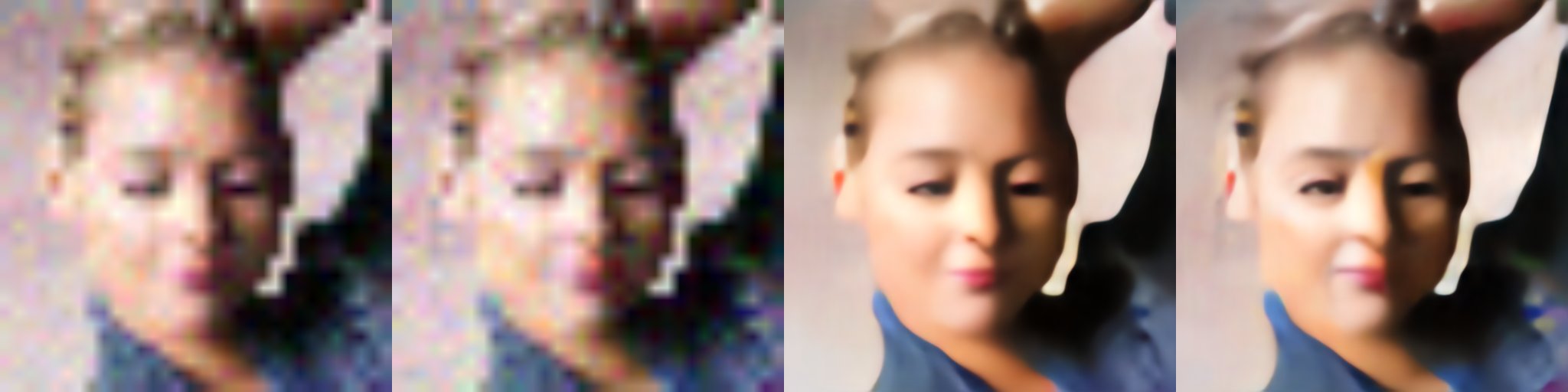}} \\
    \multicolumn{4}{l}{\hskip-0.19cm\includegraphics[width=\linewidth]{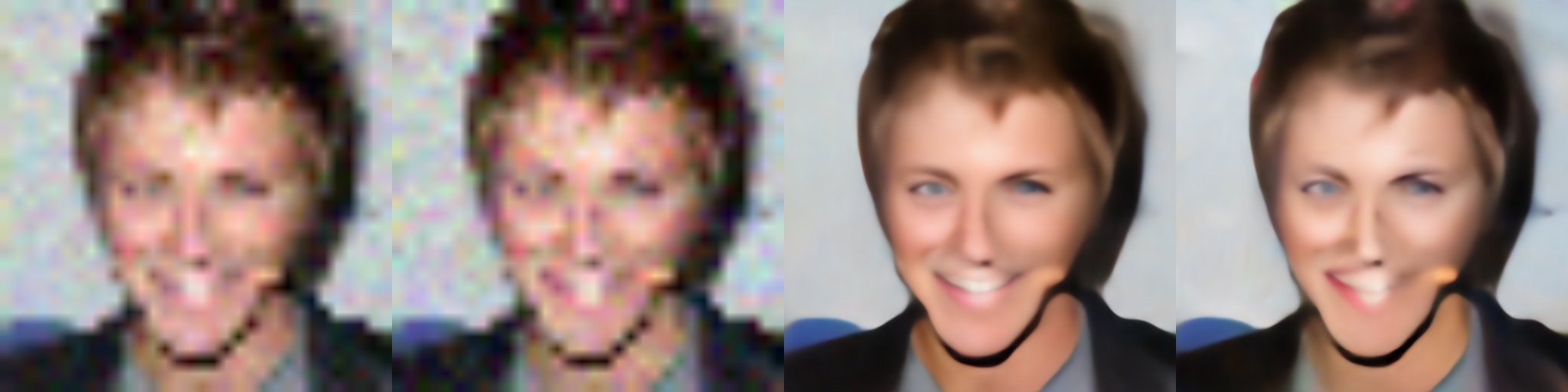}} \\
    \multicolumn{4}{l}{\hskip-0.19cm\includegraphics[width=\linewidth]{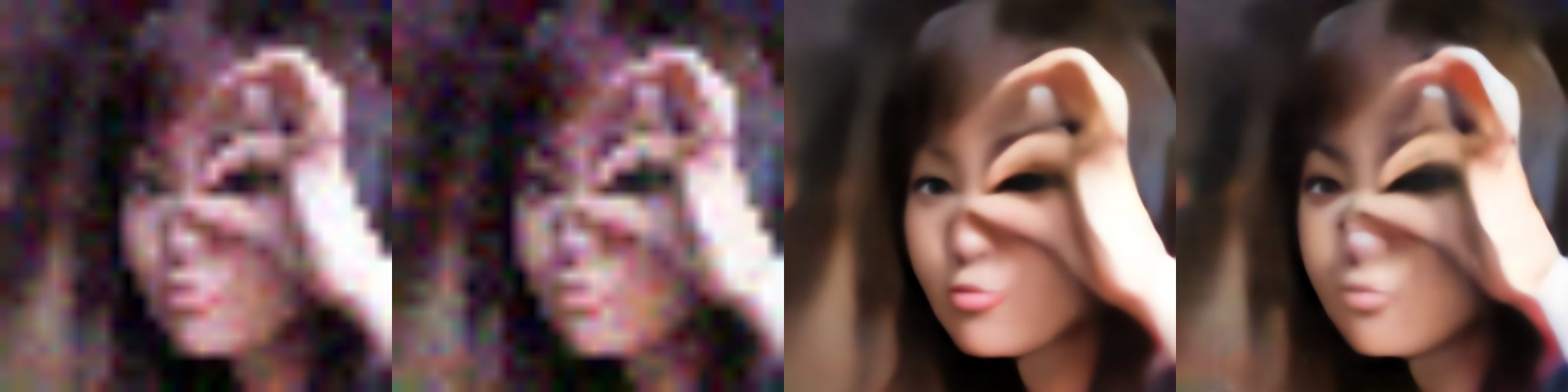}} \\
    \multicolumn{4}{l}{\hskip-0.19cm\includegraphics[width=\linewidth]{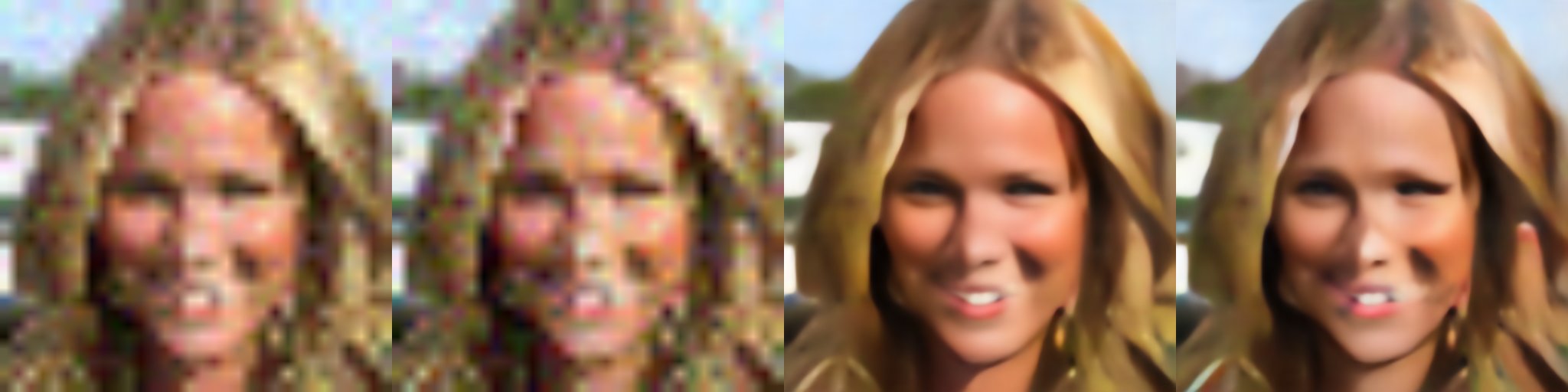}} \\
    \multicolumn{4}{l}{\hskip-0.19cm\includegraphics[width=\linewidth]{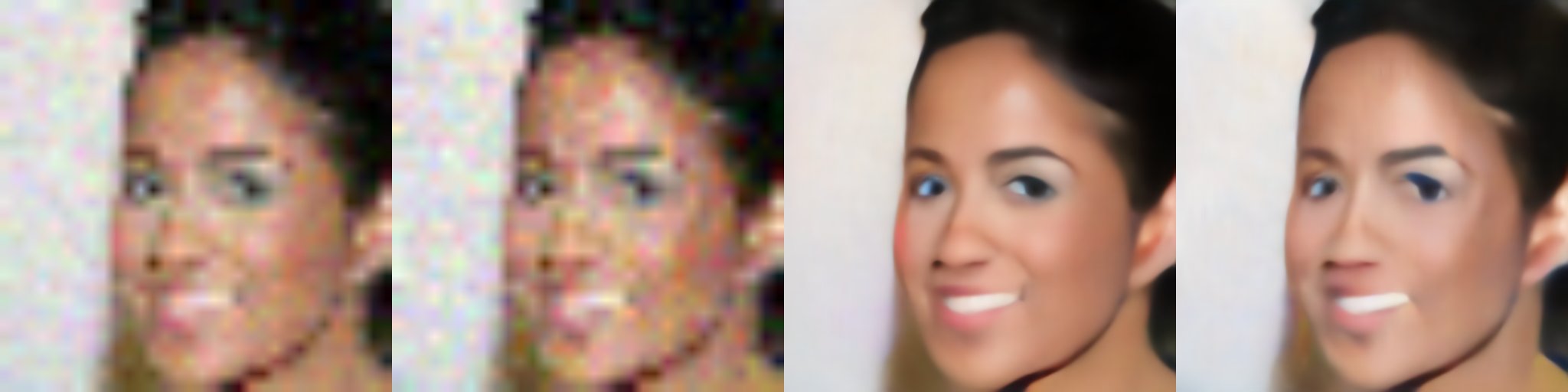}} \\
    \multicolumn{4}{l}{\hskip-0.19cm\includegraphics[width=\linewidth]{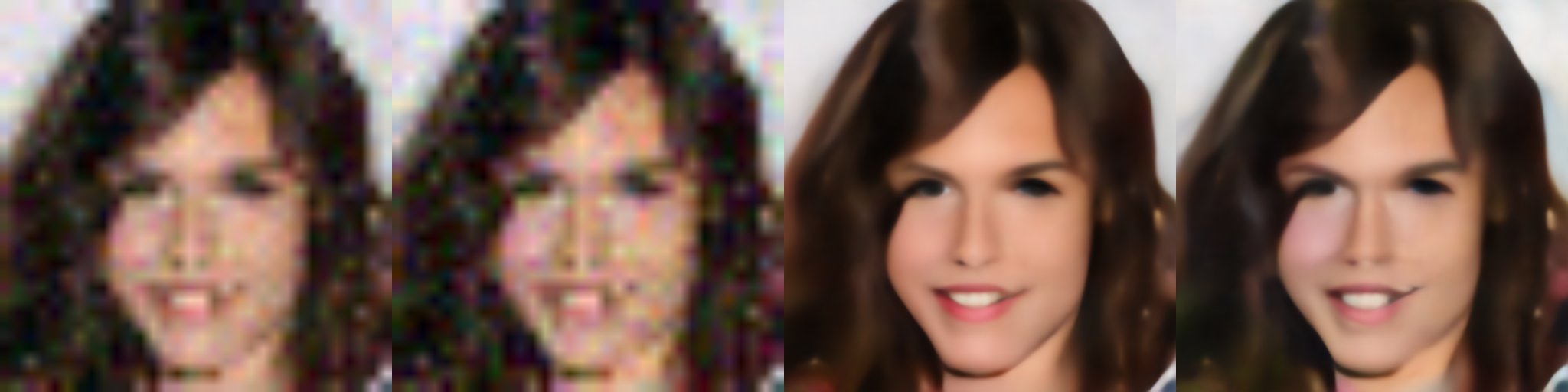}} \\
    \multicolumn{4}{l}{\hskip-0.19cm\includegraphics[width=\linewidth]{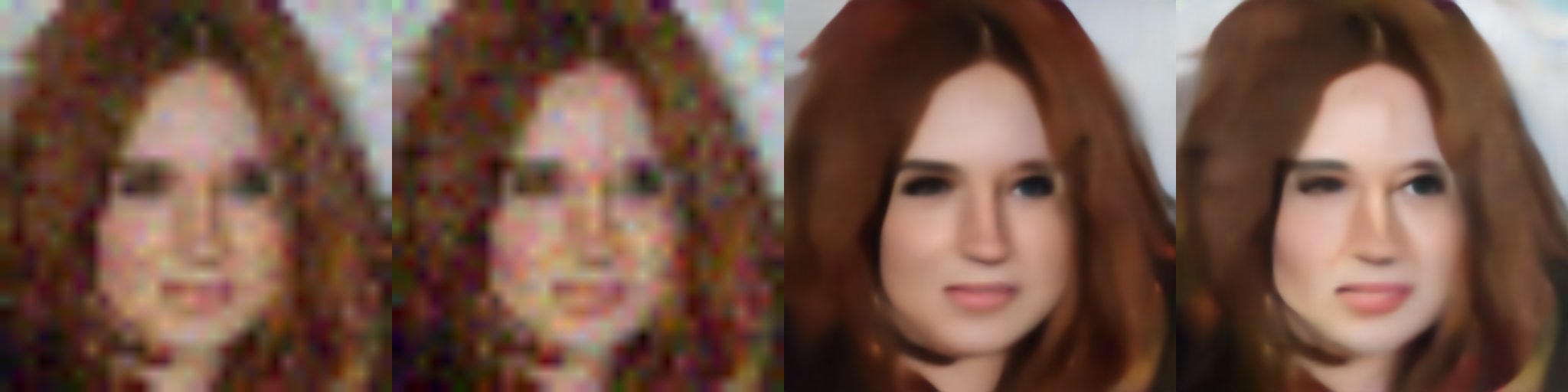}} \\
    \end{tabular}
\end{subfigure}
    \caption{Adversarial attacks on low-resolution face images intended to alter the output of the RRDB~\cite{wang2018esrgan} model to produce a female face.
    \textbf{Left}: Successful results, where $f(y)$ is classified as ``male'' and $f(y_{adv})$ is classified as ``female'' by the gender classification model we use for evaluation. The attacks successfully change the classified gender of 5.3\% of the images from ``male'' to ``female''.
    \textbf{Right}: Unsuccessful results, where $f(y)$ is classified as ``female'' and $f(y_{adv})$ is classified as ``male''. 7.5\% of the images that are originally classified as ``female'' are being classified as ``male'' after the adversarial attack.
    }
    \label{fig:rrdb_gender_switch}
\end{figure}

In~\cref{fig:gfpgan_gender_switch} and~\cref{fig:rrdb_gender_switch} we present more output examples of the gender adversarial attacks performed on GFPGAN and the trained RRDB model, respectively.
Note that the outputs of the RRDB model are blurry, as expected from a restoration model with high PSNR, so one might question whether the face gender prediction models we use in~\cref{section:adv_attacks} (for attacking the model and for evaluation) can even properly handle such images.
First, notice that the probability of the ``female'' class is almost identical for the source images (61.2\%) and the images reconstructed by the RRDB model (59.1\%).
To verify that this is not a coincidence, we compute the fraction of restored images whose class, as determined by the gender prediction model, is the same class as that of the corresponding source images.
We discover that the result is above 90\% for both of the face gender prediction models we use. Hence, according to the face gender classifiers, the RRDB model almost always preserves the true gender of the source image.

Moreover, note that the fraction of the ``female'' class for the attacked RRDB output images is \emph{lower} than that of the original restored images.
This means that some of the output images produced by RRDB are originally classified as ``female'', but their class is switched to ``male'' after performing the adversarial attack, even though the attack is intended to make the opposite effect.
In~\cref{fig:rrdb_gender_switch} we show several examples where the original class is ``female'' and the class after the attack is ``male'', and vice versa.
Subjectively, we do not see any bold qualitative reason for such an unexpected outcome.
It is perhaps related to the sensitivity of classification models on examples where the class probabilities are almost identical, e.g., when the probabilities of both classes (in a binary classification problem) are around 0.5.
We leave the study of this phenomenon for future work.
\subsection{Adversarial Attacks to Produce ``Old''-Classified Faces}\label{appendix:aging_attacks}
Like in the gender adversarial attacks, we perform similar attacks to alter the outputs of GFPGAN and RRDB to produce faces of older people.
To do so, we use the same adversarial attack procedure as described in~\cref{section:adv_attacks}, but instead of maximizing the log-probability of the class ``female'', we maximize the log-probability of the class that corresponds to the oldest age category in a facial age classification model.
Specifically, we maximize the log-probability of the 70+ age category as predicted by the model in~\cite{nate_raw_2023} (this model was trained on the FairFace data set~\cite{karkkainenfairface}).
Like in~\cref{section:adv_attacks}, we use here $\alpha=16/255$ and $T=100$ in the I-FGSM update rule as well.
We use the age prediction model in~\cite{serengil2021lightface} to evaluate the both GFPGAN and RRDB, which predicts the age as a scalar.

\begin{figure}
\centering
\begin{subfigure}{.37\textwidth}
  \centering
\renewcommand{\arraystretch}{0.0}
    \begin{tabular}{P{0.185\linewidth}P{0.185\linewidth}P{0.185\linewidth}P{0.185\linewidth}}
    {\small $y$} & 
    {\small $y_{adv}$} & 
    {\small $f(y)$} & 
    {\small $f(y_{adv})$} \\
    \vspace*{0.1cm} \ & & & \\
    \multicolumn{4}{l}{\hskip-0.19cm\includegraphics[width=\linewidth]{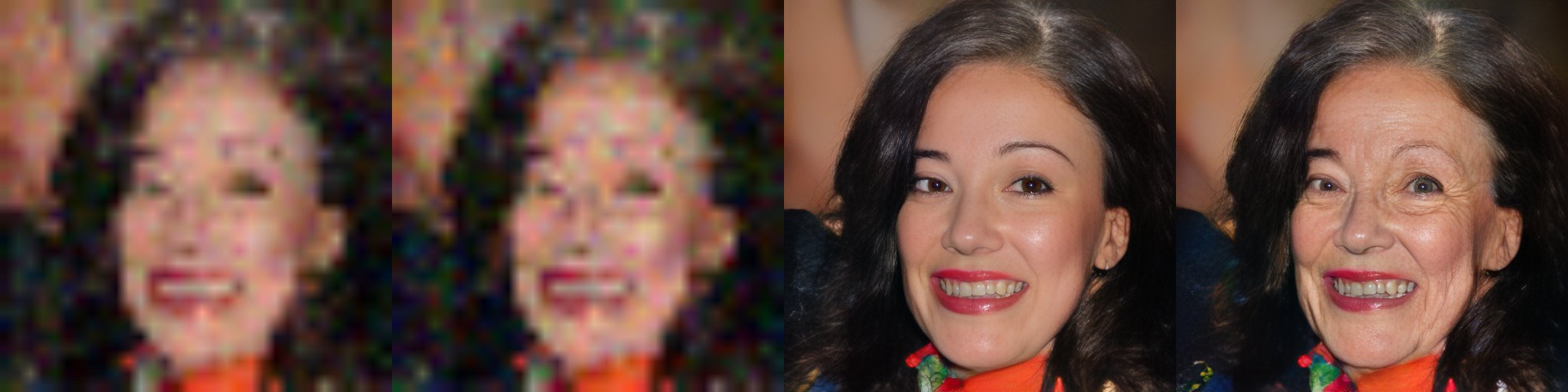}} \\
    \multicolumn{4}{l}{\hskip-0.19cm\includegraphics[width=\linewidth]{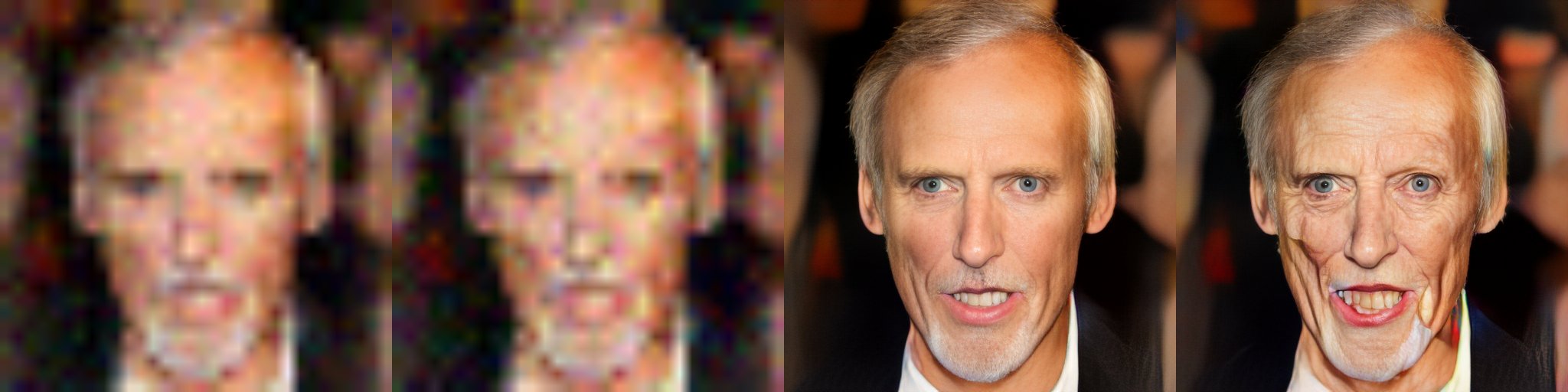}} \\
    \multicolumn{4}{l}{\hskip-0.19cm\includegraphics[width=\linewidth]{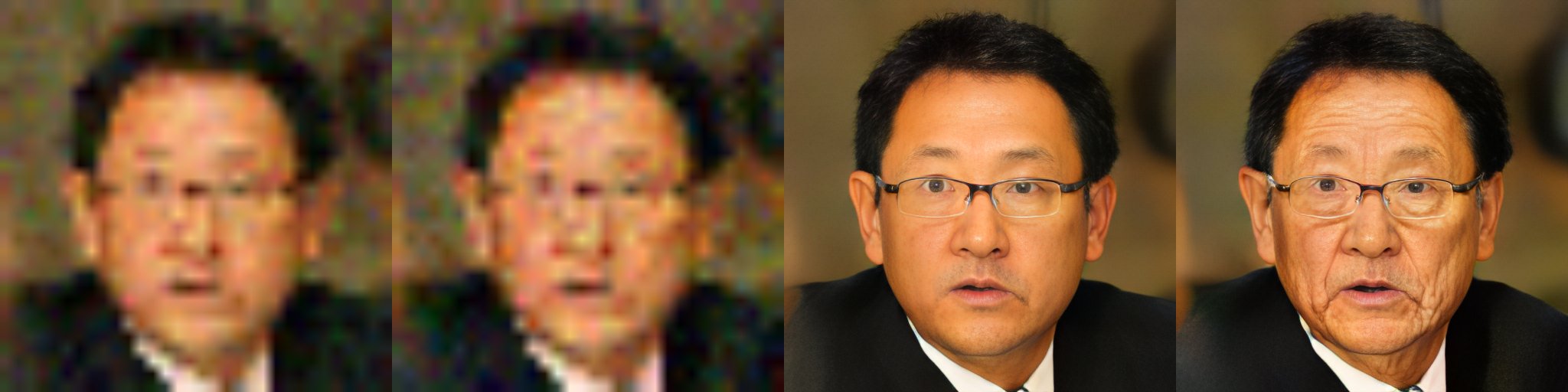}} \\
    \multicolumn{4}{l}{\hskip-0.19cm\includegraphics[width=\linewidth]{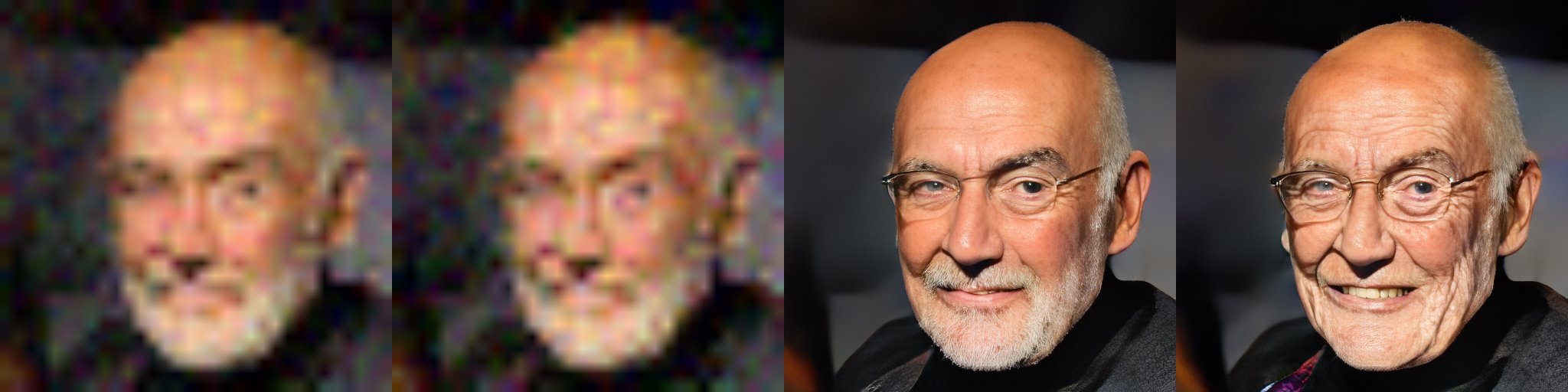}} \\
    \multicolumn{4}{l}{\hskip-0.19cm\includegraphics[width=\linewidth]{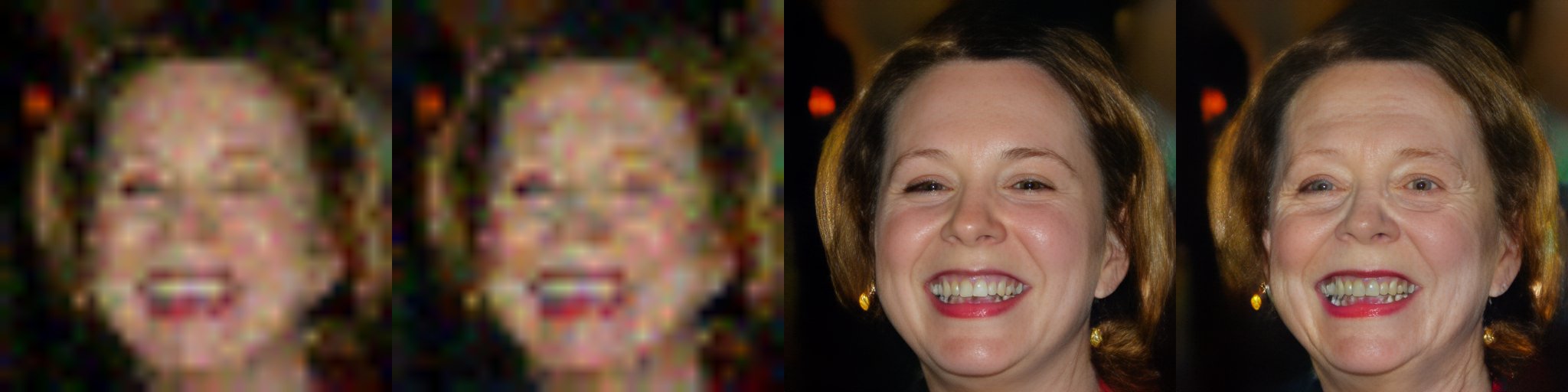}} \\
    \multicolumn{4}{l}{\hskip-0.19cm\includegraphics[width=\linewidth]{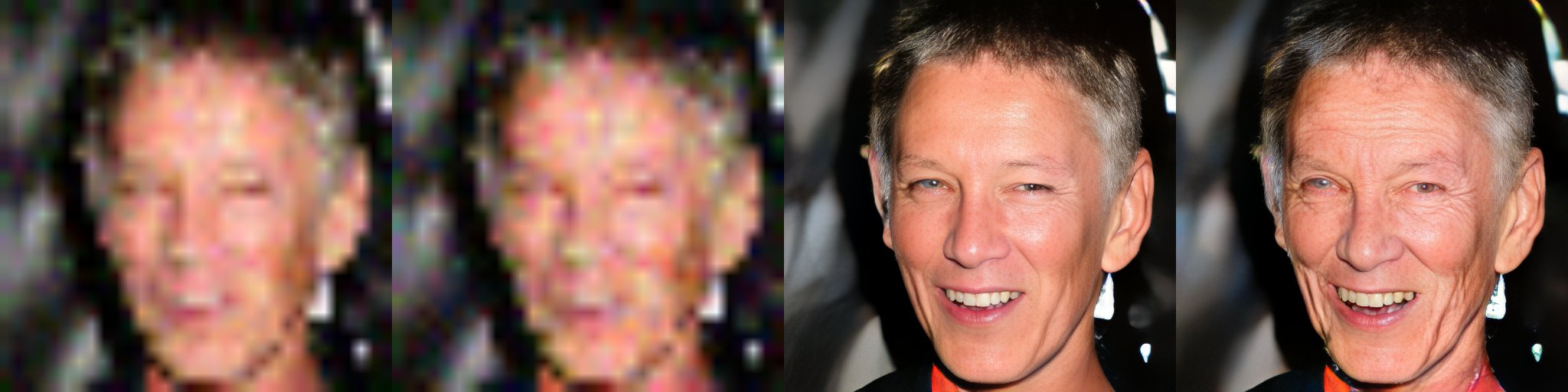}} \\
    \multicolumn{4}{l}{\hskip-0.19cm\includegraphics[width=\linewidth]{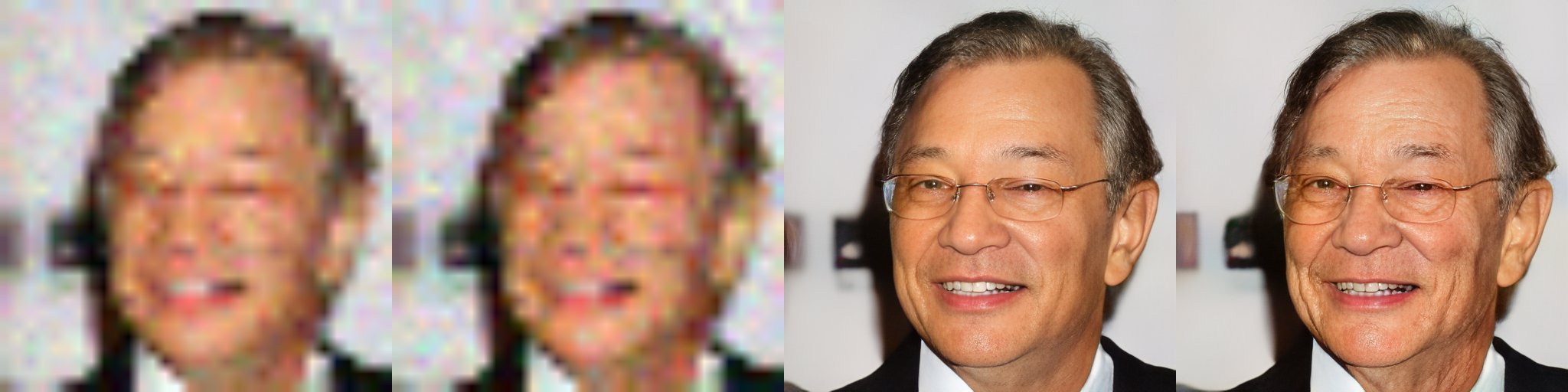}} \\
    \multicolumn{4}{l}{\hskip-0.19cm\includegraphics[width=\linewidth]{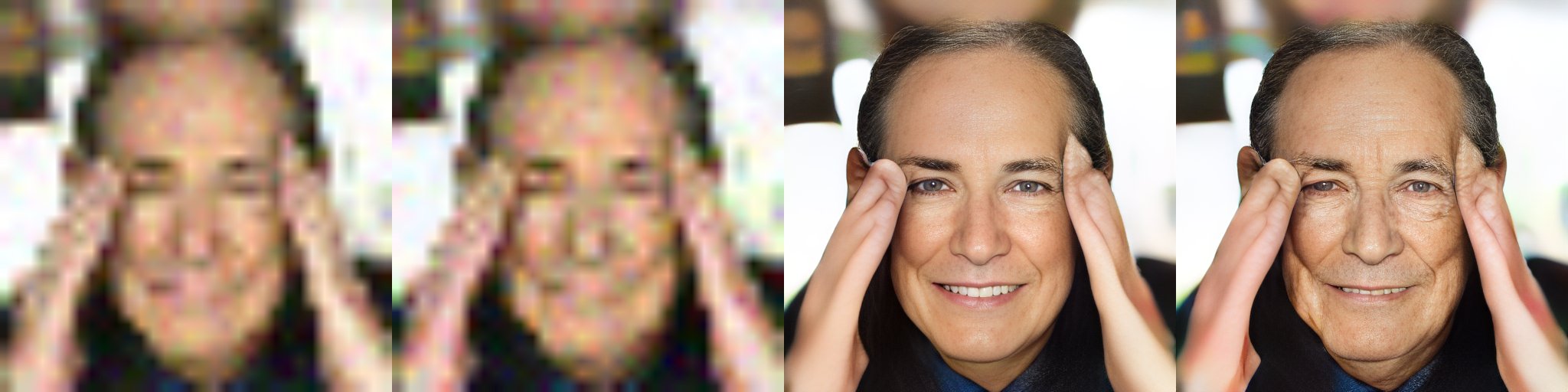}} \\
    \multicolumn{4}{l}{\hskip-0.19cm\includegraphics[width=\linewidth]{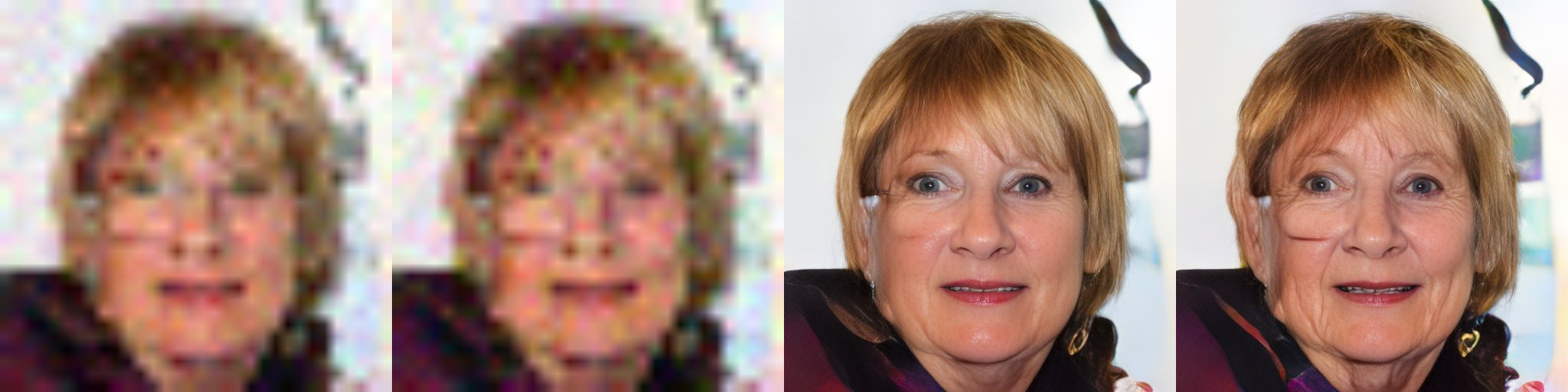}} \\
    \multicolumn{4}{l}{\hskip-0.19cm\includegraphics[width=\linewidth]{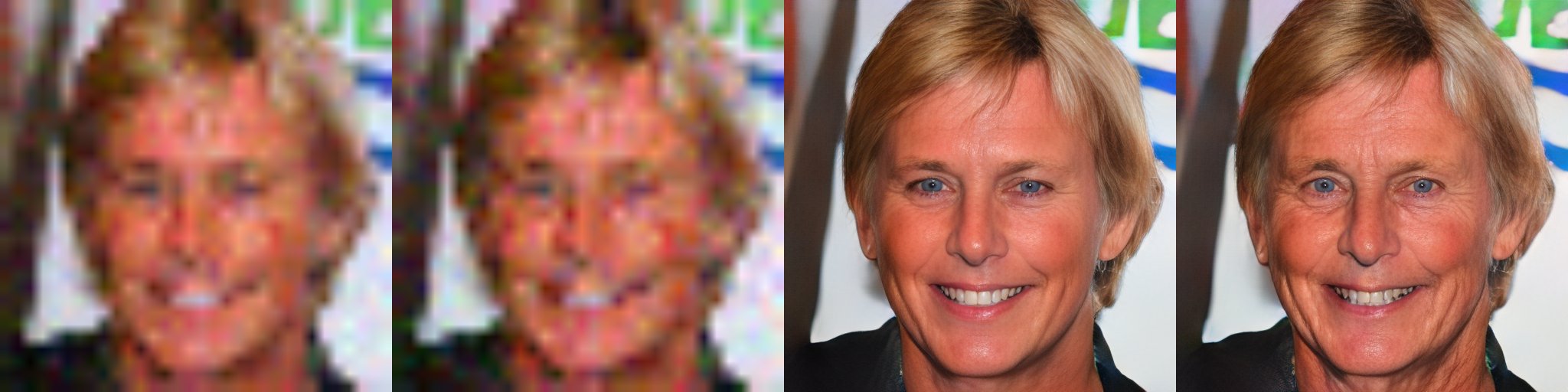}} \\
    \multicolumn{4}{l}{\hskip-0.19cm\includegraphics[width=\linewidth]{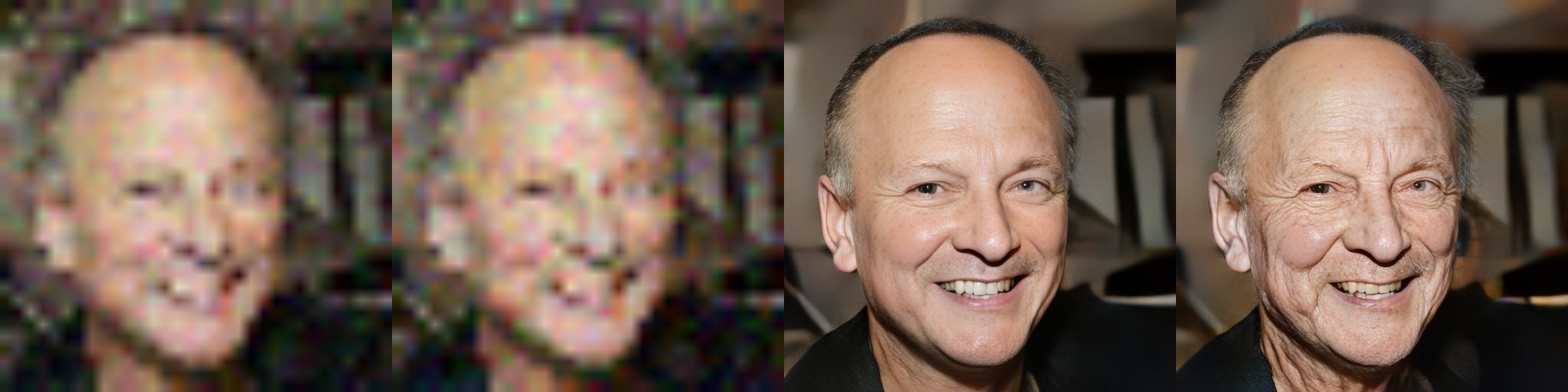}} \\
    \multicolumn{4}{l}{\hskip-0.19cm\includegraphics[width=\linewidth]{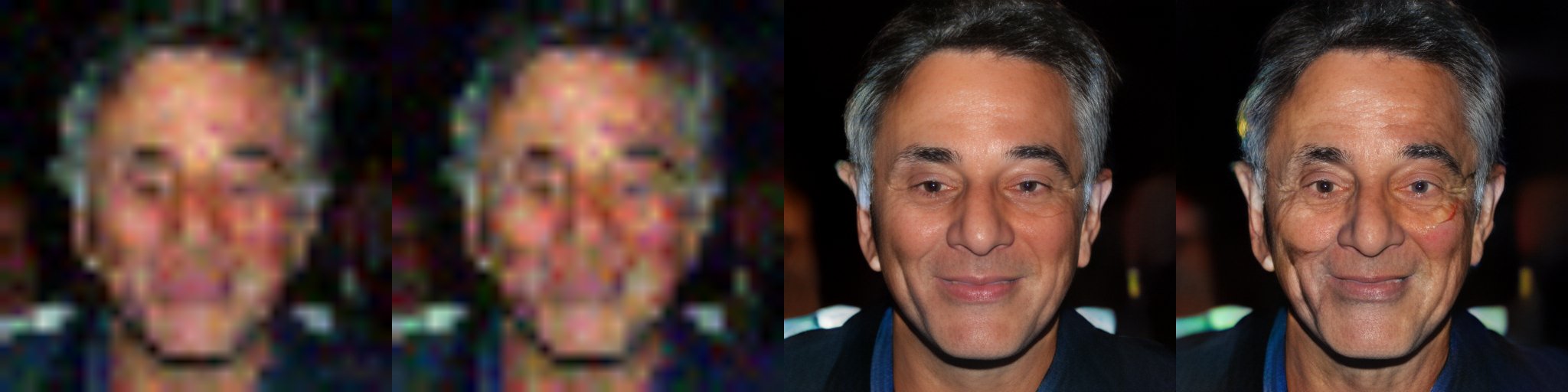}} \\
\end{tabular}
\end{subfigure}\hspace*{2cm}
\begin{subfigure}{.37\textwidth}
  \centering
  \renewcommand{\arraystretch}{0.0}
    \begin{tabular}{P{0.185\linewidth}P{0.185\linewidth}P{0.185\linewidth}P{0.185\linewidth}}
    {\small $y$} & 
    {\small $y_{adv}$} & 
    {\small $f(y)$} & 
    {\small $f(y_{adv})$} \\
    \vspace*{0.1cm} \ & & & \\
    \multicolumn{4}{l}{\hskip-0.19cm\includegraphics[width=\linewidth]{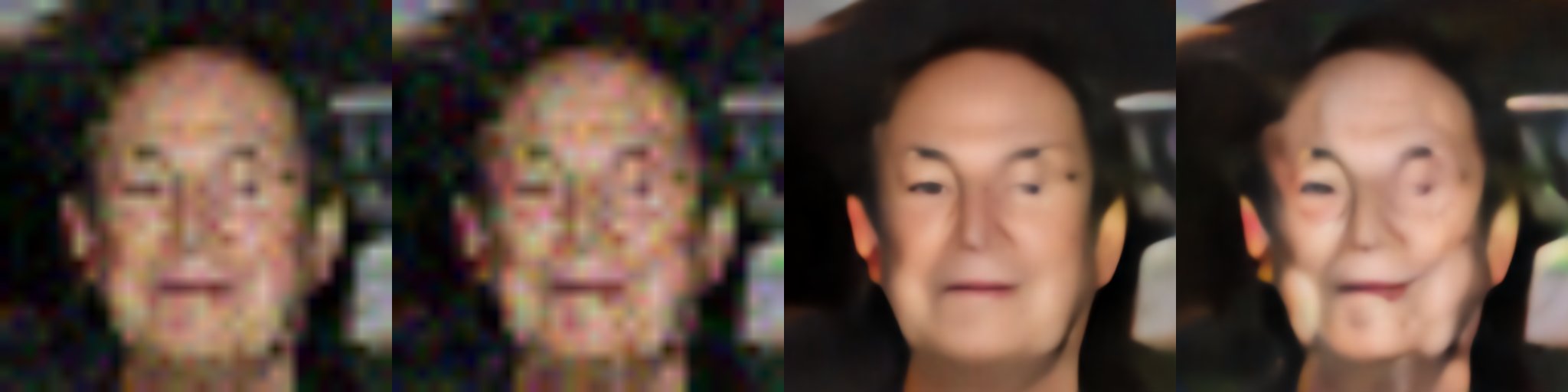}} \\
    \multicolumn{4}{l}{\hskip-0.19cm\includegraphics[width=\linewidth]{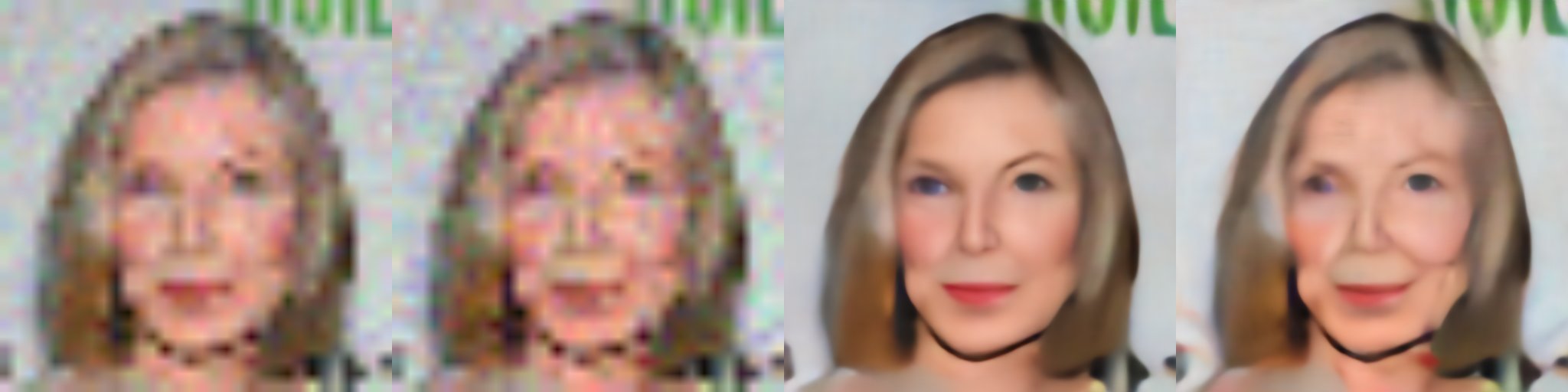}} \\
    \multicolumn{4}{l}{\hskip-0.19cm\includegraphics[width=\linewidth]{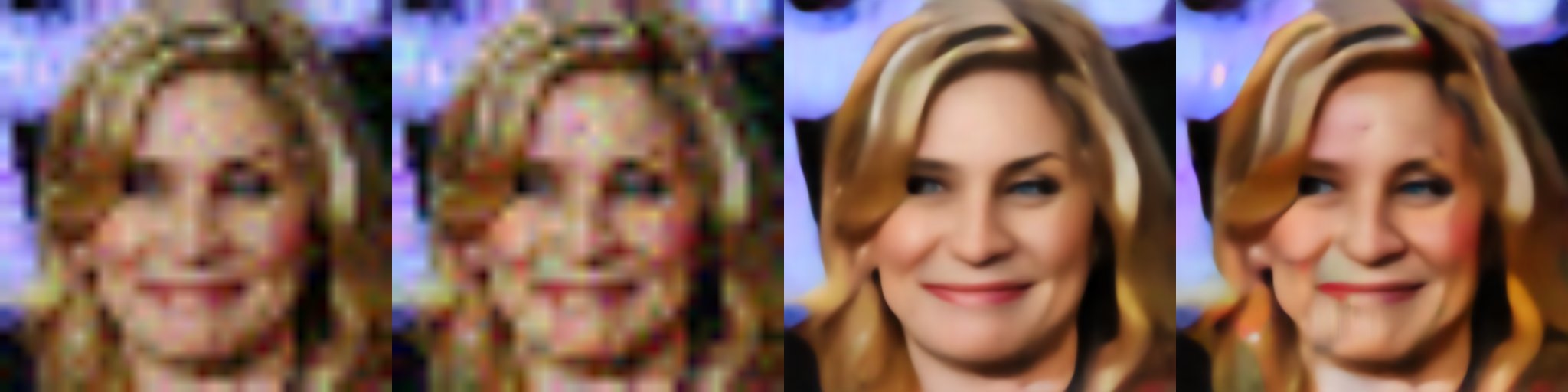}} \\
    \multicolumn{4}{l}{\hskip-0.19cm\includegraphics[width=\linewidth]{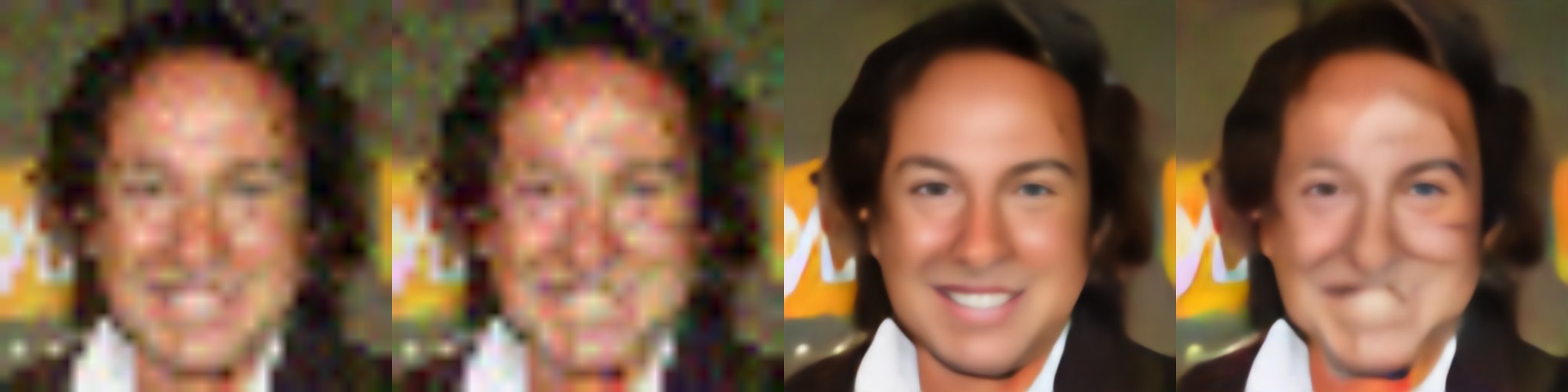}} \\
    \multicolumn{4}{l}{\hskip-0.19cm\includegraphics[width=\linewidth]{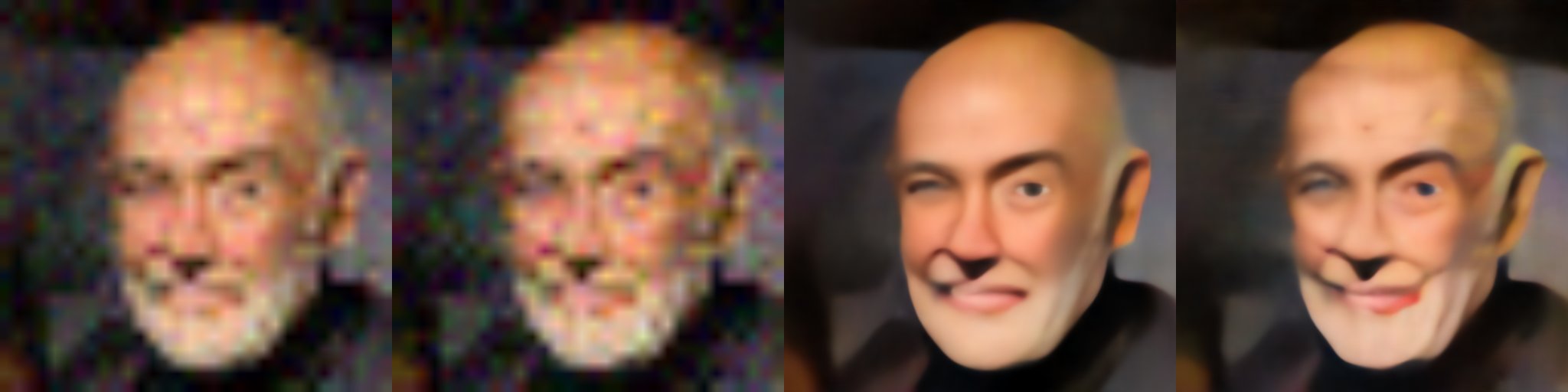}} \\
    \multicolumn{4}{l}{\hskip-0.19cm\includegraphics[width=\linewidth]{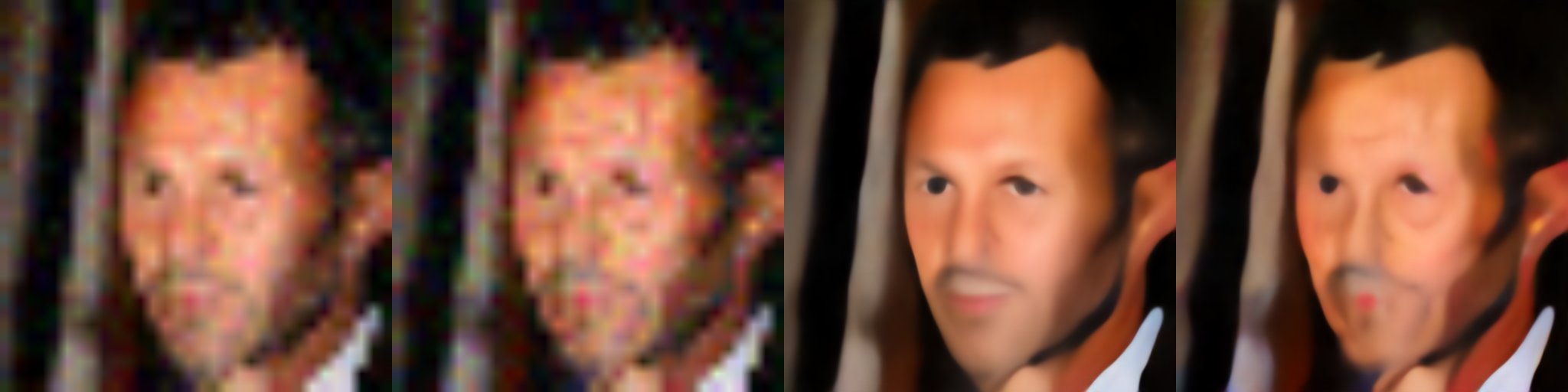}} \\
    \multicolumn{4}{l}{\hskip-0.19cm\includegraphics[width=\linewidth]{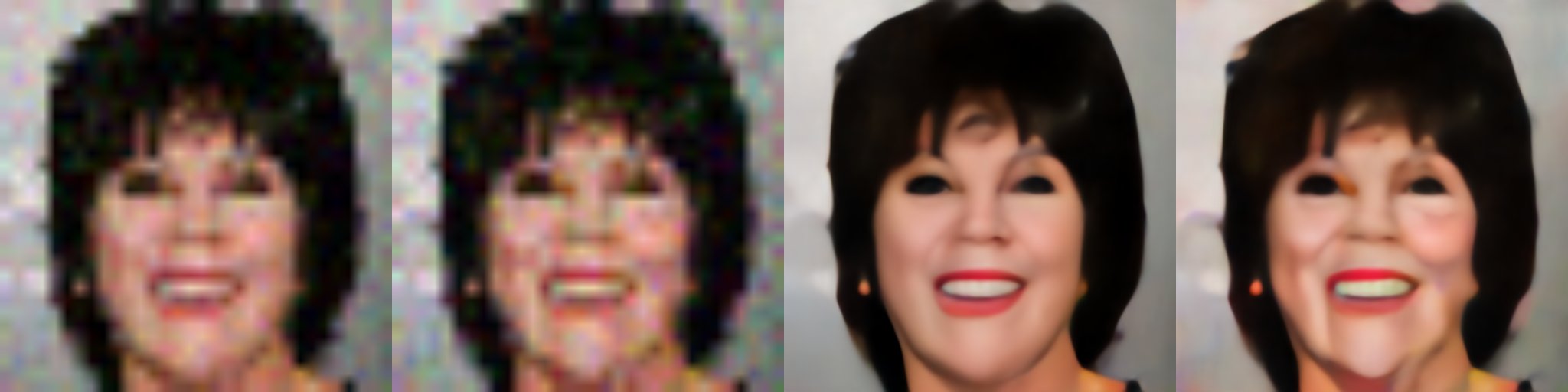}} \\
    \multicolumn{4}{l}{\hskip-0.19cm\includegraphics[width=\linewidth]{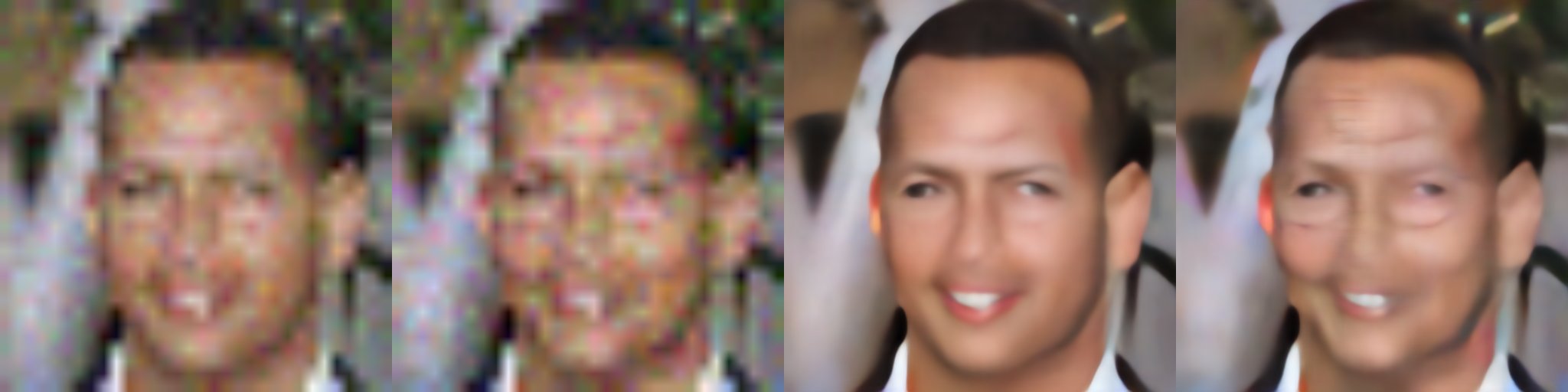}} \\
    \multicolumn{4}{l}{\hskip-0.19cm\includegraphics[width=\linewidth]{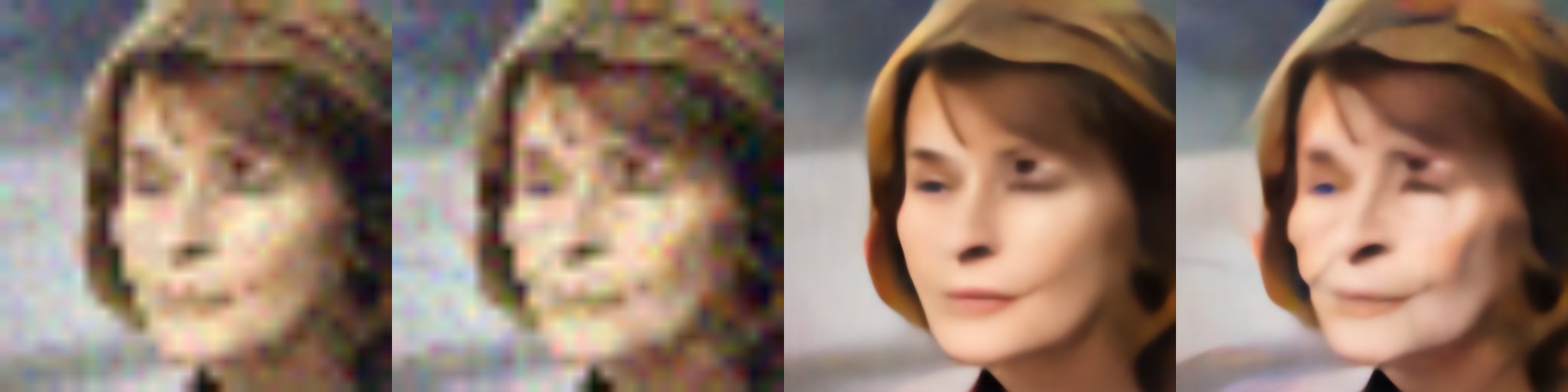}} \\
    \multicolumn{4}{l}{\hskip-0.19cm\includegraphics[width=\linewidth]{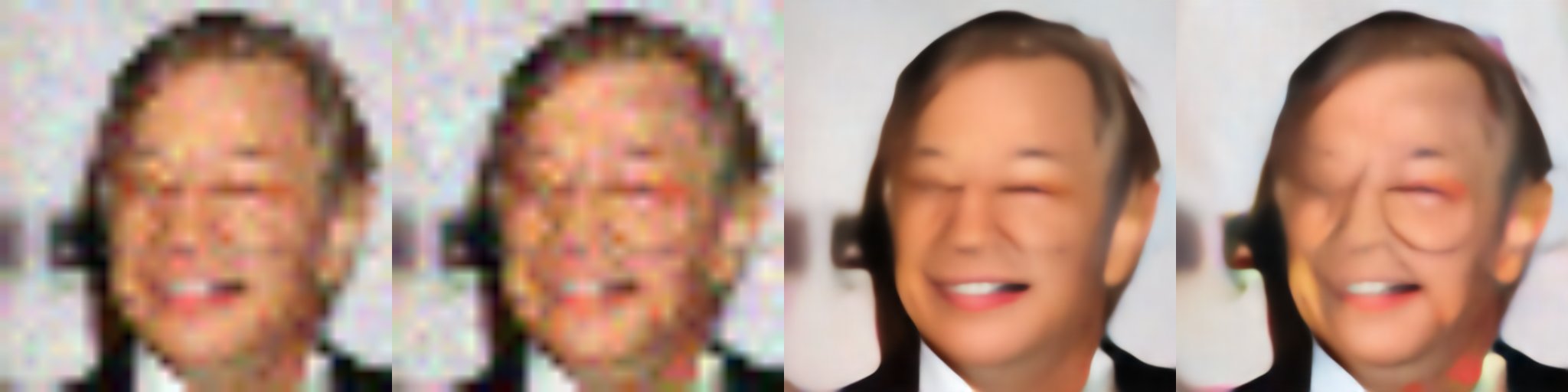}} \\
    \multicolumn{4}{l}{\hskip-0.19cm\includegraphics[width=\linewidth]{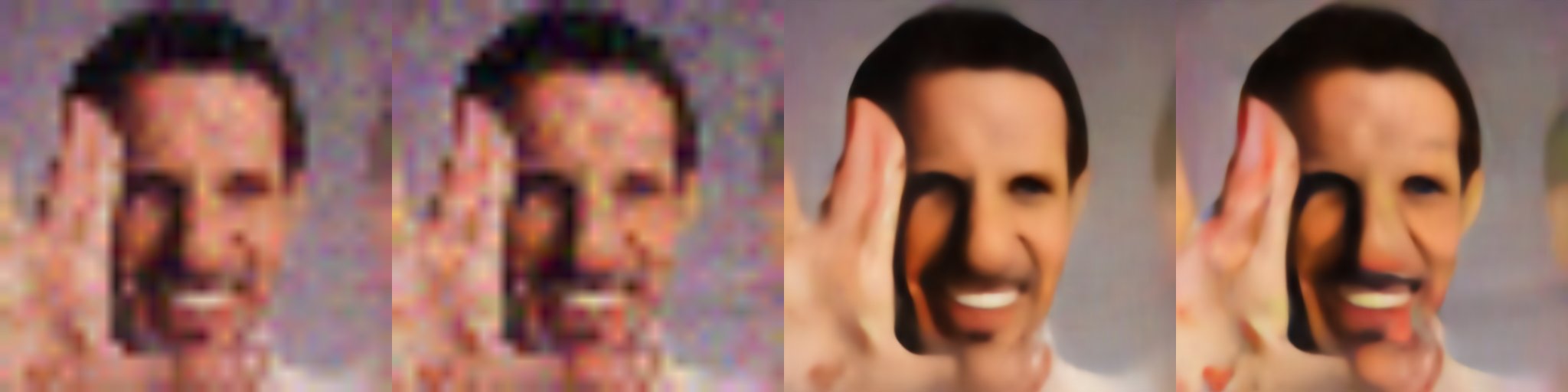}} \\
    \multicolumn{4}{l}{\hskip-0.19cm\includegraphics[width=\linewidth]{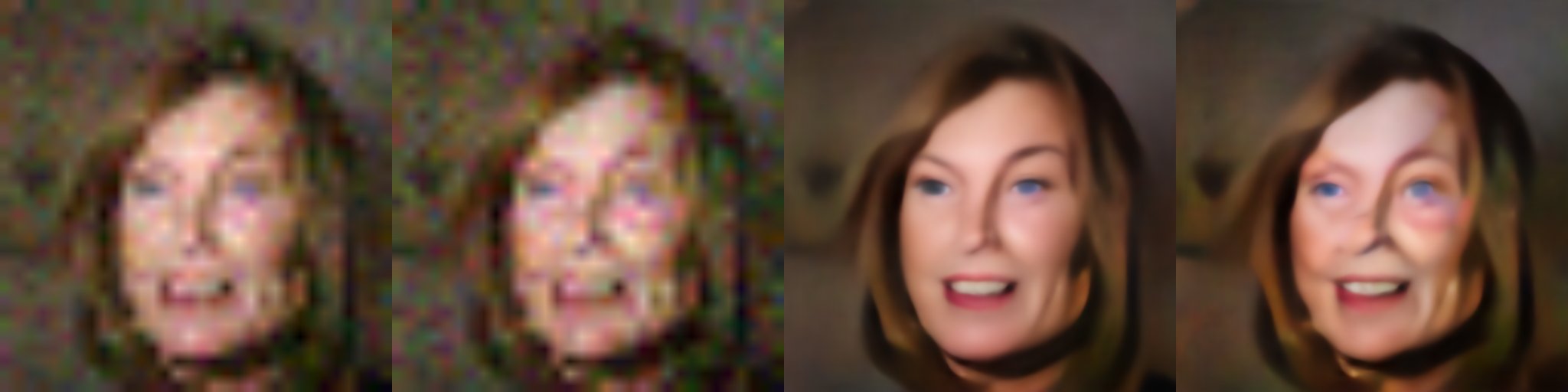}} \\
    \end{tabular}
\end{subfigure}
    \caption{Adversarial attacks on low-resolution face images intended to alter the output of the GFPGAN and RRDB models to produce a face of an older person.
    \textbf{Left}: Results using the GFPGAN model.
    \textbf{Right}: Results using the RRDB model.}
    \label{fig:age_attacks}
\end{figure}

We present visual results of both GFPGAN and RRDB in~\cref{fig:age_attacks}.
The input and output images we choose to present for each model correspond to the 12 attacked output images $f(y_{adv})$ with the oldest predicted age according to the evaluation model.
The attacked outputs of GFPGAN contains features that may be associated with older people (e.g., wrinkles).
It seems that RRDB attempts to produce the same effect but is unsuccessful due to its lower perceptual quality, and perhaps its lower sensitivity to such adversarial attacks.
In~\cref{tab:quantitative_age_attacks} we confirm these results quantitatively.
\begin{table}[h]
    \centering
    \begin{tabular}{c|c|c|c|c|c}
         & Source images & GFPGAN & \makecell{Attacked\\GFPGAN}& RRDB & \makecell{Attacked\\RRDB} \\
         \hline
         Predicted age average&32.7&32.3&38.8&29.98&33.87 \\
         Predicted age standard deviation&6.6&5.9&8.9&4&5.4 \\
         Maximum predicted age&60&62&66&48&53 \\
    \end{tabular}
    \caption{Quantitative analysis of the adversarial attacks described in~\cref{appendix:aging_attacks}, intended to alter GFPGAN and RRDB to produce faces of older people.
    The average, standard deviation, and maximum are taken over all the 1000 test images chosen from CelebA-HQ.
    The columns denoted by GFPGAN and Attacked GFPGAN correspond to the original outputs of GFPGAN (originating from the original degraded inputs), and the attacked outputs of GFPGAN (originating from the attacked inputs), respectively.
    The same notation holds for RRDB and Attacked RRDB.}
    \label{tab:quantitative_age_attacks}
\end{table}
\newpage
\section{Additional Details of~\cref{section:posterior-sampling-imitation} (Exploring the Posterior Distribution)}\label{appendix:posterior-imitation-details}
\subsection{Why Should We Explore the Posterior Distribution?}
Exploring the posterior distribution may hold a practical significance in several scenarios.
First and foremost, it enables taking into account the uncertainty inherent to the restoration task at hand. For example, in CT-reconstruction from partial measurements, radiologists may make better-informed decisions if they have a clear understanding of the uncertainty of the images obtained. Having access to the posterior distribution, one can either quantify this uncertainty or enable visualizing it by traversing through different samples from the posterior. This would allow medical practitioners to assess the reliability of their diagnoses and potentially avoid misinterpretations or misdiagnoses. See~\citep{belhasin2024principal} for more details about such possibilities. More broadly, when there is a variety of possible solutions to our inverse problem, exploring the posterior gives the ability to the user to view the possible outcomes and choose the more favored solution.
\subsection{Supplementary Details on the Experiments}
\cref{algorithm:posterior-sampling-imitation} is the formal description of the FPS approach we use to explore the posterior distribution.
We perform $T=150$ I-FGSM update steps to obtain each sample, and choose $\alpha=8/255$ to be the scale of the adversarial attack.
With such a value of $\alpha$, the visual difference between $y$ and each attacked input is negligible (see, e.g., \cref{fig:emotion_adv_attack}, where we use $\alpha=16/255$).
The images we present in~\cref{fig:posterior-imitation} are the $S=5$ samples resulting from the algorithm.
Notice that the first sample resulting from the algorithm is simply the output originating from the original input $y$, without performing any attack.
\begin{algorithm}
\caption{Farthest Point Sampling approach to explore the posterior distribution with a deterministic estimator \mbox{$\hat{X}=f(Y)$} that attains high joint perceptual quality.}\label{algorithm:posterior-sampling-imitation}
\begin{algorithmic}
\STATE {\bfseries Input:} Degraded measurement $y$, number of output samples $S$, adversarial attack scale $\alpha$, number of adversarial attack update steps $T$.

\STATE {\bfseries Output:} List of samples $\mathcal{X}=[\hat{x}_{1},\hat{x}_{2},\hdots,\hat{x}_{S}]$.
 
\STATE $\hat{x} \gets f(y)$ \algorithmiccomment{First output of the estimator.}

\STATE $\mathcal{X} \gets [\hat{x}]$

\FOR{$i=1$ {\bfseries to} $S$}

\STATE $y_{adv}\gets y$

\FOR{$t=1$ {\bfseries to} $T$}

\STATE $\hat{x}_{t}\gets f(y_{adv})$
 
\STATE $\mathcal{L}=\frac{1}{i}\sum_{s=1}^{i}\norm{\mathcal{X}[s]-\hat{x}_{t}}_{2}^{2}$\;

\STATE $y_{adv}\gets \text{I-FGSM}(\mathcal{L}, \alpha, y)$ \algorithmiccomment{The basic I-FGSM update rule in~\cite{Choi_2019_ICCV}, but with our loss $\mathcal{L}$.}
\ENDFOR
\ENDFOR

\STATE $\mathcal{X}\gets \mathcal{X}\cup \{f(y_{adv})\}$ \algorithmiccomment{Append the new output to the list of output samples.}
\end{algorithmic}
\end{algorithm}


\end{document}